\documentclass[11pt]{article}
\usepackage{changes}
\usepackage{hyperref}
\usepackage{amsmath, amsfonts, amssymb, mathrsfs}
\usepackage{dcolumn}
\usepackage{caption}
\usepackage{subcaption}
\usepackage{filemod}
\usepackage{natbib}
\usepackage{floatrow}
\usepackage{setspace}
\usepackage{verbatim}
\usepackage{graphicx}
\usepackage{url}
\usepackage{tikz}
\usepackage{bbm}
\usepackage{algorithm}
\usepackage{algpseudocode}
\usepackage{graphbox}
\usepackage{makecell}
\usetikzlibrary{shapes.geometric, arrows}

\title{\vspace{-0.3cm} Modular Jump Gaussian Processes}
\author{Anna R. Flowers\thanks{Corresponding author: Department of Statistics, 
Virginia Tech, \texttt{arflowers@vt.edu} }
\and Christopher T. Franck\thanks{Department of Statistics, Virginia Tech}
\and Micka\"{e}l Binois\thanks{Universit\'{e} C\^{o}te d’Azur, Inria, CNRS, LJAD, France}
\and Chiwoo Park\thanks{Department of Industrial and Systems Engineering, University of Washington}
\and Robert B. Gramacy\footnotemark[2]}
\date{\today}

\begin{document}

\vspace{-0.5cm}
\maketitle

\singlespacing

\vspace{-1.5cm}
\begin{abstract} 
Gaussian processes (GPs) furnish
accurate nonlinear predictions with well-calibrated uncertainty. However, the
typical GP setup has a built-in stationarity assumption, making it ill-suited
for modeling data from processes with sudden changes, or ``jumps'' in the output 
variable. 
The ``jump GP'' (JGP) was developed for modeling data from such processes,
combining local GPs and latent ``level'' variables under a joint inferential
framework.  But joint modeling can be fraught with difficulty.
We aim to simplify by suggesting a more modular setup, eschewing joint
inference but retaining the main JGP themes: (a) learning optimal neighborhood
sizes that locally respect manifolds of discontinuity; and (b) a new
cluster-based (latent) feature to capture regions of distinct output levels
on both sides of the manifold. We show that each of (a) and (b) separately
leads to dramatic improvements when modeling processes with jumps. In tandem
(but without requiring joint inference) that benefit is compounded, as
illustrated on real and synthetic benchmark examples from the recent
literature.
\end{abstract}

\noindent \textbf{Keywords:}  computer experiment, nonstationarity, latent variable, local neighborhoods, mixture model

\singlespacing

\section{Introduction}\label{sec:1}

We consider processes $Y(x): \mathbb{R}^d \rightarrow \mathbb{R}$ that are
smooth for most of the surface except along a manifold of discontinuity which
spatially\footnote{Throughout the paper, we use ``spatial'' with
reference to the input space, not ``geospatial'' locations on Earth.} partitions
the response $Y(x)$ into two levels depending on $x$. One example of such a
process is a computer simulation of an automated material handling
vehicle (AMHV) transport system measuring the average transport time $Y(x)$
of material transport devices given their characteristics $x$
\citep{park2023active}. Usually, transport time is  low, but when the devices
are working at or near capacity transport time suddenly increases -- it
``jumps''. Other examples of jump dynamics exist in geostatistics
\citep{kim2005analyzing}, physiology \citep{vieth1989fitting}, ecology
\citep{toms2003piecewise}, and materials science \citep{park2022sequential}.
We narrow our focus to jumps exhibited by computer simulation experiments, where
$Y(x)$ models complex phenomena as a deterministic function of inputs $x$ at
potentially great computational expense, although we expect our idea to 
be applicable to any data with jumps in the response. In this
paper we consider jump processes with only two levels.

When designing computer simulation experiments, one often meta-models a 
planned experiment of runs of the computer model.  That is, gather a collection 
of $N$ data pairs $(x_1, y_1),
\dots, (x_N, y_N)$, where $y_i = Y(x_i)$, and obtain a fit that can predict
$Y(x)$ at new inputs.  Such {\em surrogate models}
\citep{santner2003design, gramacy2020surrogates} are instrumental to many
applications where ready access to runs of the computer model are unavailable
or expensive. The canonical surrogate is a Gaussian process \citep[GP;
e.g.,][]{sacks1989design} because GPs furnish accurate nonlinear predictions
with good uncertainty quantification. For this reason they are also popular
for many regression applications in machine learning
\citep[e.g.,][]{williams2006gaussian} and geospatial settings
\citep[e.g.,][]{banerjee2004hierarchical}, where GPs are known as 
``kriging'' \citep{schabenberger2017statistical, zimmerman2024spatial}. 
There are two downsides to
GP modeling in many contexts.  One is that computational demands for inference
scale poorly as $N$ gets large; another is that the typical GP formulation
assumes stationarity, in that only relative distances between input
locations matter. This means GPs struggle to capture dynamics where modeling
fidelity varies in the input space, e.g., where one part is flat and another
is wiggly. GPs also struggle if there are jumps of discontinuity in which 
the level of the response drastically changes.

There is a sizable literature on nonstationary GP modeling.
\citet{sauer2023non} provide a recent review.  Many such approaches involve
imposing hard partitions in the input space \citep[e.g.,][]{kim2005analyzing,
pope2021gaussian,
payne2020conditional,gramacy2008bayesian,luo2023nonstationary}, which could
potentially cope with jumps that are axis aligned (like with trees) or are
polygonal (Voronoi tessellations).  Other approaches involve smoothly warping
the inputs into a stationary regime
\citep[e.g.,][]{sampson1992nonparametric,schmidt2003bayesian,damianou2013deep,sauer2023vecchia},
which is often better but would struggle with discontinuity. GPs exist which can 
handle discontinuities \citep[e.g.,][]{garnett2010sequential,picheny2019ordinal}, 
but they do not assume the response contains distinct levels.
\citet{park2022jump} created the ``jump GP'' (JGP) specifically for data from
processes that are piecewise continuous in order to address this gap.
JGPs have three main modeling ingredients that are tightly coupled
together. The first two are local (GP) modeling
\citep[e.g.,][]{gramacy2015local}, and latent component (i.e., output level)
membership variables. Inference is joint over all unknowns -- in
particular for the high dimensional latent variable -- via modern variants of
expectation maximization \citep[EM;][]{dempster1977maximum}, comprising the
third ingredient. JGP is a clever, but complex, modeling apparatus that works
well out-of-sample when trained on data from response surfaces exhibiting
jumps.

\begin{figure}[ht!]
\centering
\begin{tikzpicture}
\node(1A) at (0,6) {\includegraphics[scale=0.55,trim={0 10 30 50}, clip]{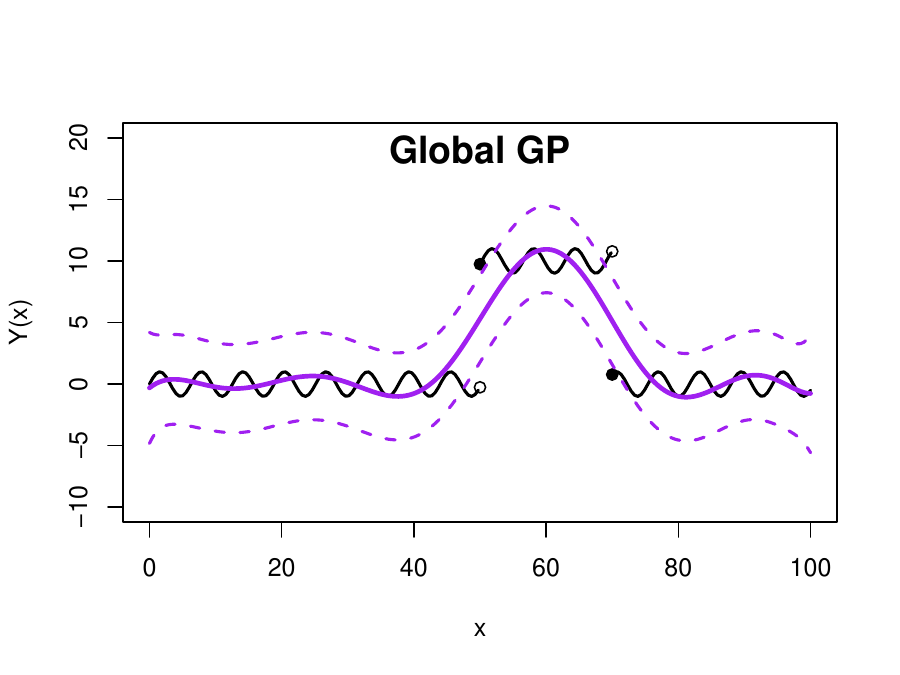}};
\node(1B) at (9.5,6) {\includegraphics[scale=0.55,trim={0 10 30 50}, clip]{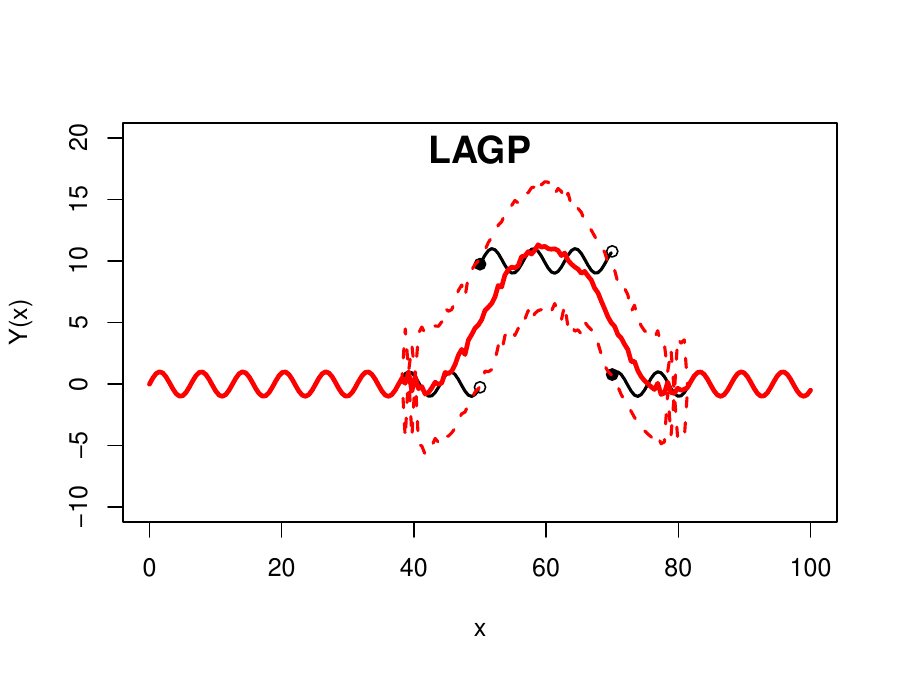}};
\node(1C) at (0,0) {\includegraphics[scale=0.55,trim={0 10 30 50}, clip]{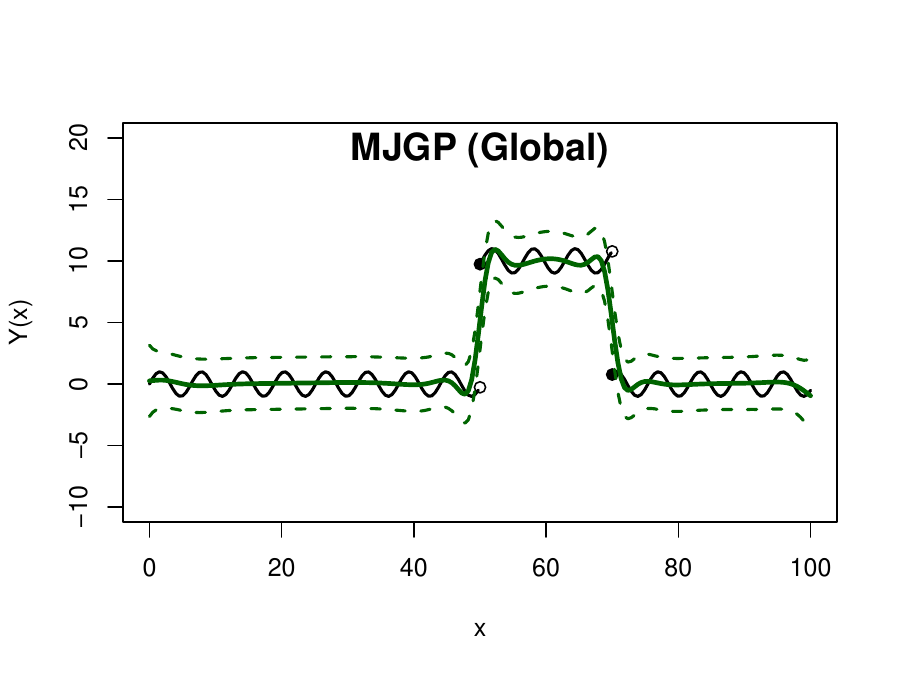}};
\node(1D) at (9.5,0) {\includegraphics[scale=0.55,trim={0 10 30 50}, clip]{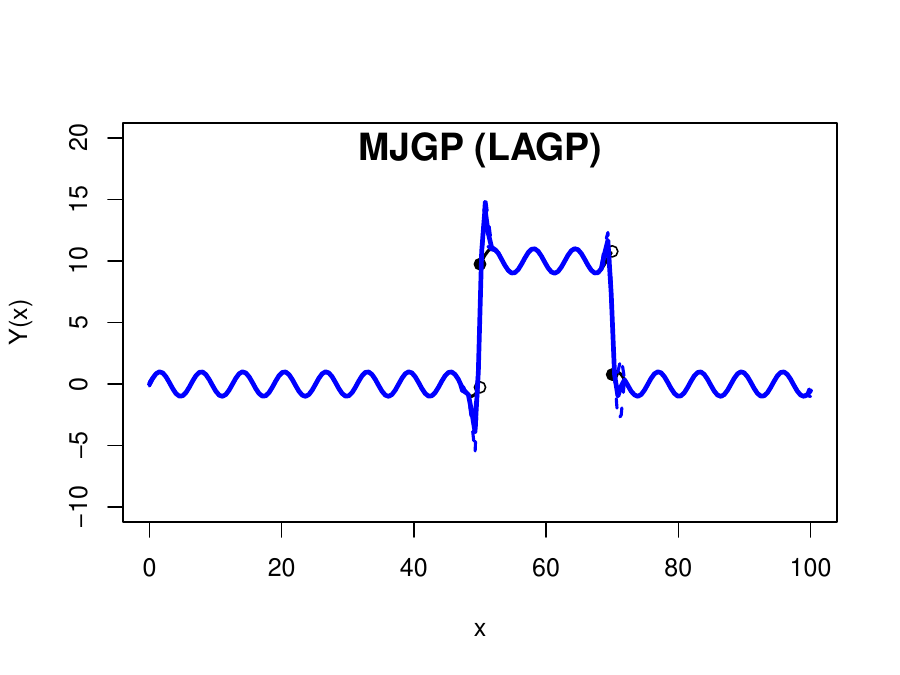}};
\draw[-latex,thick] (1A) -- (1B) node[midway,above,text width=1.5cm]{+Local};
\draw[-latex,thick] (1A) -- (1C) node[midway,right,text width=1.5cm]{+Feature};
\draw[-latex,thick] (1C) -- (1D) node[midway,above,text width=1.5cm]{+Local};
\draw[-latex,thick] (1B) -- (1D) node[midway,right,text width=1.5cm]{+Feature};
\end{tikzpicture}
\caption{All panels show true function $Y(x)$ from Eq.~(\ref{eq:1ex}) in
black. Colored lines represent the mean (solid) and 95\% PI
(dashed) via (top left) a global GP, (top right) LAGP, (bottom left)
MJGP with global GP, and (bottom right) MJGP with LAGP.
Training data not shown.}
\label{fig:plot1}
\end{figure}
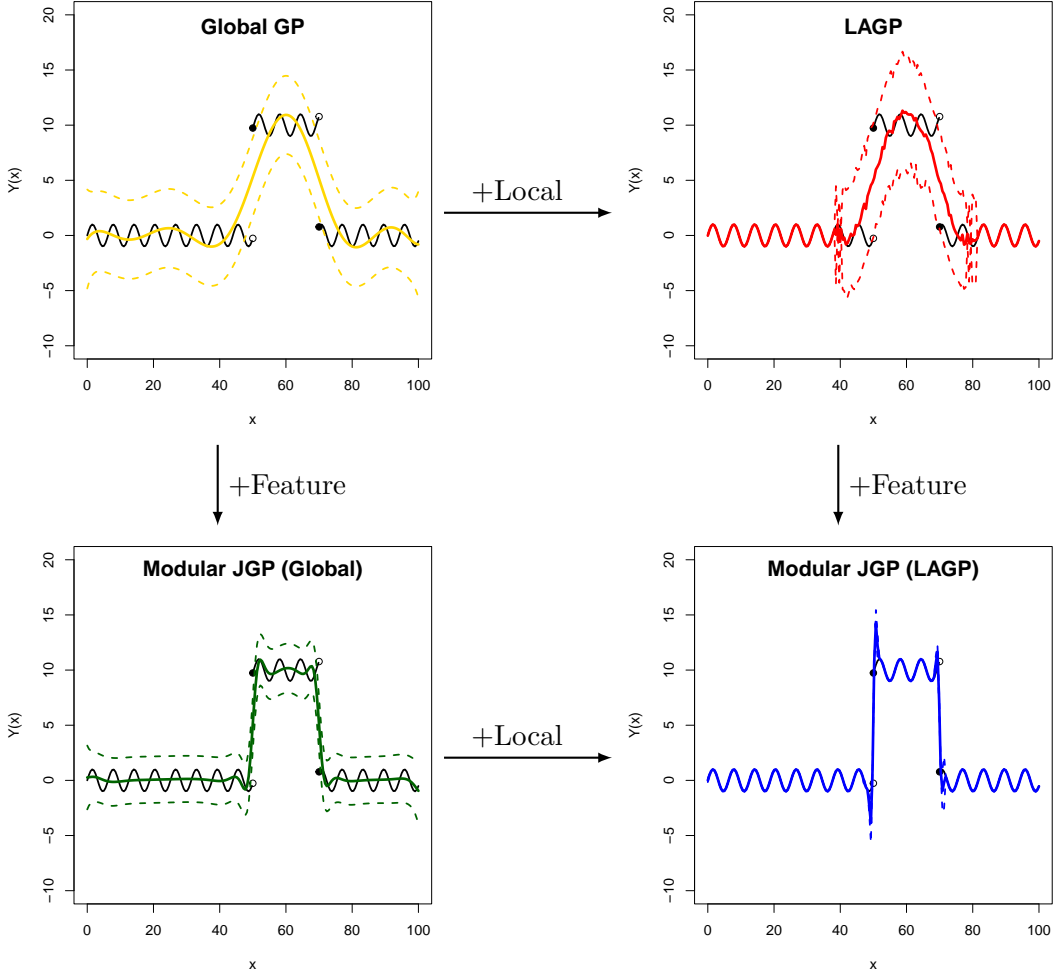

Here we suggest that the essence of a JGP can be extracted by appropriating
its main ingredients in a more modular setup.  We still use local
modeling, latent variables, and clustering via EM, but we avoid joint
inference and instead do things one step at a time by daisy-chaining
off-the-shelf subroutines.  To explain, consider a simple illustrative example
of a process with jumps in 1d.
\begin{equation}
Y(x) = 
\begin{cases}
\sin(x)  &x < 50 \\
\sin(x) + 10 & 50 \leq x < 70\\
\sin(x)  &x \geq 70
\label{eq:1ex}
\end{cases}
\end{equation}
So $Y(x)$ follows a simple sinusoid but jumps up by 10 between $x=50$ and $x=70$. 
Each panel of Figure \ref{fig:plot1} shows
$Y(x)$ as a solid black line with a different ``fit'' from data $(x_1, y_1),
\dots, (x_N, y_N)$.  We use $N=100$ inputs $x_i$ chosen via Latin hypercube
sample \citep[LHS;][]{mckay:1979}. These are not shown to reduce clutter.
Figure \ref{fig:plot1} was produced with standard libraries and
default settings, detailed in Section \ref{sec:results}.

The top left panel of Figure \ref{fig:plot1} shows an ordinary/stationary GP fit, as
reviewed in Section \ref{sec:gp}.  Three issues are of note here:
the GP is mistaking signal for noise, overestimating variance, and smoothing
over the jumps.   It seems to only have detected signal from the jumps, not
from the sinusoid.  Uncertainty is uniform because inputs are uniform and the
GP assumes stationarity.  Now turn to the top right panel, showing a local
(approximate) GP (LAGP), reviewed in Section \ref{sec:lagp}. For now we
remark that a LAGP is basically an ordinary GP where training data far from
the testing location(s) is not used.  Less data means faster
calculations and increased ``reactivity'' if dynamics change in the input
space. Notice how LAGP shows an improvement over the ordinary GP (top left)
for $x<40$ and $x>80$. Signal is detected there, and variance is much
reduced. But this comes at the expense of accuracy in the middle 
region, and near the ``jumps'' on either side. Uncertainty here is inflated
relative to the stationary GP fit. Importantly, LAGP uses a fixed-sized
neighborhood of locality ($n=25$) everywhere in the input space which we shall explain
in more detail in Section \ref{sec:lagp}. Our proposed approach improves 
upon this by dynamically selecting the optimal neighborhood size throughout 
the input space as described in Section \ref{sec:opt}.

The lower panels of Figure \ref{fig:plot1} highlight our main contribution.
Focus first on the lower-left.  Suppose we were to cluster the observations
marginally. We ignore $(x_1, \dots, x_N)$ and use an unsupervised
learning scheme to assign (one of two) labels to each of $(y_1,
\dots, y_N)$. In this 1d example, most clustering methods will easily
partition these $y$-values into components around 0 and 10.  We use
model-based clustering via EM  for this, as described in Section
\ref{sec:feat}, but suggest several variations and other approaches that often
work just as well. Using cluster memberships, e.g., $c_i=1$ for $y_i$ around
0 and $c_j=2$ for $y_j$ around 10, suppose we were to learn a classification
rule spatially for pairs $(x_1, c_1), \dots ,(x_N, c_N)$, i.e., reincorporating 
the inputs $(x_1,\dots, x_N)$ with supervised learning. There are again many choices for
this; we prefer a classification GP (CGP).  This allows us to predict a
probability of class 2 at $x$, in-sample at ($x_1,
\dots, x_N$) yielding $(p_1,\dots, p_N)$, or out-of sample for testing as
$p(x)$.  We then treat the class predictions $(p_1 \dots p_N)$, derived from the training data, 
as an input feature in our GP fit.  That is, take the same GP
from the top-left panel, but train it on input tuple $(x_i, p_i)$ and output
$y_i$, for $i=1,\dots N$.  When predicting, augment any testing $x$ with
$p(x)$ for tuple $(x, p(x))$. This global GP with the new feature is visualized 
in the lower left panel of Figure \ref{fig:plot1}.  Notice how
the fit tracks the jumps better (compared to the top left), and it is more
confident everywhere, but it still struggles to pick up on the sinusoid.  The
idea of learning an auxiliary input feature to accommodate nonstationarity is
not new \citep{bornn2012modeling} but, like \citet{park2022sequential} for
JGP, \citeauthor{bornn2012modeling} perform joint inference for the GP and the
latent quantity, which can be cumbersome. A key feature of our approach is
that we do everything in a pre-processing step with off-the-shelf libraries. 
This both increases the flexibility of our approach and makes it more widely 
accessible.

Finally, the bottom right panel of Figure \ref{fig:plot1} combines the feature
with local modeling: LAGP applied on $(x_i, p_i)$ and output $y_i$, but
otherwise the same as the feature-expanded GP on the left.  We call this the
Modular JGP (MJGP). Except for some elevated uncertainty near the jump points, which
is sensible, the fit is nearly perfect: high accuracy, low variance.  In the
remainder of the paper our goal is to flesh out these ideas, making them more
concrete, especially as regards implementation.  We begin with review in
Section \ref{sec:review}.  We then describe our main methodological
contributions in Sections \ref{sec:opt} and  \ref{sec:mod}:~optimizing local
neighborhoods and cluster-based features.  Throughout, we provide
illustrations on examples that have the same spirit as Figure \ref{fig:plot1},
but are more realistic.  Section \ref{sec:results} provides empirical
benchmarking exercises in several variations including on a real simulation
from manufacturing.  Finally, Section \ref{sec:discuss} concludes with a brief
discussion of limitations and possible future work. All code is provided via
Git: \url{https://bitbucket.org/gramacylab/jumpgp}.

\section{Review}\label{sec:review}

Here we provide a brief review of GPs and a popular variant, LAGP.

\subsection{Gaussian Processes}\label{sec:gp}

Let $X_N$ be an $N \times d$ matrix of input features and $Y_N$ be a
corresponding $N \times 1$ vector of outputs. We assume $X_N$ has 
been pre-scaled to the unit $d$-cube. Additionally, let $x_i^\top$
denote the $i^\mathrm{th}$ row of $X_N$. We consider estimating an unknown
function via training data $X_N$ and $Y_N$ using a GP
\citep{williams2006gaussian, gramacy2020surrogates}. A typical GP prior
implies that $Y_N \sim \mathcal{N}_N \bigl( \mu, \Sigma(X_N) \bigr)$, where $\mu$ is
often assumed to be 0 so that all of the modeling ``action'' is in $\Sigma(X_N)$. 
This setup induces a likelihood function:
\begin{equation}
L(Y_N \mid X_N) \propto \bigl\vert \Sigma(X_N) 
\bigr\vert^{-\frac{1}{2}} \mathrm{exp} \left\{ -\frac{1}{2}Y_N^\top \Sigma(X_N)^{-1} Y_N \right\}.
\label{eq:2lik}
\end{equation}
The likelihood (\ref{eq:2lik}) can be used to learn $\Sigma(X_N)$ under some
restrictions, such as positive definiteness. Often, covariance is built from a
kernel specifying $\Sigma(X_N)_{ij} = \Sigma(x_i, x_j)$. Kernels that depend
only on the distance between inputs yield stationary processes, eliminating
all input position information from the model. One example is squared
exponential kernel
\begin{equation}
\Sigma(X_N)_{ij}=\Sigma(x_i, x_j) = \tau^2 \mathrm{exp} 
\left\{ - \sum_{k=1}^d \frac{\left\Vert x_{ik}-x_{jk} \right\Vert ^2}{\theta_{k}} \right\},
\label{eq:2gp}
\end{equation}
which we use throughout, although none of our contribution is explicitly tied
to this choice.  Other stationary kernels, such as Mat\'{e}rn
\citep{stein2012interpolation}, are similarly appropriate. Above,
$\tau^2$ and $\theta = [\theta_1,\dots,\theta_d]$ are hyperparameters
representing the scale and lengthscale, respectively. These
quantities may be inferred via maximum likelihood estimation (MLE) with
Eq.~(\ref{eq:2lik}). 

The same structure may be extended from training to testing for prediction.
Let $\mathcal{X}$ denote an $m \times d$ matrix of new/testing locations, and
$\tilde{x}^{\top}_i$ be the $i^{\text{th}}$ row of $\mathcal{X}$. Then, using
any hyperparameters learned from training data,  $Y(\mathcal{X}) \mid X_N, Y_N
\sim \mathcal{N}_m \left(\mu_N({\mathcal{X}}), \Sigma_N({\mathcal{X}})
\right)$ where
\begin{align}
\mu_N({\mathcal{X}}) & = \Sigma(\mathcal{X}, X_N) \Sigma(X_N)^{-1} Y_N, & \mbox{using } \; \Sigma(\mathcal{X},X_N)_{ij} &=\Sigma(\tilde{x} _i, x_j), \nonumber \\
\mbox{and } \quad \Sigma_N({\mathcal{X}}) & = \Sigma(\mathcal{X}) - \Sigma(\mathcal{X}, X_N) \Sigma(X_N)^{-1} \Sigma(\mathcal{X},  X_N)^\top.
\label{eq:2newpred}
\end{align}
As a shorthand for later, let $\mathrm{GP}\left(\mathcal{X};X_N, Y_N \right)$
indicate the use of Eqs.~(\ref{eq:2lik}--\ref{eq:2newpred}) to represent
predictions at $\mathcal{X}$ via MLE hyperparameters using training data $X_N$
and $Y_N$. We refer to this as a ``global GP'' since all training data are
used. Such a fit is valid for prediction at any $\mathcal{X}$ in the input
space. This is in contrast to a GP fit using a subset of the training data,
which we shall review next. The GP shown in the top left panel of Figure
\ref{fig:plot1} is a global GP, where the solid purple line is
$\mu_N(\mathcal{X})$ and the dashed purple line is a 95\% prediction interval
(PI) calculated using the diagonal entries of $\Sigma_N(\mathcal{X})$.

A global GP has two drawbacks: computation becomes prohibitive as $N$
increases, and it does not perform well when training data come from 
nonstationary processes. The first can be seen by noting $\Sigma(X_N)^{-1}$
and $|\Sigma(X_N)|$ in Eqs.~(\ref{eq:2lik}) and (\ref{eq:2newpred}), both
$\mathcal{O}(N^3)$ operations. To see the second, refer back to the discussion
of Figure \ref{fig:plot1} (top-left), illustrating fits on a process with
jumps.  Consider a single data point at $x=45$, which is part of the level-0
region. The issue with the global GP setup is that two equidistant points, say
$x=35$ and $55$, generate the same covariance, even though one point is in the
level-0 region and the other is in the level-10 region. However notice that
the data is locally stationary, in three spatial regimes, which suggests a
divide-and-conquer approach may work well.

\subsection{Local Approximate Gaussian Processes}\label{sec:lagp}

LAGP \citep{gramacy2015local} tackles both drawbacks at once: computational
expense and stationarity. Observe from Eq.~(\ref{eq:2gp}) that for two
locations $x_i$ and $x_j$, $\Sigma(x_i,x_j)
\rightarrow 0$ as $\left\Vert x_i-x_j \right\Vert \rightarrow \infty$.
Consider a single new location $x$ where we wish to make a prediction, e.g., a
single row of $\mathcal{X}$ in Section \ref{sec:gp}. Note in
Eq.~(\ref{eq:2newpred}) that $\Sigma(x,X_N)$ serves as a weight for
$\mu_N(\mathcal{X})$; thus, any training $y_i$ whose $x_i$ is far from $x$ has
little influence on $\mu_N(\mathcal{X})$. Recognizing this, LAGP chooses a
subset of $n \ll N$ ``useful" data points $X_{n}(x) \subset X_N$ to predict
$x$. There are many ways to define ``useful", such as nearest neighbors (NN),
mean-squared predictive error, and Fisher information.  Throughout we consider
LAGP via NN, although it could be worth entertaining other criteria as we
discuss briefly in Section \ref{sec:discuss}. LAGP is not unique in
its use of a small selection nearby points. Other, similar methods
include nearest neighbor GPs \citep{datta2016hierarchical, datta2016nearest},
inducing points \citep{snelson2005sparse, banerjee2008gaussian}, or kernels
and basis functions with compact support \citep{nychka2015multiresolution,
kaufman2011efficient}.  However, LAGP is unique in its focus on points nearby
to the predictive location $x$, as opposed to globally or for all $X_N$.


The core idea behind LAGP is that fitting a GP to $X_{n}(x) \subset X_N$ with
$n \ll N$, along with corresponding $Y_n$, is both faster, and more reactive
to local dynamics around $x$.  We encapsulate that in shorthand as
$\text{LAGP}_n(x) = \text{GP}\left(x;X_n(x), Y_n \right)$ after optimizing any
hyperparameters in $\Sigma(X_n(x))$, and
$\text{LAGP}_n(\mathcal{X})$ for $m$ separate applications
covering all of $\mathcal{X}$. The top right panel of Figure
\ref{fig:plot1} shows $\text{LAGP}_{25}(\mathcal{X})$, where the red solid
line denotes $\mu_N(\mathcal{X})$ and the red
dashed line denotes a 95\% PI. Since calculations are computationally exclusive 
for each ${x} \in \mathcal{X}$, we do not get a full covariance
structure $\Sigma_N(\mathcal{X})$, only a diagonal one
with entries $\sigma^2_{x}$.  The upside, though, is that the calculations are
massively parallelizable \citep{gramacy2014massively}.

LAGP reduces computational costs from $\mathcal{O}(N^3)$ to
$\mathcal{O}(n^3)$, offering substantial speedups if $n \ll N$. Much work has
been done to further decrease the computation time for fixed neighborhood size
$n$ \citep{gramacy2016speeding,sung2018exploiting,cole2021locally}. Besides
being faster, LAGP may also be more accurate for data from a nonstationary
process. To see this, refer again to Figure \ref{fig:plot1} (top-right). By
excluding points that are far from each $x$, LAGP has captured local trends
(at least for the $x<40$ or $x>80$ regions) better than the global GP (top-left).
However, LAGP does not perform well in the $40<x<80$ region. This is because these
predictive locations have neighbors from across the jumps.

To explore a potential remedy, consider a higher-dimensional illustration.
\begin{figure}[ht!]
\centering
\includegraphics[scale=0.45, trim = 20 10 50 20, clip]{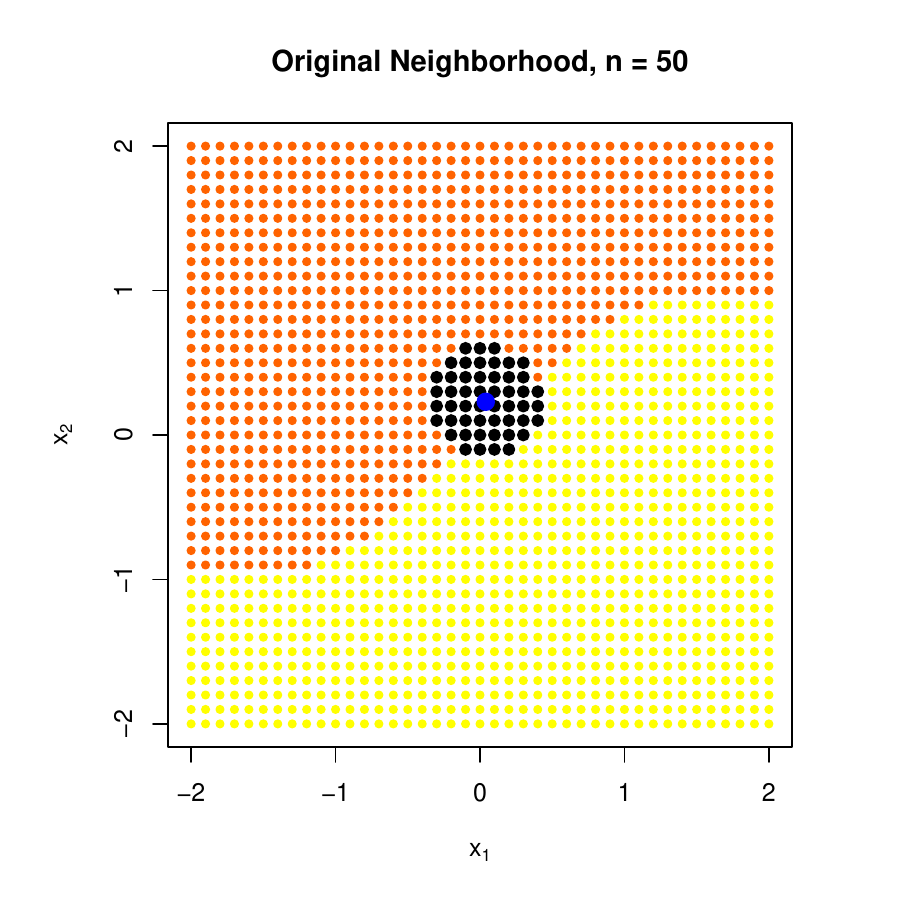}
\hfill
\includegraphics[scale=0.45, trim = 40 10 50 20, clip]{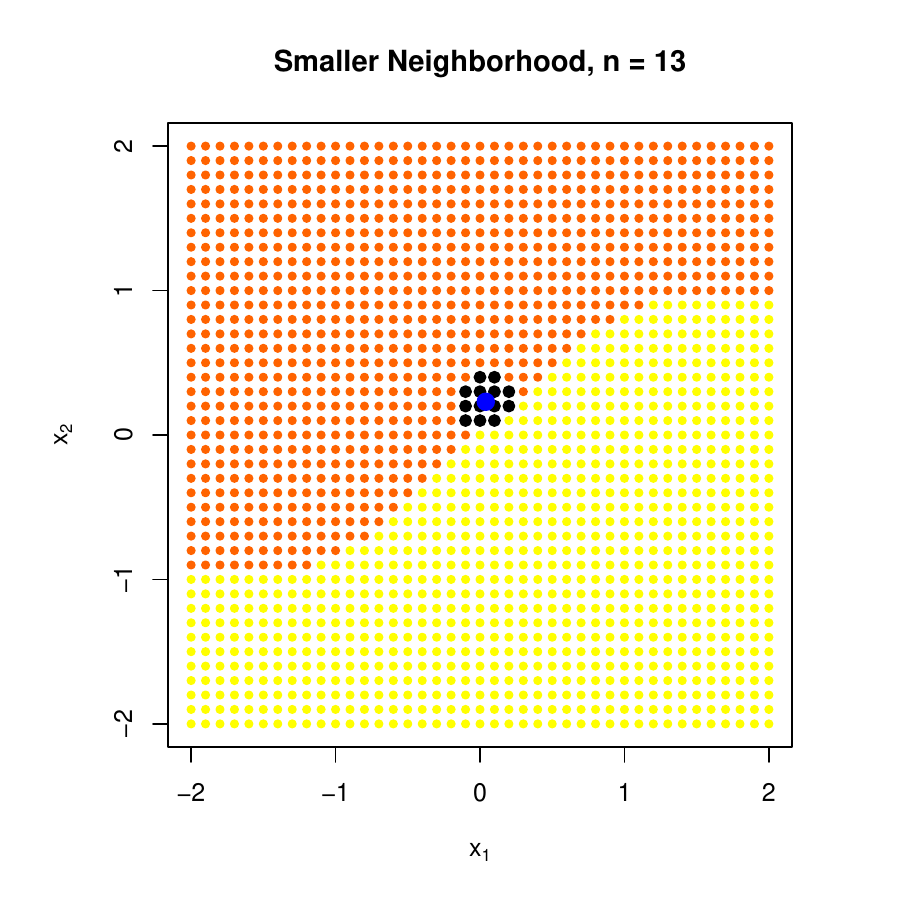}
\hfill
\includegraphics[scale=0.45, trim = 40 10 50 20, clip]{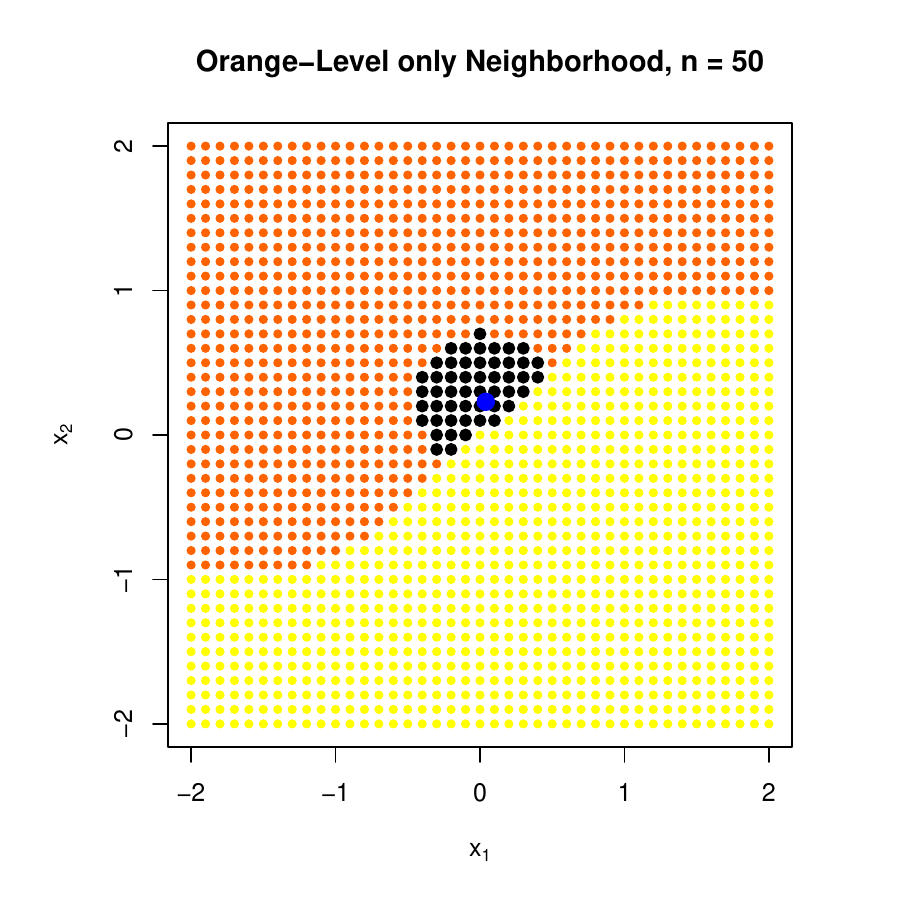}
\caption{Three possible neighborhood choices for $x = (0.04,0.023)$ (the blue dot). 
The two colors (orange and yellow) are two levels of the response, with a
manifold of discontinuity at $\sin(x_1) = x_2$. The orange and yellow dots are
$X_N$ and the black dots are three different choices for $X_n(x)$.}
\label{fig:plot2}
\end{figure}
Figure \ref{fig:plot2} depicts a simple manifold of discontinuity with two
output values, orange and yellow, observed on an $X_N$ grid. Now consider
predicting at $x=(0.04, 0.23)$, in blue, using LAGP.  Some potential
neighborhoods $X_n(x)$ are depicted in black, one in each panel.  The left
panel uses Euclidean NNs with $n=50$, but many such neighbors come from the
opposite side of the manifold. One workaround is to shrink the neighborhood
size, e.g., $n=13$ in the center panel, which is the largest $n$ that only
contains neighbors from the orange region. 
So the choice of $n$ implies a tradeoff between accuracy and reactivity. It
could be advantageous to have a larger neighborhood (e.g., $n=50$), but
comprised of only orange points, as shown in the right panel.  

\section{Optimal Neighborhood-size LAGP}\label{sec:opt}

We turn now to the neighborhoods depicted in the center [this section] and
right  [Section \ref{sec:mod}] panels of Figure \ref{fig:plot2}, respectively.
The idea is to make adjustments near a manifold of discontinuity without
affecting modeling dynamics elsewhere. To aid illustrations, Figure
\ref{fig:phantom} borrows an example from \citet{park2022jump}, known
as ``Phantom'' data, with details in Appendix \ref{sup:phantom}.
\begin{figure}[ht!]
\centering
\includegraphics[scale=0.6, trim = 25 10 50 38, clip]{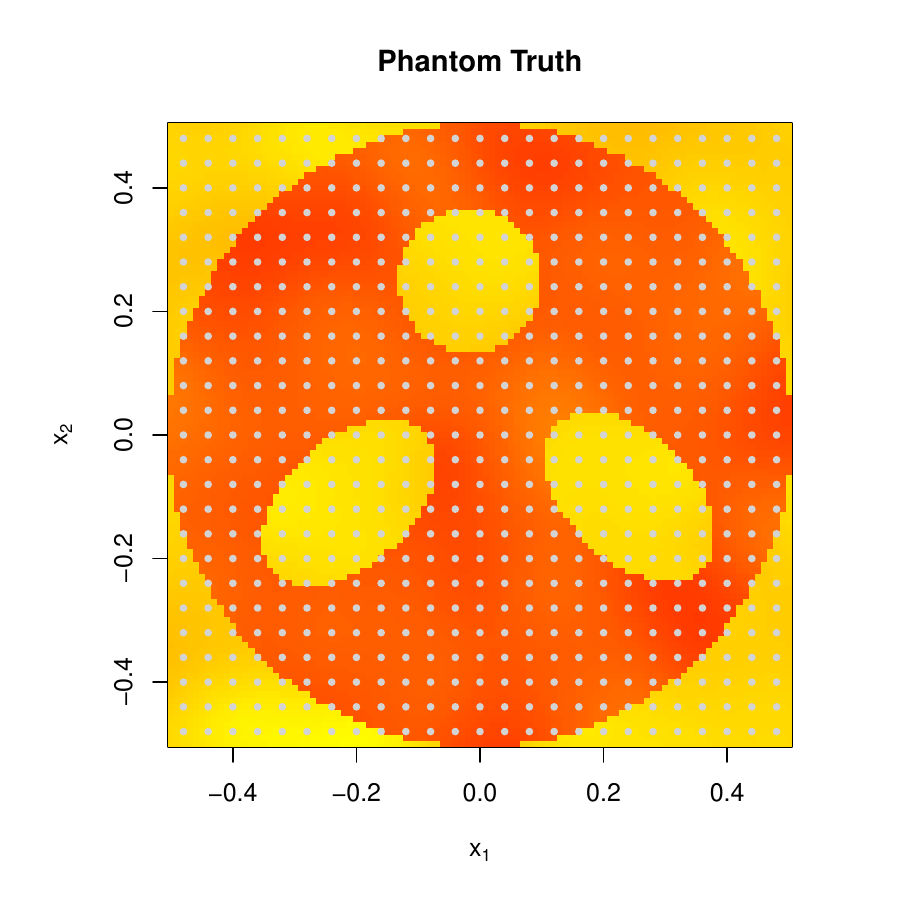}
\hspace{1cm}
\includegraphics[scale=0.6, trim = 25 10 50 38, clip]{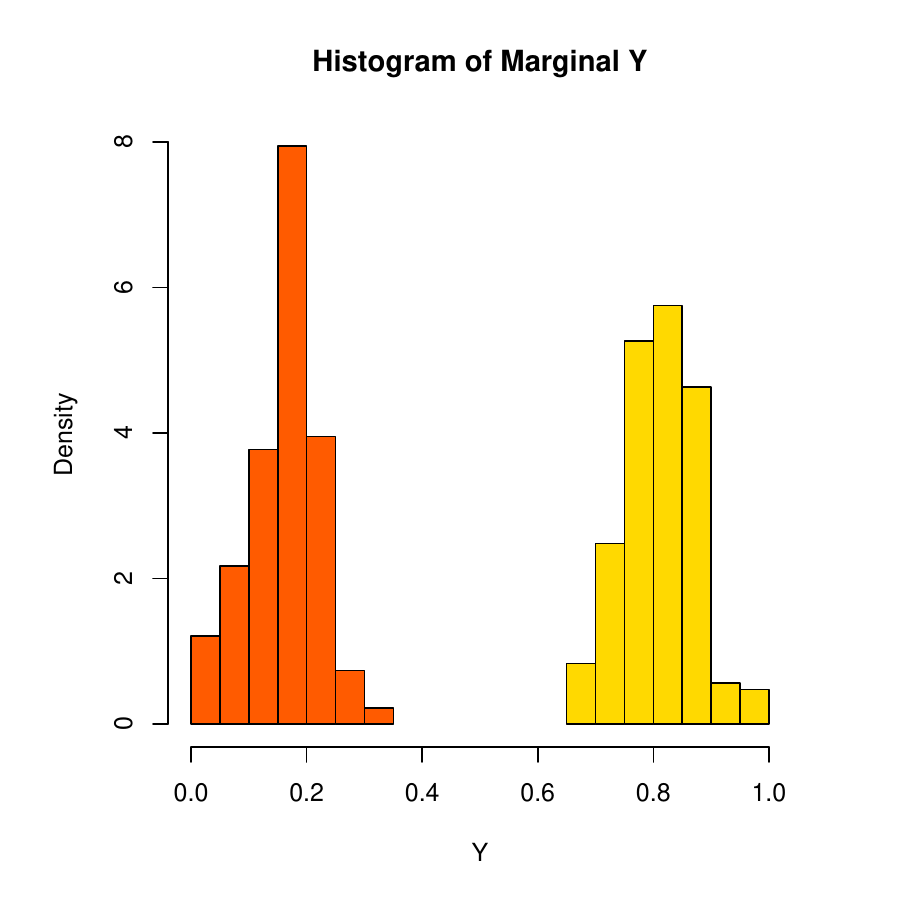}
\caption{{\em Left}: True ``Phantom'' with training grid 
in gray; {\em Right:}
Histogram of the marginal response. }
\label{fig:phantom}
\end{figure}
The image on the left is formed by evaluating the test function on a dense
$101 \times 101$ grid in $[-0.5,0.5]^2$.  All responses ($y$-values) are
deterministic.  Observe that there is smooth color variation within the
(mostly) red and yellow regions, revealing a strong signal locally (light reds
and dark yellows, respectively), even away from the ``jumps''.  The histogram
on the right helps understand the magnitudes of these signals. Notice that red
and yellow regimes have comparable within-color variability and they are
well-separated, with no values between 0.4 and 0.6.

The visuals in the panels of Figure \ref{fig:plot2} could represent zoomed-in
versions of any of the jumps separating red from yellow in the left panel of
Figure \ref{fig:phantom}.  Here, in Section \ref{sec:opt}, we present the
first prong, which targets finding the optimal $n(x)$ for LAGP [Figure
\ref{fig:plot2}, middle] using its default, radial topography. The right panel
of Figure \ref{fig:plot2}, morphing that topography, represents the second
prong and is the subject of Section \ref{sec:mod}. While jumps naturally focus
interest on inputs near the manifold, the value of dynamically determining
$n(x)$ for LAGP (other than the software default of $n=50$) is not limited to
those areas only. Intuitively, regions of slowly varying response should
benefit from larger neighborhoods, whereas ones where it is changing more
rapidly (via jump or not) should benefit from the greater reactivity provided
by a smaller one.

\subsection{Searching via validation}\label{sec:val}

Consider a single predictive location $x$ and a set of possible neighborhood
sizes $n \equiv n(x) \in \{n_{\min},\dots, n_{\max}\}$. We wish to estimate an
optimal $\hat{n}(x)$ by some criterion, where $X_{n_{\min}}(x)
\subseteq X_{\hat{n}}(x) \subseteq X_{n_{\max}}(x) \subseteq X_N$, as solved
by LAGP [Section \ref{sec:lagp}]. Let $\mu_{n}(x)$ denote the predictive mean
under some $X_{n}(x)$, i.e., following Eq.~(\ref{eq:2newpred}), but
conditional on $(X_n(x), Y_n)$.  This prediction has some accuracy, say
measured by squared error $\mathrm{SE}(y, \mu_{n}(x))$, where $y$ is the true
but unknown value of the response at $x$. Generically, $\mathrm{SE}(y,\hat{y})
= (y - \hat{y})^2$. One of the main contributions of \citet{gramacy2015local}
involves arguing that NN-constructed $X_{n}(x)$, conditional on a neighborhood
size $n$, approximately minimizes the expectation of this quantity (i.e.,
the mean-squared error) for $y$ following a stationary GP.

Here, we wish to extend that to choosing the best $\hat{n}$.  However, we
cannot use MSE under a stationary GP for $y$.  There are two reasons for this.
One is practical: this is not the case we are interested in -- our jump
processes are nonstationary.  The other is more fundamental.  If the process
is stationary, then MSE is a decreasing function of the subset size
$n$. In other words, you want to condition on as much data as possible.
Since large $n$ is generally intractable, \citeauthor{gramacy2015local}
recommended choosing $n$ as large as computational constraints allow.
However, they observed that performance varied with $n$ when the data
exhibited telltale signs of nonstationarity, leaving an open question about
what to do in practice.  So instead of averaging SE, we create a validation
exercise which allows us to plug-in an actual $y_i$ value from the training
data as a proxy.

Let $x_*$ be the input in $X_N$ that is closest (in terms of Euclidean
distance) to $x$.  Note that this is also the closest point to $x$ in any
$X_n(x)$ for $n \geq 1$ and that $x_* \equiv X_1(x)$.
For ease of notation, we shall henceforth refer to $X_1(x)$ as $x_*$.  
Analogously, let $y_*$ be
the element in $Y_N$ corresponding to $x_*$. We use $y_*$ as a proxy for the
value of $y(x)$, the mean response at new location $x$, which is unknown.
Next, let $X_n(x_*)$ denote an $n$-sized neighborhood for predicting at $x_*$,
i.e., the $n$-nearest points to $x_*$ in $X_N$, excluding $x_*$.   Using these
quantities, and for any neighborhood size $n$, one may fit
$\text{LAGP}_n\left(x_*; X_n(x_*), Y_n\right)$, furnishing a prediction for
$y_*$.
Specifically, let $\mu_n(x_*)$ denote the predictive mean of $x_*$ 
calculated with neighborhood size $n$.  Suppose we do
this for each $n$ of interest, and take the best one by squared error:
\begin{equation}
\hat{n}(x_*) = \underset{n \in \{n_{\min}, \dots, n_{\max}\}}{\arg \min} 
\mathrm{SE}\left(y_*,\mu_n(x_*) \right).
\label{eq:3optest}
\end{equation}
This $\hat{n}(x_*)$ is the best neighborhood size for $x_*$, which is nearby
the location of interest $x$.  What remains is to refit the local model at $x$
using the $\hat{n}(x_*)$-nearest neighbors to $x$,
$\mathrm{LAGP}_{\hat{n}(x_*)}(x; X_{\hat{n}}(x), Y_{\hat{n}})$.

\begin{figure}[ht!]
\centering
\includegraphics[scale = 0.45, trim =  25 10 40 10, clip]{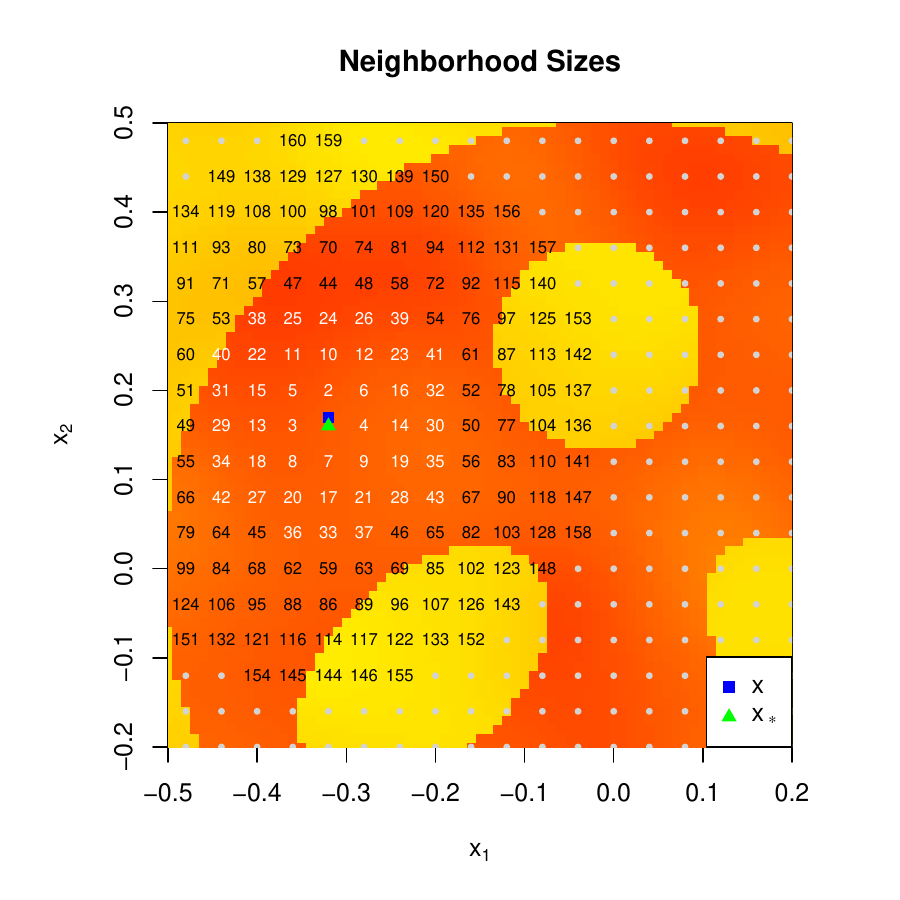}
\includegraphics[scale = 0.45, trim =  25 10 40 10, clip]{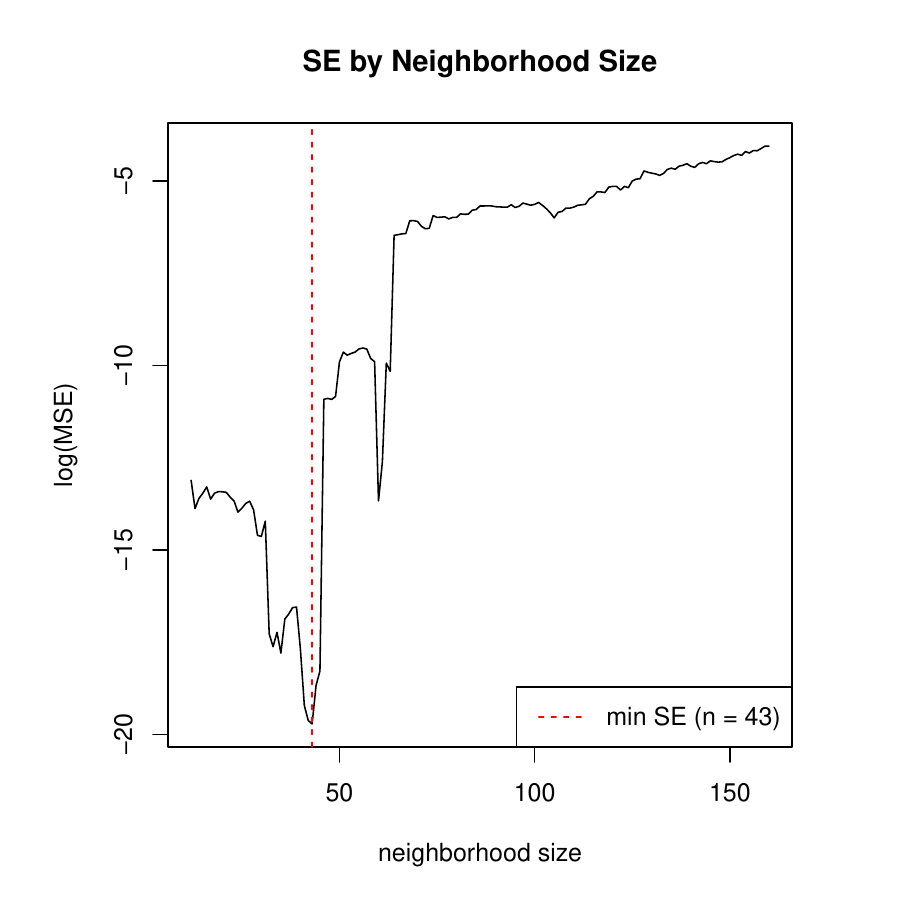}
\includegraphics[scale = 0.45, trim =  25 10 40 10, clip]{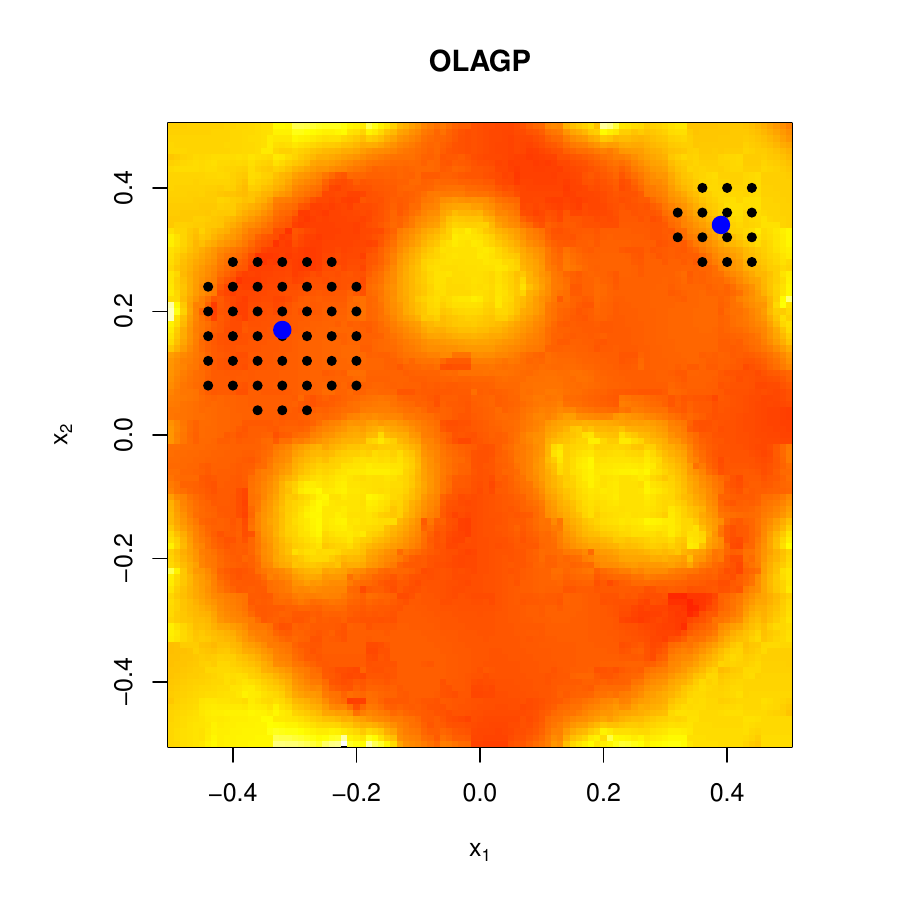}
\caption{{\em Left}: Order of LAGP neighborhood for point $(-0.32,0.17)$;
{\em Center}: SE by neighborhood size for point $(-0.32,0.17)$; {\em Right}:
Optimal neighbors LAGP (OLAGP) fit. Blue dots indicate two predictive sites
and the surrounding black dots are the respective optimal neighborhoods.}
\label{fig:olagp}
\end{figure}

Without care, fitting $(n_{\max} - n_{\min})$-many GPs in pursuit of
Eq.~(\ref{eq:3optest}) could be prohibitively expensive. One option is
parallelization, since the calculation for each $n$ is computationally
exclusive of the others.  However, we prefer to parallelize instead over the
predictive location(s), $x \in
\mathcal{X}$, which will be described in Section \ref{sec:results}.  A computationally
thrifty solution takes advantage of the nested structure of neighborhoods as
$n$ increases: $X_{n_{\min}}(x_*)
\subseteq X_{n_{\min}+1}(x_*) \subseteq \dots \subseteq X_{n_{\max}}(x_*)$.
Observe that any two neighborhoods $X_n(x_*)$ and $X_{n-1}(x_*)$ are identical
except for one point: the $n^{\mathrm{th}}$ furthest point from $x_*$. Let
$x_n = X_{n}(x_*) \setminus X_{n-1}(x_*)$ denote this new location as the
neighborhood increases by one.  A visual of these $x_n$, selected from $X_N$
on a $25 \times 25$ grid, is provided in the left panel of Figure
\ref{fig:olagp}.
Each number indicates the position of $x_n$ for some $n$ relative to $x_* \in
X_N$, the closest point to $x=(-0.32, 0.17)$.

Now, proceeding in order for $n=n_{\min}, \dots, n_{\max}$, suppose we had
already calculated $\mathrm{LAGP}_{n-1}(x_*)$, beginning initially with $n-1
\equiv n_{\min}$ via the description in Section \ref{sec:lagp}, yielding
$\Sigma_{n-1}^{-1} \equiv \Sigma(X_{n-1}(x_*))^{-1}$ and similarly for $\log |
\Sigma_{n-1} |$. Incorporating the next point, $x_n$, involves adding a new
row/column to $\Sigma_{n-1}$ to get $\Sigma_n$, which may be calculated without
new cubic cost matrix decompositions by following the partition inverse
equations \citep{barnett1979matrix}. Specifically,
\begin{align}
\label{eq:inveq}
\Sigma_n^{-1} &= \begin{bmatrix}
[\Sigma_{n-1}^{-1} +h h^{\top }v] & h \\
h^{\top} & v^{-1}
\end{bmatrix}  
&& \mbox{and } &
\log \vert \Sigma_n \vert &= \log \vert \Sigma_{n-1} \vert + \log(v), \quad \text{where} \\
h &= -v^{-1} \Sigma_{n-1}^{-1}\Sigma(X_n^{-1}(x_*), x_n), \nonumber
&& \mbox{and} &
v &= \Sigma(x_n) - \Sigma(x_n, X_{n-1}(x_*)) \Sigma_{n-1}^{-1} \Sigma(X_{n-1}(x_*),x_n).
\end{align}
Each update is quadratic in $n$, so that the entire process requires
$c\sum_n^{n_{\max}} n^2 \in \mathcal{O}(n_{\max}^3)$ flops.  In other words,
the computational expense is no greater than that involved in computing the
largest LAGP in the search. Given these quantities, $\mu_n(x_*)$ required as
in Eq.~({\ref{eq:2newpred}) is again at worst cubic in $n$.

These updates, and the computational cost analysis, presume that
$\Sigma_{n-1}$ and $\Sigma_n$ utilize the same hyperparameters, like
lengthscales $\theta_k$, which may not be ideal.  In particular, shorter
lengthscales will generally be preferred for smaller $n_{\min}$-sized
neighborhoods whose squared distances (\ref{eq:2gp}) are short.  Whereas when
$n$ is closer to $n_{\max}$, including longer squared distances, longer
lengthscales may be preferred. However, an MLE calculation (\ref{eq:2lik})
after each update, as new $x_n$ arrives, would incur cubic costs and result in
a quartic overall computational complexity, which is not ideal either.
Recognizing that an additional data point is of greater value to the
likelihood, say via its Fisher information, when $n$ is small as opposed to
large, we deploy a logarithmic MLE-updating schedule so that cubic MLE costs
are incurred much less often as $n$ increases.  Any sub-linear updating
schedule will keep computation cubic.
 \begin{algorithm}[ht!]
\caption{OLAGP (Optimal Neighbors LAGP)}
\label{alg:opt}
\begin{algorithmic}[1]
\Require{$X_N, Y_N, x, n_{\min}, n_{\max}$}

\State{$x_* = X_1(x)$}
\Comment{nearest neighbor of $x$}
\State{$\{ \Sigma_{n_{\min}}^{-1}, \log \vert \Sigma_{n_{\min}} \vert\} = \mathrm{chol}(\Sigma(X_{n_{\min}}(x_*)))$}
\Comment{calculate $\Sigma_{n_{\min}}^{-1}$ and $\log \vert \Sigma_{n_{\min}} \vert$ using (\ref{eq:2gp})}
\State{$\{\tau^2, \theta\}=\mathrm{MLE}(\Sigma_{n_{\min}}^{-1}, \log \vert \Sigma_{n_{\min}} \vert, Y_{n_{\min}})$}
\Comment{calculate starting MLE using (\ref{eq:2lik})}
\State{$\mu_{n_{\min}}(x_*) =  \Sigma(x_*, X_{n_{\min}}(x_*)) \Sigma_{n_{\min}}^{-1} Y_{n_{\min}}$}
\Comment{make prediction for $x_*$ using (\ref{eq:2newpred})}
\For{$n \in \{n_{\min} + 1, \dots, n_{\max}\}$}
\State{$x_n = X_n(x_*) \setminus X_{n-1}(x_*)$}
\Comment{newest point in neighborhood}
\State{$\{\Sigma_n^{-1}, \log \vert \Sigma_n \vert\}=\mathrm{update}(\Sigma_{n-1}^{-1}, \log \vert \Sigma_{n-1} \vert, x_n)$}
\Comment{calculate new $\Sigma_n^{-1}$ and $\vert \Sigma_n \vert$ using (\ref{eq:inveq})}
\If{$\mathrm{log}_2(n) \mod 1 = 0$}
\State{$\{\tau^2, \theta\}=\mathrm{MLE}(\Sigma_n^{-1}, \log \vert \Sigma_n \vert, Y_n)$}
\Comment{recalculate MLE using (\ref{eq:2lik})}
\EndIf
\State{$\mu_n(x_*) =  \Sigma(x_*, X_n(x_*)) \Sigma_n^{-1} Y_n$}
\Comment{make prediction for $x_*$ using (\ref{eq:2newpred})}
\EndFor
\State{$\hat{n}(x_*) = \underset{n \in \{n_{\min}, \dots, n_{\max}\}}{\mathrm{argmin}} 
\mathrm{SE}\left(y_*,\mu_n(x_*) \right)$}
\Comment{choose best neighborhood size using (\ref{eq:3optest})}
\State{\Return $\mathrm{LAGP}_{\hat{n}(x_*)}(x; X_{\hat{n}}(x), Y_{\hat{n}})$}
\Comment{make prediction for $x$ with $\hat{n}(x_*)$}
\end{algorithmic}
\end{algorithm}
The details, summarizing the entirety of our approach to solving
Eq.~(\ref{eq:3optest}), is provided by Algorithm \ref{alg:opt}.  Notice that
the procedure ends after re-fitting LAGP with the chosen $\hat{n}(x_*)$
with $x$ and all of $X_N$.

Continuing our illustration in Figure \ref{fig:olagp}, the center panel
provides a view of the SE-values encountered via Eq.~(\ref{eq:3optest}) using
the setup on the left, for $x=(-0.32, 0.17)$, as $n$ ranges from $n_{\min} =
12$ to $n_{\max} = 160$. The result is $\hat{n}(x) = 43$, and the indices for
these locations are indicated with white text. Notice how $\hat{n}(x)$ is
chosen such that $X_{\hat{n}}(x)$ is as big as possible while residing
entirely in the red-response area.  This process must be repeated for $x \in
\mathcal{X}$ of interest, whereas only one is illustrated in the figure.  Some
of the details of how we recommend doing this are deferred to Section
\ref{sec:results}.  A \texttt{for} loop around Algorithm \ref{alg:opt} is one
simple option, and is the basis of the illustration described next.

\subsection{An Illustration}\label{sec:illust}

Now consider applying Alg.~\ref{alg:opt} independently on each $x
\in\mathcal{X}$, e.g., testing on a $101 \times 101$  grid, extending our
running ``Phantom'' example.  The right panel of Figure \ref{fig:olagp} uses
color to indicate the predictive means obtained at each $x$ on the last line
of the algorithm.  Compared to the truth [Figure \ref{fig:phantom}] the fit
looks good, but not perfect. We show in Appendix
\ref{sup:phanfits} that this fit is more accurate by
root mean square prediction error (RMSPE) than an
ordinary GP fit.  Here we aim to illustrate the optimal choice of neighborhood
size, $\hat{n}(x)$, in two ways.  We begin with focus on two particular
predictive locations: $(-0.32, 0.17)$ and $(0.39, 0.34)$, far from and near to
the manifold of discontinuity, respectively. Observe how the optimal
neighborhood size (OLAGP) adjusts as appropriate ($n=43$ far from the
manifold; $n=14$ near).

 \begin{figure}[ht!]
\centering
\begin{subfigure}[c]{.45\textwidth}
\centering
\includegraphics[scale=0.6, trim=0 10 30 45,clip]{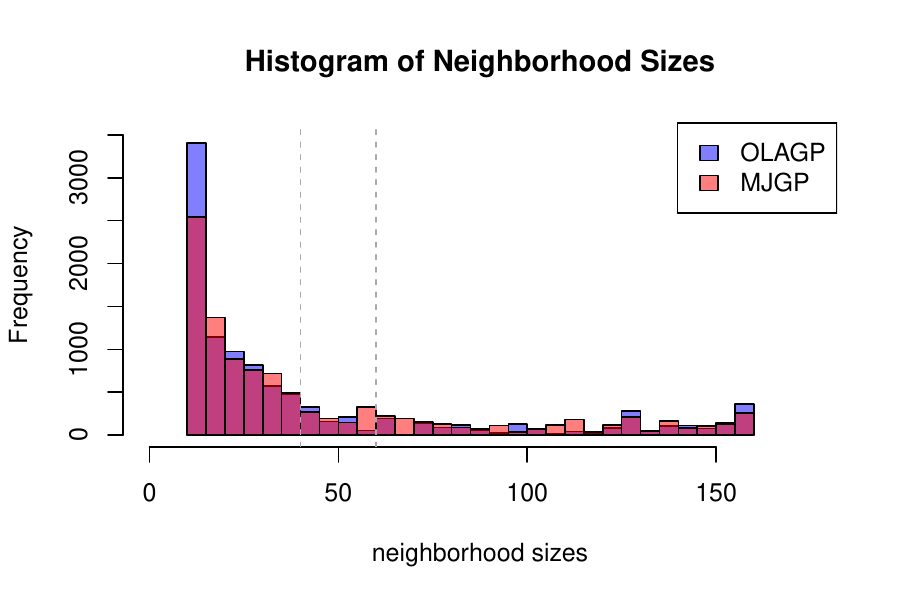}
\end{subfigure}
\hspace{1cm}
\begin{subfigure}[c]{.45\textwidth}
\centering
\footnotesize
\begin{tabular}{ c|r|r} 
 \textbf{Size} & \textbf{OLAGP} & \textbf{MJGP}  \\ 
 \hline
12--20 & 4548 & 3914  \\ 
 \hline
 21--40 & 2838 & 2853\\ 
 \hline
 41--60 & 745 &933\\
 \hline
 61--80 & 414&697\\
 \hline
 81--100 & 319&298\\
 \hline
  101--120 & 139&396\\
 \hline
  121--140& 511&544\\
 \hline
  141--160& 687& 566\\
 \hline
  \end{tabular}
\end{subfigure}
\caption{Distribution of OLAGP and MJGP [Section \ref{sec:mod}] estimated neighborhood sizes.}
\label{fig:neighbors}
\end{figure}

Now consider how all 10,201 neighborhoods, over the whole grid $\mathcal{X}$,
compare to the \texttt{laGP} software default of $n=50$.  Figure
\ref{fig:neighbors} shows how that choice (between the vertical
lines) is not advantageous for the Phantom data. Many smaller,
but also many much larger neighborhoods are estimated by OLAGP, and
with much greater frequency.  The figure also provides neighborhoods for
the new method (``MJGP'') discussed next in Section \ref{sec:mod}.

Before turning to that, we note anecdotally that OLAGP performs worst at or
near the manifold of discontinuity (See Appendix \ref{sup:phanfits}). This is
because it cannot choose $\hat{n}(x)$ small enough to exclude points from the
opposite side of the manifold without maintaining a viable minimum ($n_{\min}
= 12$) for a GP fit. Observe how the neighborhood for $(0.39, 0.34)$ contains
several points in the red region.   At $\hat{n}(x)=14$ for that location, 
the neighborhood cannot shrink much further. Consequently, the
resulting mean predictions (\ref{eq:2newpred}) are weighted averages of red-
and yellow-region response values. This will yield predictions between 0.4 and
0.6, which we know from Figure \ref{fig:phantom} are inaccurate.  Shrinking
neighborhoods farther, even below $n_{\min} = 12$, is a double-edged sword:
less data means greater reactivity, but also lower information content.  It
would be better to allow a more flexible notion of neighborhood, beyond simple
Euclidean distance in $X_N$. Next, we suggest how that can be done without
fundamentally altering the LAGP setup by introducing a latent variable.  To
foreshadow, observe that the ``MJGP'' neighborhood sizes summarized in
Figure \ref{fig:neighbors} show fewer small neighborhoods, {\em and} fewer
large neighborhoods compared to OLAGP.

\section{Modular Jump Gaussian Processes}\label{sec:mod}

Next we target disparate levels of data from a jump process: first identifying
 marginally, then predicting spatially, ultimately constructing a feature
that can be used as a predictor in a (LA)GP.

\subsection{Identifying and predicting levels}\label{sec:lev}

Data $Y_N$ from a jump process can often be partitioned marginally (i.e., $y$
without $x$), which can inform a more complex spatial partition over $x$. To
see this, consider again the histogram of the marginal $Y_N$ for the
Phantom data [Figure \ref{fig:phantom}, {\em right}]. There are
two, roughly symmetric, well-separated densities, centered at 0.2
and 0.8. These modes are so well separated you can partition it visually, 
say at $y=0.5$.

Visual partitioning of the marginal response will not always be possible, as
we shall show momentarily, and it is unsatisfying to require a human in the
loop of an otherwise easily automatized procedure.  We assume generally that
data from jump processes can be captured, marginally, as a mixture
distribution with two components. Under this assumption, each observation $y_i
$ is associated with a latent variable $\ell_i \in \{1,2\}$ describing the
component $y_i$ belongs to. Of interest to us is the quantity
$\mathbb{P}(\ell_i = j)$ for $j =1,2$, the probability that observation $y_i$
belongs to component $j$.

A model-based clustering method can be used to identify the level labels of
each observation in $Y_N$. We prefer EM
\citep{dempster1977maximum,mclachlan1997algorithm} under Gaussian mixtures
\citep[][Chapter 2]{mclachlan1988mixture}, extracting $\mathbb{P}(\ell_i=j)
\equiv
\tau_{ji}$ in the notation of that paper. Then, let
$c_i = \underset{j \in \{1,2\}}{\arg \max} \, \mathbb{P}(\ell_i = j)$, the
highest probability component for $y_i$. One view of the results of such a
clustering is indicated via the marginal histogram in Figure
\ref{fig:phantom}, where color indicates each $c_i$, $i=1, \dots, n$; bars
colored red are sorted into component 1, yellow into component 2.
 
\begin{figure}[ht!]
\centering
\includegraphics[scale=0.43, trim =  0 10 35 20, clip]{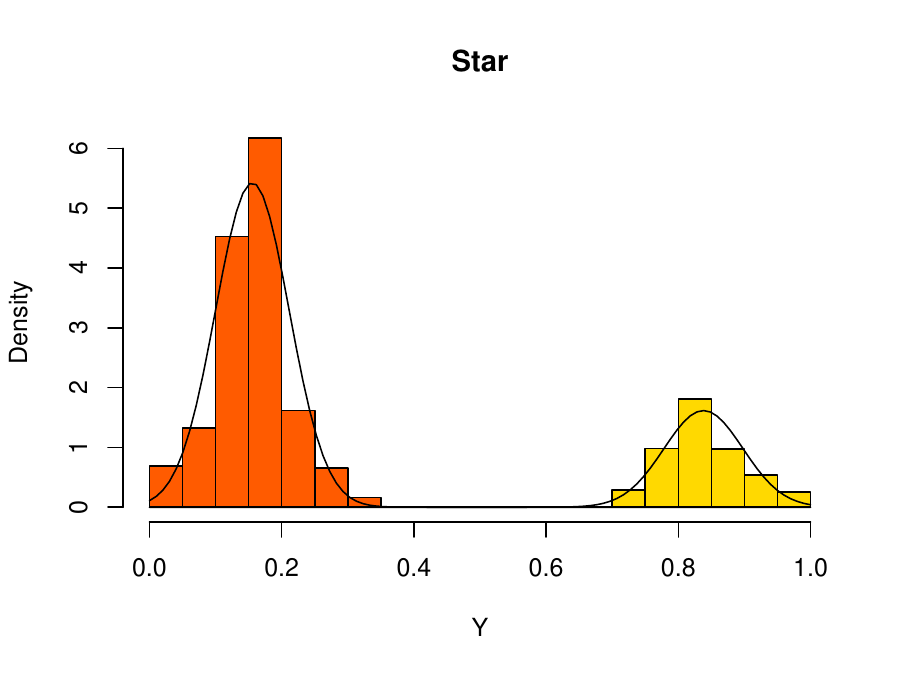}
\hfill
\includegraphics[scale=0.43, trim =  30 10 35 20, clip]{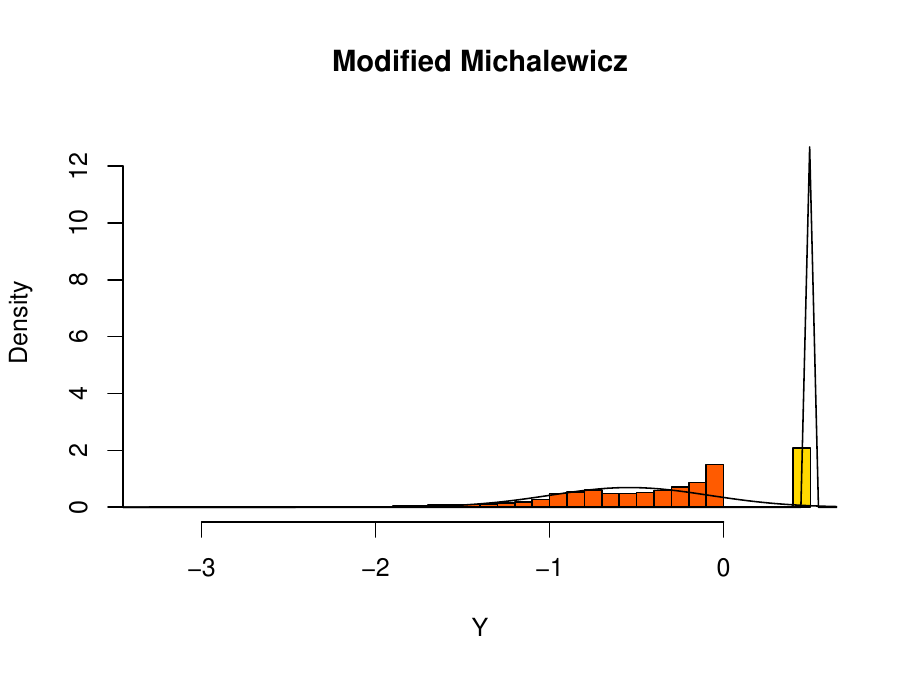}
\hfill
\includegraphics[scale=0.43, trim =  30 10 35 20, clip]{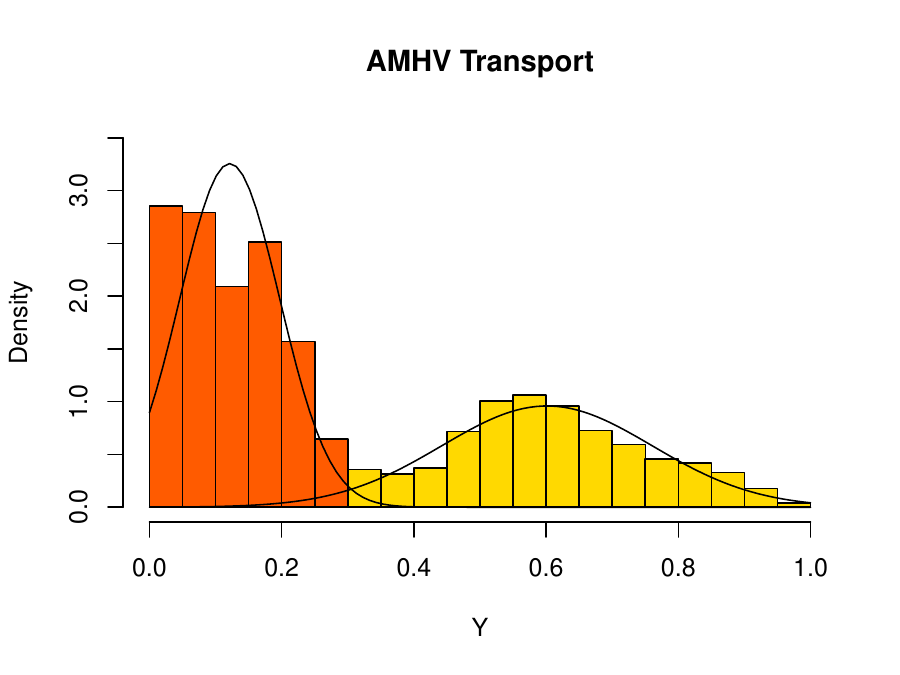}
\caption{{\em Left:} Star 
histogram; {\em Middle:} Modified Michalewicz function histogram; {\em Right:} AMHV transport  
histogram. Colors represent the two levels identified by each clustering algorithm. }
\label{fig:hists}
\end{figure}

Figure \ref{fig:hists} provides similar views for three other examples
entertained in our empirical work in Section \ref{sec:results}, including our
motivating real-data set (right panel). In these plots, we also include lines
showing the two-component Gaussian mixture estimated by EM. Like the Phantom,
the Star example (left panel) is relatively simple.  Visual partitioning would
have sufficed. The other two benefit from a deliberate, model-based approach.
In both of these examples (middle and right panels), it may well be that more
than two components (or non-Gaussian ones) would be preferable.  However, we
aim to show that a crude marginal ``partition'' is powerful when this
information is reincorporated spatially, which we describe next.

Let $c_i \in \{1, 2\}$, $i=1,\dots,N$ denote the level learned for each
$y_i \in Y_N$ via model based clustering, with vector $C_N$ holding
all $N$ levels. For example, we could have $c_i =
2$ correspond to ``yellow'' and 1 for ``red'', or vice-versa. Then, pair these
with inputs $X_N$ to form $(X_N, C_N)$, comprising ``data'' for a spatial
classification task.  Here we shall write $\hat{f}(\cdot) =
\mathrm{class}(X_N, C_N)$, generically, for the predictive equations of a
fitted, spatial classification.  One option is (basis-expanded) linear
logistic regression, or logistic GP classification \citep[e.g][Chapter
3]{williams2006gaussian}. Any classification method works, although 
the accuracy of a particular method on a particular data set should be 
tested, for example by evaluating 
classification accuracy on $X_N$, before proceeding.  Our preferred
implementation is detailed in Section \ref{sec:results}.  We may then use
$\hat{f}$ to predict component membership at testing locations $\mathcal{X}$:
$\hat{f}(x)=\mathbb{P}(y(x) = 1)$ for any $x \in \mathcal{X}$. Predicted
probabilities for the Phantom data are shown in the left panel of Figure
\ref{fig:mjgp} with the manifold of discontinuity overlaid.
\begin{figure}[ht!]
\centering
\includegraphics[scale = 0.45, trim =  5 10 20 20, clip]{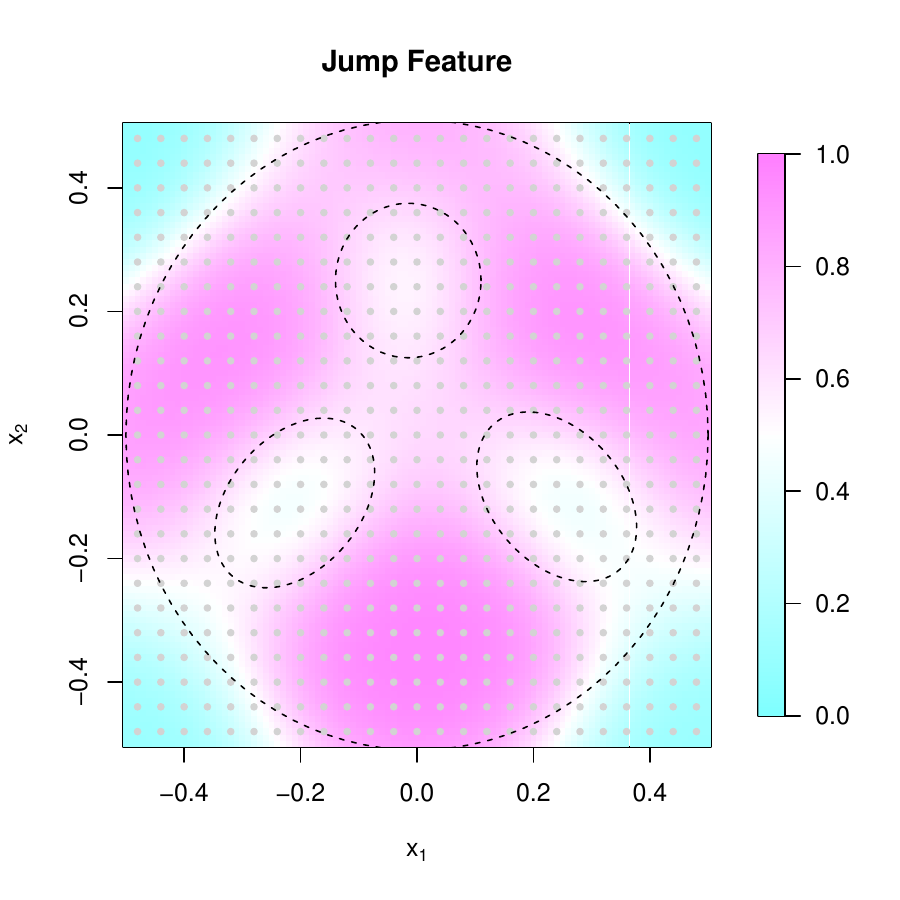}
\hfill
\includegraphics[scale = 0.45, trim =  45 10 40 20, clip]{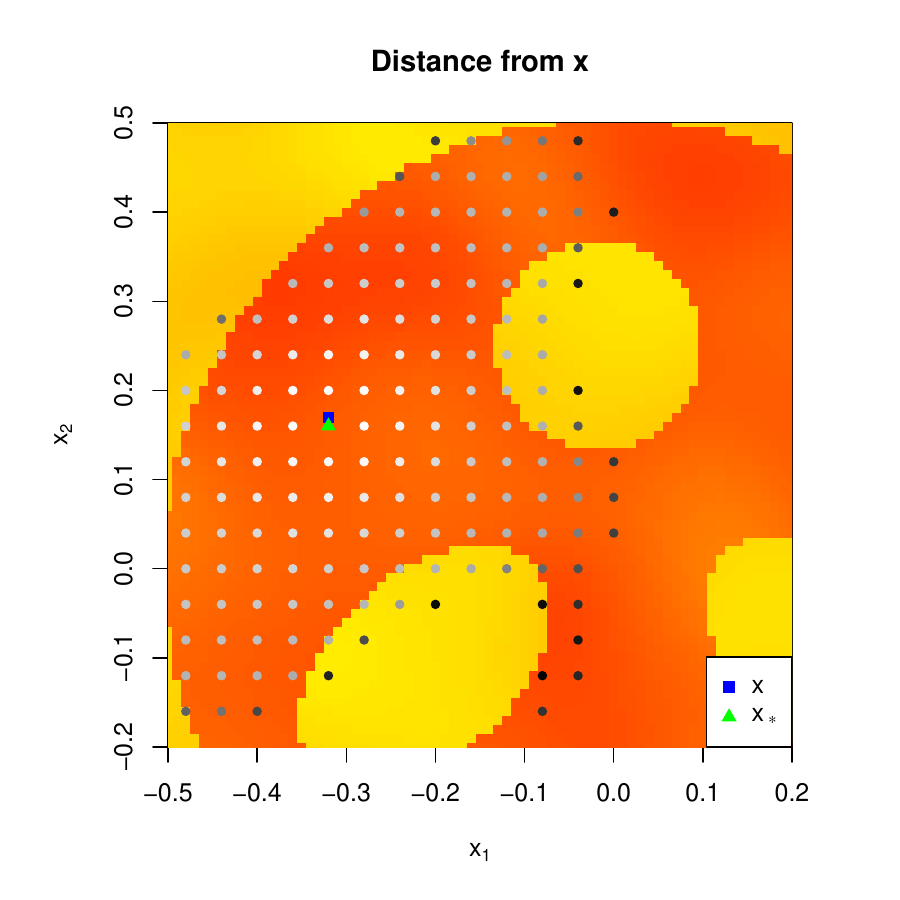}
\hfill
\includegraphics[scale = 0.45, trim =  45 10 40 20, clip]{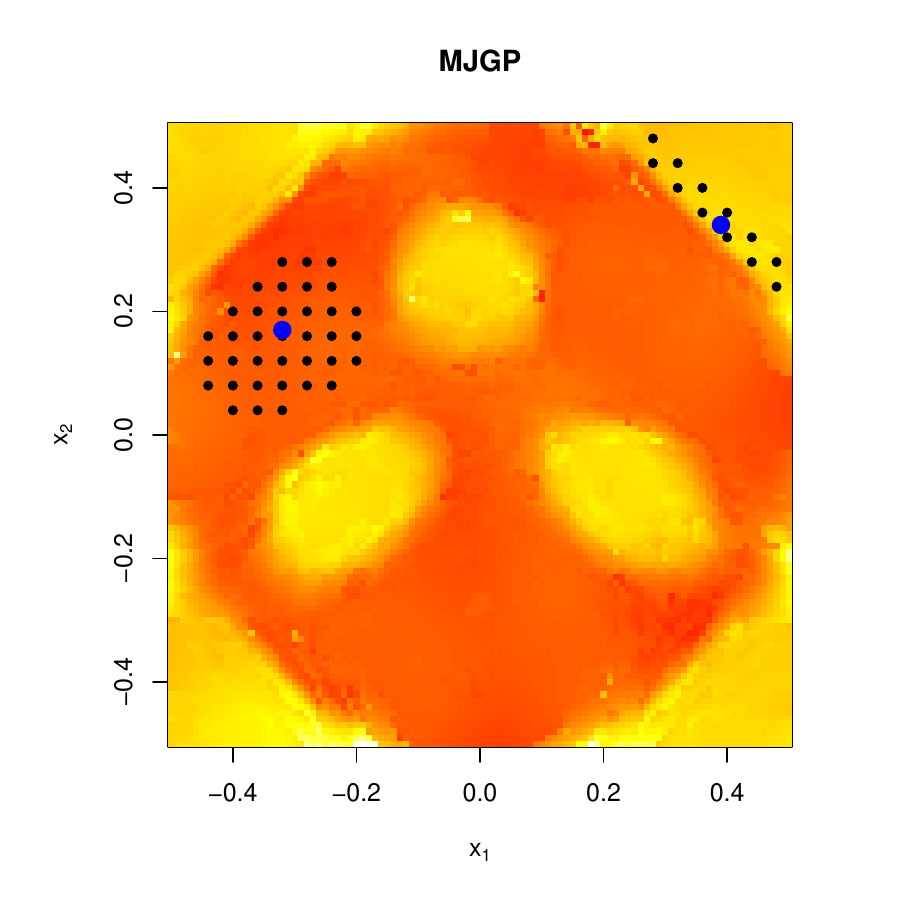}
\caption{{\em Left}: Jump feature.
{\em Center}: Nearest 160 neighbors to $x=(-0.32,0.17)$ after adding the jump feature. 
White points are closest to $x$, and black points are furthest away.
{\em Right}: MJGP fit. Blue dots are two predictive
locations and the surrounding black dots are the selected neighborhood.}
\label{fig:mjgp}
\end{figure}
The other panels will be discussed momentarily, in Section \ref{sec:feat}.
Observe how the contours in this plot are similar in shape to the original
data [see Figure \ref{fig:phantom}], with values close to 0 in deeply yellow
regions and values close to 1 in deeply red regions. In particular, note that
$\hat{f}(x) = 0.5$ for locations $x$ near the manifold of discontinuity, where 
component membership is least certain.

\subsection{A new feature for GP regression}\label{sec:feat}

Now recall that stationary kernels enforce a covariance structure that is
based only on the pairwise distances between points (\ref{eq:2gp}). As the
distance between two points increases, their covariance decreases. This
covariance is then used in the GP mean prediction (\ref{eq:2newpred}), where
it serves as a weight. This means that points that are close to the predictive
location have a lot of influence on its mean, and points that are far away
have little influence. Therefore, if we can find a way to modify the distances
between points so that points from the same component are closer together than
those from opposite components, this covariance structure would naturally
place more influence on points from the same component than those from the
opposite one. A simple way to modify these distances without sacrificing the
original distance structure is by augmenting with a new input feature
\citep{bornn2012modeling}.

This new variable should be related to component membership. We propose using
the probability of component membership [Section \ref{sec:lev}], which should
be high for observations likely to be in one component and low for the other.
In our example, classifier $\hat{f}(x)$ predicts the probability each location $x$
belongs to the red region; it is large (close to 1) in deeply red
regions and small (close to 0) in deeply yellow regions. Suppose we treat
those values as a new feature, comprising a new column of $X_N$ for GP modeling.
In the new, higher dimensional input space, points in the red region will be
closer together than those from the yellow region, and vice versa. Under the
likelihood (\ref{eq:2lik}) and when predicting (\ref{eq:2newpred}), points
from the same component will have more influence on each other than on points
from opposite components.

For this to work, we need to know the probability of component membership for
both $Y_N$ and $\mathcal{Y}$. Section \ref{sec:lev} provides estimates of
$C_N$, for $Y_N$, and $\hat{f}(\mathcal{X})$, for $\mathcal{Y}$. The issue is
that $C_N$ is not a probability; it is a vector consisting of 1s and 2s
describing the most likely component a point in $Y_N$ belongs to. However,
$\hat{f}(\cdot)$ may be used to predict, in-sample, the probability that $Y_N$
belongs to a particular component. Vectors $\hat{f}(X_N)$ and
$\hat{f}(\mathcal{X})$ comprise the training and testing jump features for
$X_N$ and $\mathcal{X}$, respectively. These may be appended as new columns
$X \equiv [X, \hat{f}(X)]$ and $\mathcal{X} \equiv [\mathcal{X},
\hat{f}(\mathcal{X})]$, overloading the notation somewhat.

To illustrate the effect this new feature has, consider the center plot of
Figure \ref{fig:mjgp}, again focusing on $x=(-0.32,0.17)$ and its nearest
neighbor $x_*$. We also plot the 160 NNs to $x$, color coded by
the rank of their distance from $x$. The points that are the lightest are
closest to $x$, and those that are the darkest are furthest away. Note that
these distances are calculated prior to hyperparameter fitting. Observe first
that the neighborhood is no longer circular; it has become oval-shaped to
avoid the yellow region. Any points that are in the yellow region are
much further away than their 2d distance would suggest. Their colors are
darker than nearby points the red region, and thus farther from $x$.

We generically notate the fitting stage as $\mathrm{GP}([\mathcal{X},
\hat{f}(\mathcal{X})];[X_N,\hat{f}(X_N)], Y_N )$, where ``GP'' can stand in
for any type (e.g., global, LAGP, OLAGP) depending on the problem at hand. We
refer to this method, beginning with the clustering algorithm and ending with
the final GP fit, as a ``MJGP''.  Algorithm \ref{alg:mjgp} provides the
details for quick reference.

\begin{algorithm}
\caption{Outline of MJGP}
\label{alg:mjgp}
\begin{algorithmic}[1]
\Require{$X_N, Y_N, \mathcal{X}$}

\State{$\hat{C}_N = \text{clust}(Y_N)$} 
\Comment{perform clustering algorithm on marginal $Y_N$, yielding $C_N \in \{1,2\}^N$}
\State{$\hat{f}(\cdot) = \text{class}(X_N, C_N)$}
\Comment{build classifier mapping $X_N$ to $C_N$, yielding $\hat{f}(x) = \text{Pr}(y(x) = 1)$}
\State{$\text{GP}([\mathcal{X},\hat{f}(\mathcal{X})];[X_N,\hat{f}(X_N)],Y_N)$}
\Comment{fit any GP (e.g. OLAGP) with jump feature}
\end{algorithmic}
\end{algorithm}

The right panel of Figure \ref{fig:mjgp} provides a MJGP fit on the
Phantom data, contrasting with Figure \ref{fig:olagp}. In this case the GP
used is an OLAGP, chosen to illustrate the change in neighborhood structure
induced by the MJGP.  Here, in Figure \ref{fig:mjgp}, both
neighborhoods are smaller, but are completely contained to the same side of the
manifold. MJGP has improved prediction both visually and
via RMSPE (see Appendix \ref{sup:phanfits}) over
the global GP, LAGP, and OLAGP.

Now return to Figure \ref{fig:neighbors}, summarizing all 10,201 $\hat{n}(x)$
for OLAGP and MJGP. Again, the choice of $n=50$ (the LAGP default)
is not particularly advantageous, except for a small percentage of $x
\in \mathcal{X}$.  Notice that MJGP slightly prefers more
mid-range neighborhood sizes than the optimal neighbors GP. In particular,
many fewer small $\hat{n}(x)$ are chosen since the neighborhood shape is no
longer circular in the original input space (i.e., ignoring the new feature).
This facilitates excluding points from the opposite component without
resorting to extremely small neighborhood sizes.

Note that while we have thus far chosen examples where the magnitude of the jump 
is large compared to the overall variability of the data, that is not required 
to use MJGP. Data sets with large jumps will see the most improvement 
when compared to regular GP methods, but using MJGP when jumps are 
small or nonexistent will not lead to overfitting. See examples of MJGP 
fit to data with small or nonexistent jumps in Figures \ref{fig:smalljump} and 
\ref{fig:nojump} in Appendix \ref{sup:special}. Although the benefit of using MJGP 
lessens as the size of the jump decreases, it will not compromise results when 
jumps are not present.

We chose to use the predicted probability of component membership 
$\hat{f}(X_N)$/$\hat{f}(\mathcal{X})$ as a new jump feature. However, we could
have instead used the predicted component membership to partition the input
space into two regions, and fit separate GPs to each region. If these
partitions perfectly matched the manifold of discontinuity, this would
outperform MJGP. The problem occurs when the manifold of discontinuity is
complex, like in the Phantom example. In this case (see Appendix
\ref{sup:partition}), the classifier cannot accurately learn the manifold of
discontinuity.  MJGP improves upon a partitioned GP by using the predicted
probability of component membership, which incorporates the uncertainty of the
classifier. Accounting for this extra uncertainty in the model ultimately
improves prediction over partitioned GPs.

Our MJGP is purposefully generic: clustering, classifying, predicting, and
then fitting  with a new feature. We mentioned some of the possible choices
for each step (e.g. EM, CGP, and OLAGP), but the choice of methods for
clustering, classifying, and regression should be tailored to the context of the
problem and the data involved.
We aim to show that our preferred instance of
MJGP is an improvement on existing methods in the context of surrogate
modeling of jump processes from the recent literature.

\section{Implementation and empirical benchmarking}\label{sec:results}

Here we present three simulated and one real-world example(s) to demonstrate
the effectiveness of OLAGP and MJGP. A summary of each data
set/experiment is included in Table \ref{tab:mjgp_details}; we do not list the 
clustering algorithm because it did not vary between experiments. Input and response 
ranges are listed prior to any scaling. Details are
provided in separate sub-sections to follow. Broadly, for each data set, we
conduct a Monte Carlo (MC) experiment which repeatedly randomly partitions the
data into 90:10 training and testing sets. In each each repetition, we sample
a random subset of 90\% of the data set to be used as the training set, and
use the remaining 10\% as the testing set.  The number of MC trials depends on
the computational expense implied by the training data set size $N$, which
varies in the experiments. We present five competing methods:

\begin{itemize}
\item \textbf{Global GP} via
 \texttt{newGPsep} in \texttt{laGP} \citep{gramacy2016lagp},
using defaults. This was excluded from the Modified Michalewicz experiment due to 
size constraints.
 \item \textbf{LAGP} via \texttt{aGPsep} 
 in \texttt{laGP} \citep{gramacy2016lagp}, with 
 \texttt{method = "nn"} and all other defaults.
 \item \textbf{OLAGP} from Section \ref{sec:opt} with $n_{\min}=12$ and 
 $n_{\max}=160$.
 \item \textbf{Original JGP (OJGP)} via \citet{park2022jump}'s {\sf MATLAB} library
 using CEM and quadratic boundaries.
 \item \textbf{MJGP} from Section \ref{sec:mod} with further details provided in Table \ref{tab:mjgp_details}.
\end{itemize}
All experiments were completed in \textsf{R}. Code is provided at
  \url{https://bitbucket.org/gramacylab/jumpgp}. In this code we 
  provide functions {\tt aGP.optneigh.par()}, {\tt mjgp()}, and {\tt jgp()} to 
  run OLAGP, MJGP, and OJGP, respectively. OLAGP is parallelized over 
  predictive locations $x \in \mathcal{X}$, turning
statistical independence into computational exclusivity via  \texttt{foreach}
\citep{microsoft2022foreach}. OJGP leveraged \texttt{R.matlab} \citep{bengtsson2022rmatlab}. 
  
\begin{table}[ht!]
\begin{tabular}{ |c||c|c|c|c|c|c| c|} 
 \hline
 \textbf{Data} &  \boldmath$N$ & \boldmath$d$ & \boldmath$X_N\in$ 
 & \boldmath$Y_N\in$ & \textbf{Classifier} & \textbf{GP Type}  & \textbf{Reps} \\ 
 \hline \hline
 Phantom & 10,201& 2 & $[-0.5,0.5]^2$  & $(-17,26)$&  CGP & OLAGP & 100\\ 
 \hline
 Star & 10,201&2 & $[0,1]^2$  & $(-16,15)$& CGP & OLAGP & 100\\ 
 \hline
 Mod.~Michalewicz & 100,000&4 & $[0,\pi]^4$ & $(-3.25,0.5)$ & RF &  OLAGP & 30\\
 \hline
AMHV Transport & 9,503&4 & \makecell{$[50,400] \times [1,5] \times$ \\  $[0.25,5] \times [0.5,1.5]$} & $(100,517)$& CGP & Global GP & 100\\
 \hline
\end{tabular}
\caption{Implementation summary for each data set/experiment.}
\label{tab:mjgp_details}
\end{table}

 For MJGP, we utilize EM-based
clustering and either a CGP \citep{williams1998bayesian} or a random forest
\citep[RF;][]{breiman2001random} for classification via \texttt{mclust}
\citep{scrucca2023model}, \texttt{kernlab} \citep{karatzoglou2004kernlab} and
\texttt{randomForest} \citep{liaw2002rf}, respectively. We only
prefer a RF over a CGP when $n$ is large, for example during the
Modified Michalewicz experiment. Again, see Table \ref{tab:mjgp_details}.
Model performance is measured out-of-sample using 
$
\mathrm{RMSPE} = (\frac{1}{m}\sum_{x \in \mathcal{X}} (y(x) - \mu(x))^2)^{1/2},
\label{eq:4rmse}
$
for which lower is better. Computation times are in Appendix \ref{sup:comptime} 
and prediction interval coverage is in Appendix \ref{sup:pred}.

\subsection{Phantom}\label{sec:phantomsim}

In Sections \ref{sec:opt}--\ref{sec:mod}, we assessed model performance
(visually and with RMSPE) on the Phantom data for a single train--test
partition. The outcome of a full MC experiment is summarized by boxplots in
the left panel of Figure \ref{fig:phan_oos}.
 \begin{figure}[ht!]
\centering
\includegraphics[scale = 0.6, trim =  25 15 30 50 clip]{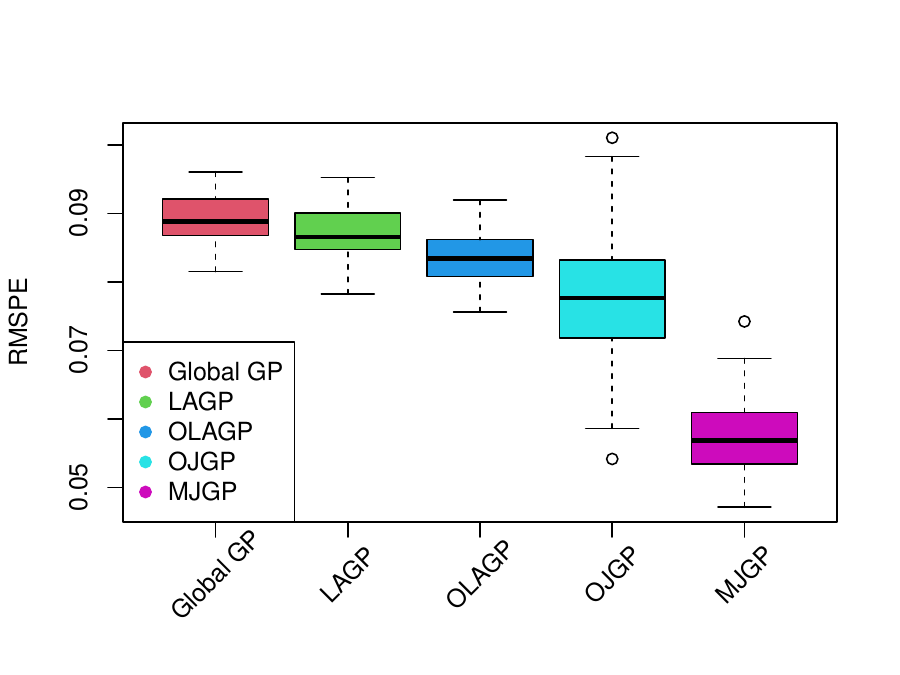}
\hspace{1cm}
\includegraphics[scale = 0.6, trim = 25 15 30 50, clip]{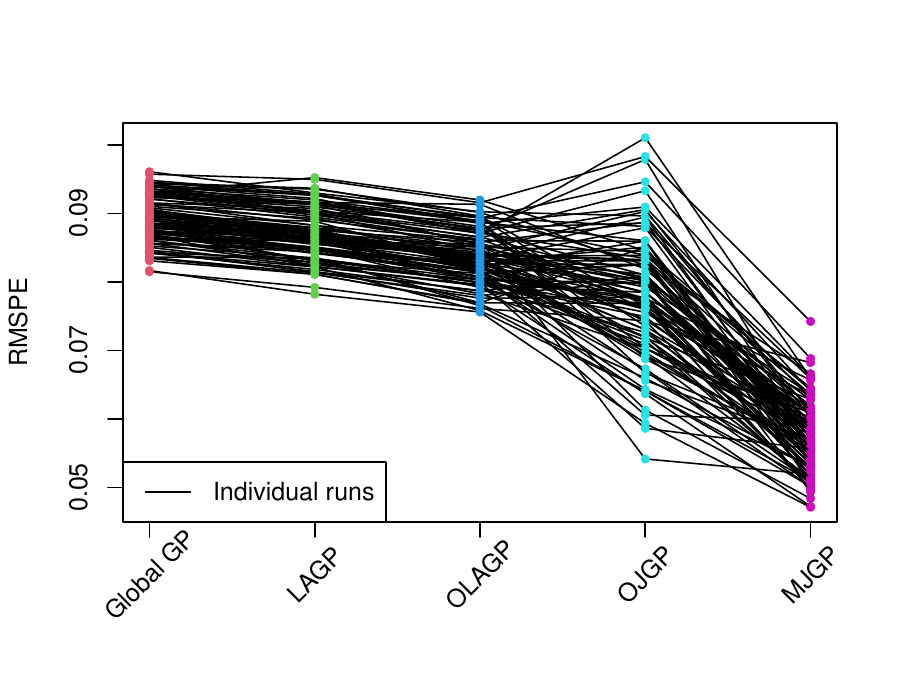}
\caption{{\em Left:} MC experiment on the Phantom data set; 
{\em Right:} A line plot showing RMSPE at individual runs for each of 
the five methods. MJGP performs best in all runs.}
\label{fig:phan_oos}
\end{figure}
For a more granular view, the right panel draws black lines connecting RMSPEs
from the same rep, allowing us to see the best performing method for every
training/testing set combination. We conclude that most of the overlap of the
boxplots is due to variability in the training/testing split.  The ordering of
accuracy from one MC rep to the next is largely consistent across methods.

In these plots (and all plots in this section), our two methods are shown
third from right (OLAGP) and furthest right (MJGP). Both show improvement over
the global GP and LAGP, although OLAGP is typically outperformed by OJGP. This
is unsurprising, since OJGP was designed specifically for data from jump
processes, and OLAGP was not. But notice that MJGP consistently outperforms
OJGP, here and broadly over the next few subsections. Looking first at the
line plot on the right of Figure \ref{fig:phan_oos}, we see that for every
data set split, predictions made by MJGP are more accurate than those from
OJGP. Additionally, the boxplot shows that OJGP has higher variability than
MJGP. OJGP makes predictions by estimating a boundary line separating
points from the same and opposite components as the predictive location. Our  
choice of quadratic boundary yields more accurate predictions, but at
the expense of higher variability. For more on this distinction, see
\citet{park2022jump}.

\subsection{Star}\label{sec:starsim}

Now consider the ``Star'' data, which we created for this paper.  The goal was
to have something like the Phantom, but more angular. Details for how the data
set was created are provided in Appendix \ref{sup:star}. See the left panel of
Figure \ref{fig:starfit}, with marginal response histogram depicted earlier in
the left panel of Figure \ref{fig:hists}. Like the Phantom, the Star has two
well-separated regimes that form a clear shape on the response surface. But
the Star is more difficult to model because all points in the yellow regime are geographically
close to red ones.  It is hard to build a neighborhood inside the yellow portion of the 
Star.

\begin{figure}[ht!]
\centering
\includegraphics[scale = 0.45, trim =  25 10 40 20, clip]{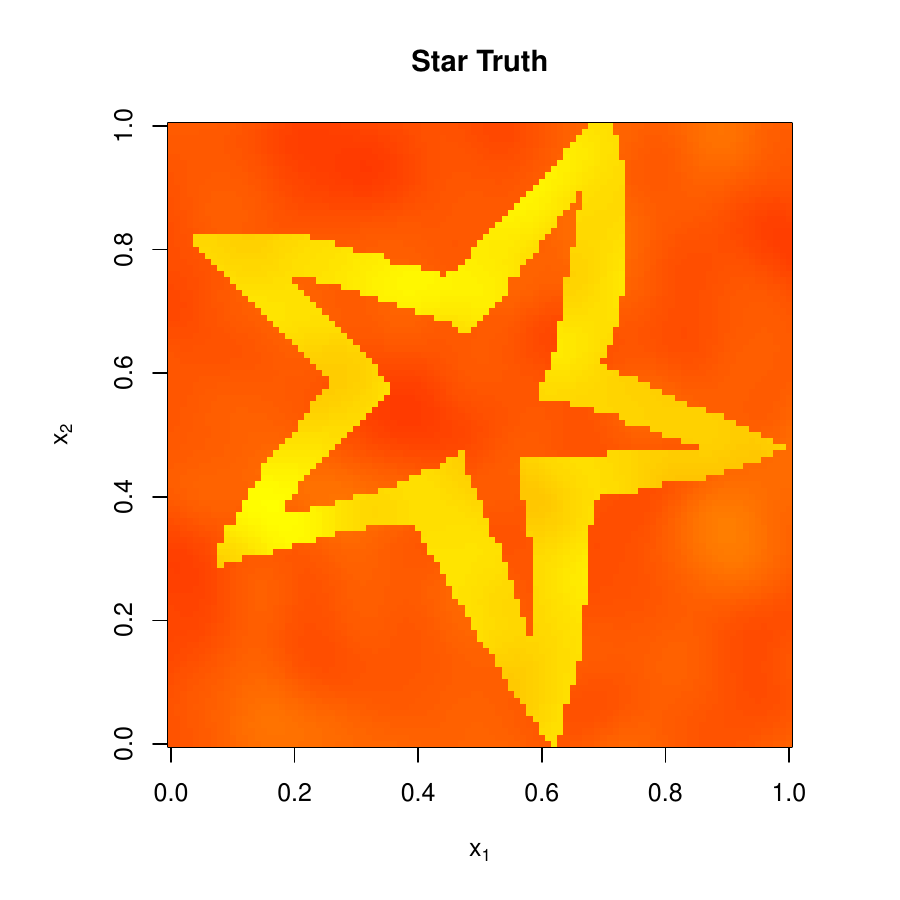}
\hfill
\includegraphics[scale = 0.45, trim =  45 10 40 20, clip]{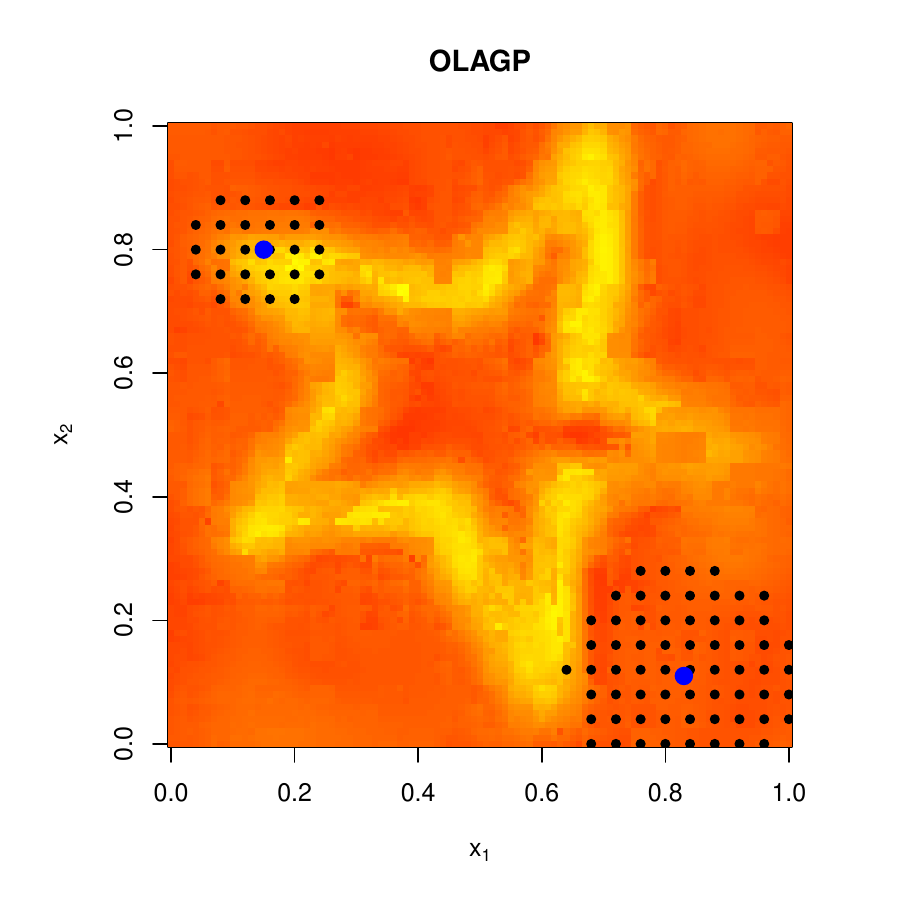}
\hfill
\includegraphics[scale = 0.45, trim =  45 10 40 20, clip]{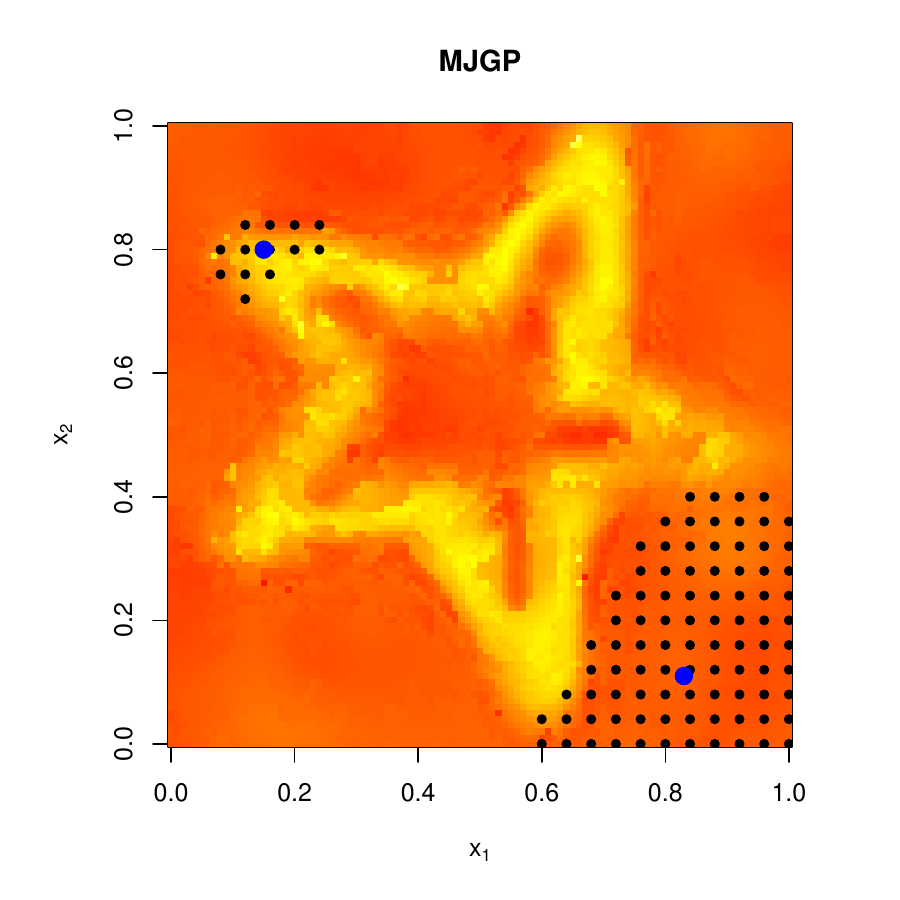}
\caption{Model fit and chosen neighborhoods for Star data set. {\em Left:} Ground truth; 
{\em Center:} OLAGP fit of Star data set; {\em Right:} MJGP fit 
of Star data set. Two predictive locations, in blue, are shown with their chosen neighborhoods,
in black.}
\label{fig:starfit}
\end{figure}

Repeating themes from the Phantom analysis, consider model performance for a
single Star train--test split. Mean fits for the global GP and LAGP are shown
in Figure \ref{fig:starorig} in Appendix \ref{sup:starfits}, whereas those
of OLAGP and MJGP are shown here in the right two panels of Figure
\ref{fig:starfit}. RMSPEs can be found in Appendix \ref{sup:starfits}, where we
see that the local methods perform well, with MJGP producing the
best predictions of the four. The scenarios depicted in right two panels
focus on $x=(0.15,0.8)$ and
$(0.83,0.11)$. Compared to LAGP, OLAGP has chosen a neighborhood of roughly
the same size for $(0.83, 0.11)$, and much smaller for $(0.15,0.8)$; however,
this smaller neighborhood still contains points from the red regime. 
MJGP provides an interesting change: a much smaller neighborhood size
for $(0.15, 0.11)$, and a {\em larger} neighborhood size for $(0.83, 0.11)$.
Both neighborhoods have adapted their shape to only include points from the
same regime, and this improves prediction overall.

\begin{figure}[ht!]
\centering
\includegraphics[scale = 0.6, trim =  25 15 30 50 clip]{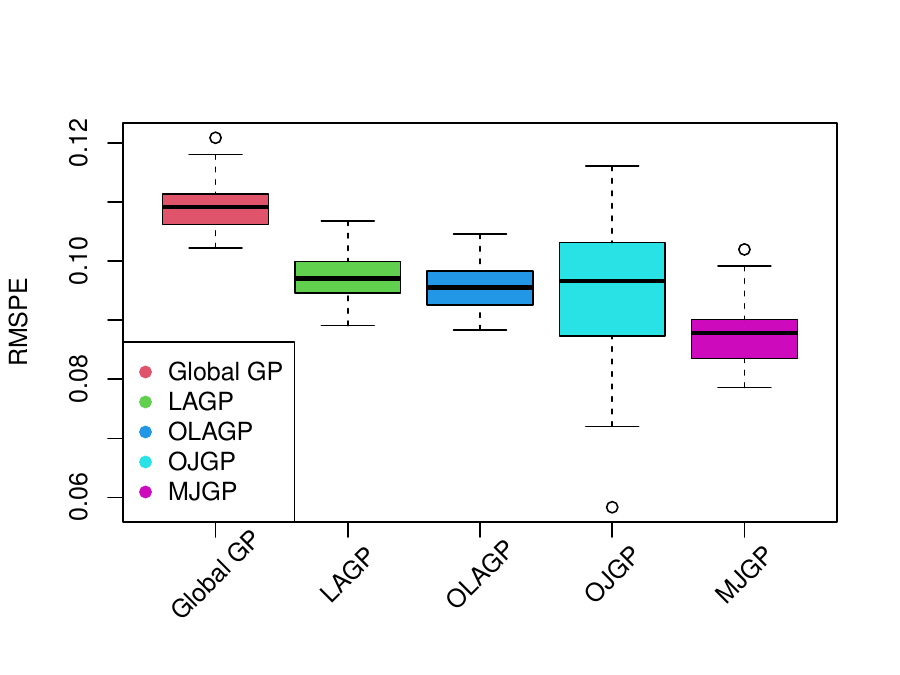}
\hspace{1cm}
\includegraphics[scale = 0.6, trim = 25 15 30 50, clip]{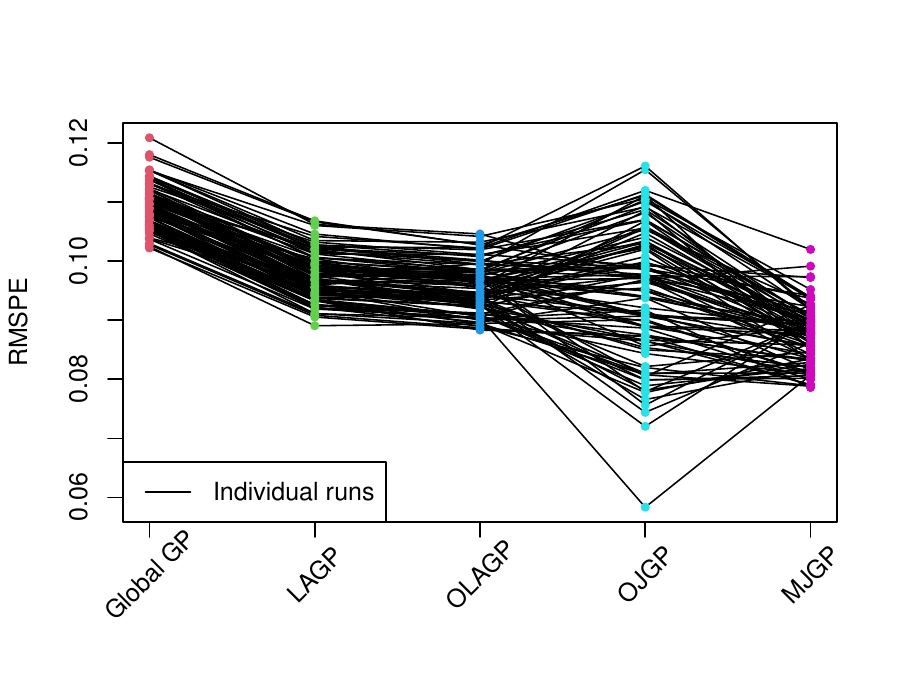}
\caption{{\em Left:} MC experiment on the Star data set;
{\em Right:} A line plot showing RMSPE at individual runs 
for each of the five methods. MJGP performs best in 82 of 100 runs.}
\label{fig:star_oos}
\end{figure}
 
Figure \ref{fig:star_oos} summarizes a MC experiment along similar lines.
Here again, OLAGP outperforms both the global GP and LAGP; the two JGP methods
typically outperform OLAGP. OJGP is again highly variable.
Although it outperforms MJGP more frequently than for the
Phantom (OJGP was better in 17 of 100 runs), 
MJGP is still the best performing method for the Star.

\subsection{Modified Michalewicz}\label{sec:michalsim}

For a higher-dimensional example, we adapted the so-called Michalewicz
function, which can be found in the Virtual Library of Simulation Experiments
\citep{simulationlib}.  Details can be found in Appendix \ref{sup:michal}; 
note that although the original function is in $[0,\pi]^4$, we scale it to fall 
in $[0,1]^4$.
Michalewicz is dimension-adaptive; here we present a 4d analysis to
match the real data analysis coming in Section \ref{sec:real}. An LHS design is used
with $N=100,000$ training inputs.  The marginal distribution of the response
is shown in the center panel of Figure \ref{fig:hists}. We cannot visualize 
the response surface in 4d, so instead we show a 2d 
version in the left panel of Figure \ref{fig:datavis}.} 
\begin{figure}[ht!]
\centering
\includegraphics[align=c,scale=0.45, trim = 25 15 0 0, clip]{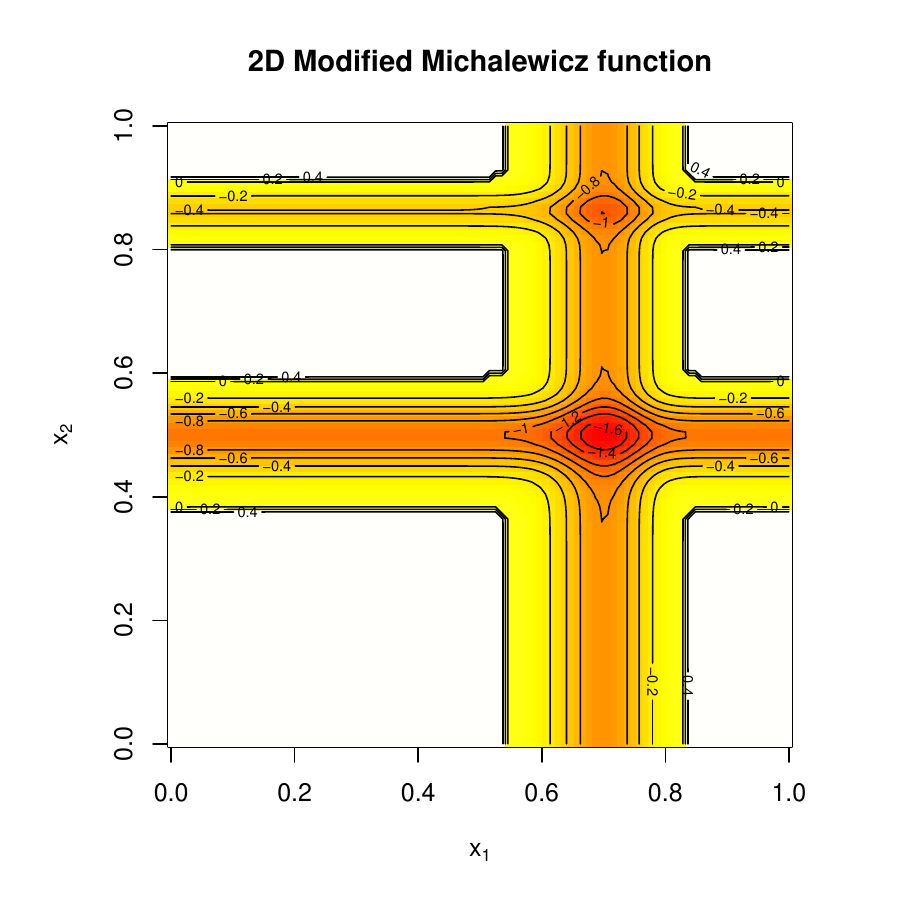}
\hspace{0.5cm}
\includegraphics[align=c,scale=0.35, trim = 0 0 0 0, clip]{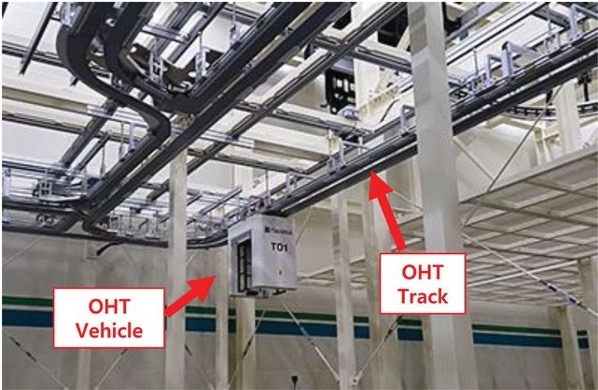}
\caption{{\em Left:} Modified Michalewicz response surface in 2d. {\em Right:} 
Image of AMHV transport device; taken from \citet{park2023active}.}
\label{fig:datavis}
\end{figure}
Unlike the previous
two data sets, the response is not plausibly composed of two Gaussian
components. Instead, it is comprised of a point mass around 0.5 with other
points scattered between in $[-4,0]$.  Nevertheless, we find that EM with two
Gaussian components works well in the MJGP setup. One goal of the
presentation here is to demonstrate how even a na\"ive marginal clustering often
works well in practice.

\begin{figure}[ht!]
\centering
\includegraphics[scale = 0.6, trim =  25 15 30 50 clip]{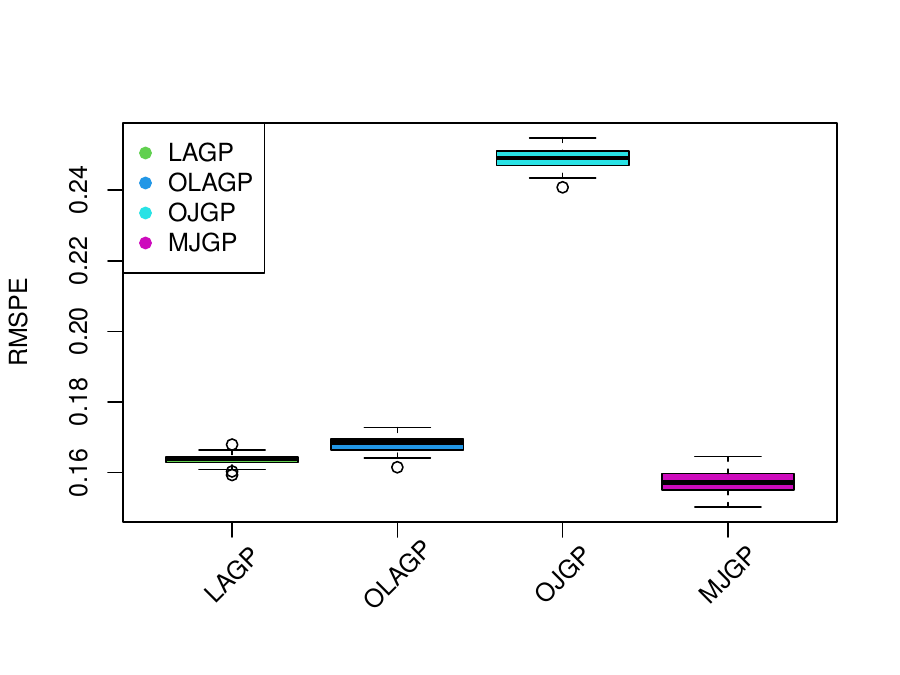}
\hspace{1cm}
\includegraphics[scale = 0.6, trim = 25 15 30 50, clip]{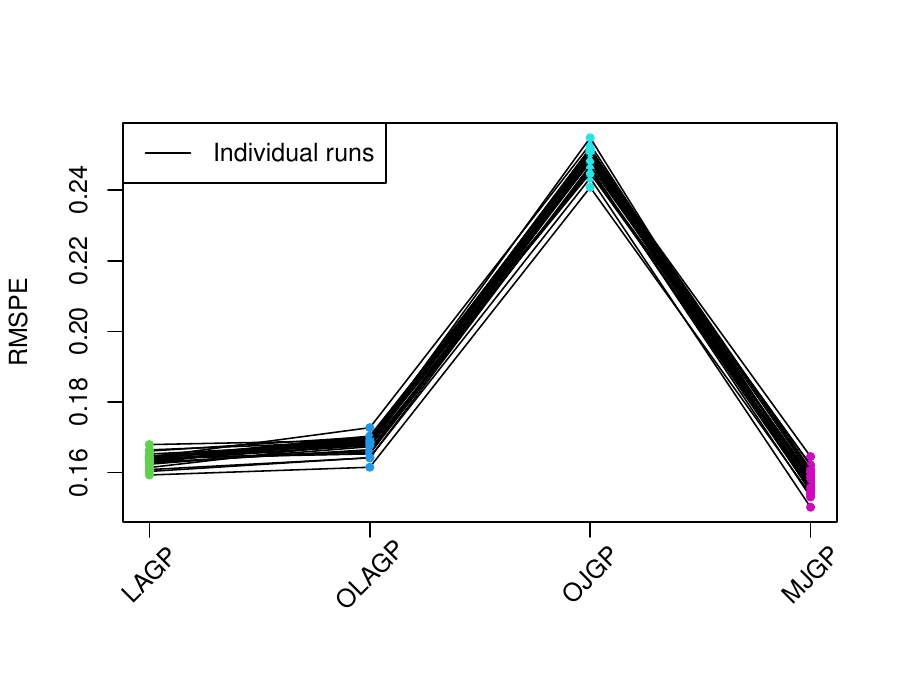}
\caption{{\em Left:} MC experiment on the Modified Michalewicz data set;
{\em Right:} A line plot showing RMSPE 
at individual runs for each of the four methods. MJGP performs best in 29 
of 30 runs.}
\label{fig:michal_oos}
\end{figure}

Since $N$ here is very large, we must omit the global GP as a competitor in this
experiment. Instead of a CGP we deploy a RF as our classifier.  Finally, we limit 
the number of MC reps to thirty. Results are
provided in Figure \ref{fig:michal_oos}. LAGP, OLAGP, and 
MJGP all considerably outperform OJGP. LAGP and OLAGP
perform comparably, although interestingly LAGP seems to slightly outperform
OLAGP in this case. We suspect this is due to the spiky nature of the response 
surface; see Figure \ref{fig:datavis}. In particular, we believe that OLAGP 
tends to overestimate the optimal neighborhood size for low values of the response 
(i.e., the spikes), yielding predictions much higher than the true response. MJGP 
performs best of all, yielding the lowest RMSPE in 29 of the 30 reps.

\subsection{AMHV transport}\label{sec:real}

Finally, consider the AMHV transport data first introduced in
Section \ref{sec:1}. Specifically, we used a larger version of the data set
described in Section 5.2 of \citet{park2023active}, scaled so that the
inputs lie in the unit hypercube; $d=4$ and $N=9{,}503$. 
A visual is provided in the right panel of
Figure \ref{fig:datavis}. Although $N$ is similar in magnitude to the
sample size of the Star and Phantom data sets $(N=10{,}201)$, the addition of
two dimensions means that there is a larger volume to fill.
In that sense, this example is the most sparse of the ones
entertained in this paper. Outputs are derived from a simulation of the
behavior of AMHVs in a particular facility, tallying the average transport
time of the AMHVs, including any time spent waiting for the vehicle to arrive.
\citet{park2023active} show that as production in the facility increases,
transport time, which depends on various design variables, increases and
becomes a major bottleneck to the entire manufacturing process. The four
variables here include vehicle speed, vehicle acceleration, minimum distance
required between vehicles, and the maximum search range of empty vehicles.

A view of the marginal distribution of the response for these data was
provided earlier, in the right panel of Figure \ref{fig:hists}. This is the
first example we have looked at where the two clusters overlap. Additionally,
the sparse nature of this data set means that local methods do not perform
nearly as well. (The closest points in $X_N$ to a predictive location $x$ may
actually be far and not representative.) This is a particular problem for
OLAGP, which uses $x_*$ as a proxy for $x$. Consequently, we prefer a global
rather than local GP in our application of MJGP for these data.

\begin{figure}[ht!]
\centering
\includegraphics[scale = 0.6, trim =  25 15 30 50 clip]{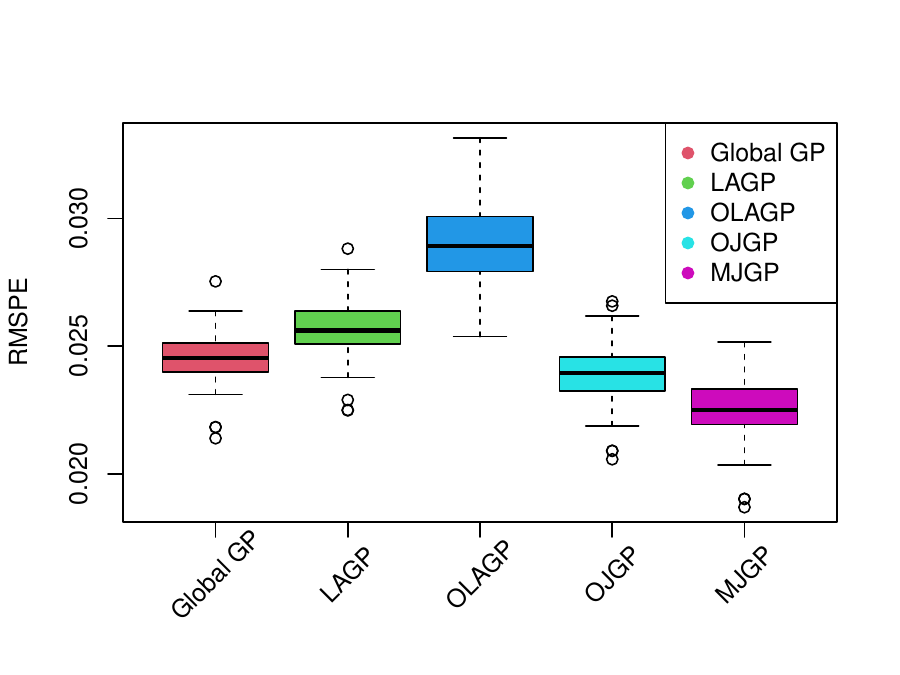}
\hspace{1cm}
\includegraphics[scale = 0.6, trim = 25 15 30 50, clip]{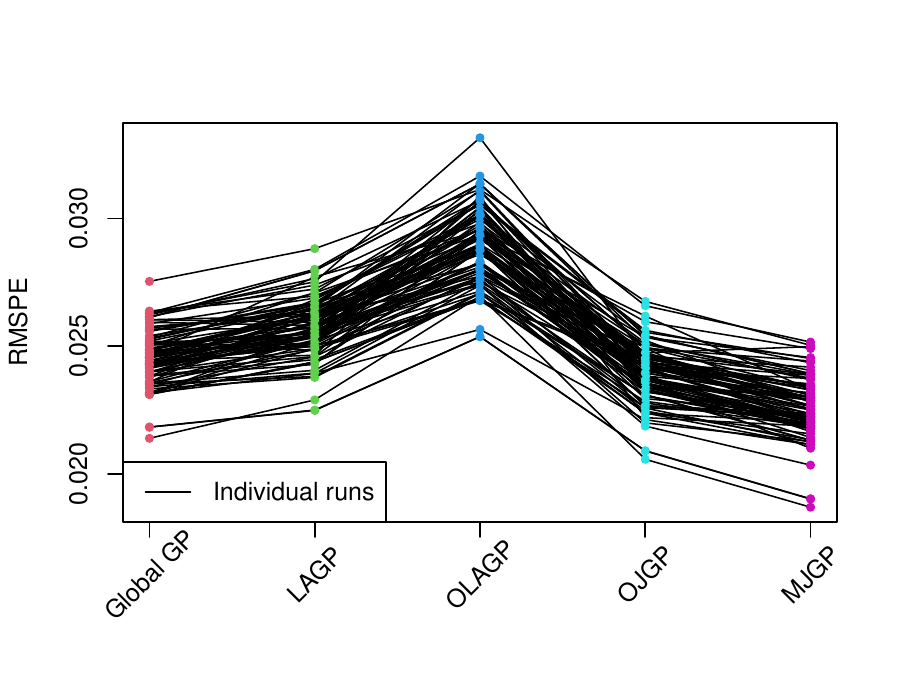}
\caption{{\em Left:} MC experiment on the AMHV transport data set; 
{\em Right:} A line plot showing RMSPE at individual runs for each of the
five methods. MJGP performs best in 99 of 100 runs.}
\label{fig:real_oos}
\end{figure}

The results of the experiment are shown in Figure \ref{fig:real_oos}. As
expected, the local methods do not perform as well as the global ones, with
OLAGP suffering most. But the global GP performs well, producing
comparable results to OJGP. MJGP produces only slightly
better results, but is consistent in doing so (lowest RMSPEs in 99 of the
100 reps).

\section{Discussion}\label{sec:discuss}

We introduced two ideas, adapting LAGP neighborhoods locally (OLAGP) and learning
a spatial feature from marginal clustering. Separately, both improve
predictions for data from jump processes.  They shine brightest when used
together.  We call this hybrid MJGP, since it involves many of the same
ingredients as OJGP, except with off-the-shelf subroutines
daisy-chained together.

One opportunity for future work pertains specifically to OLAGP. We mentioned
previously that this method has thus far only been implemented for
neighborhoods chosen using NN. However, it has been shown that NN neighborhoods can be
suboptimal compared to other subdesigns in the GP contexts
\citep{vecchia1988estimation,stein2004approximating}. \citet{gramacy2015local}
demonstrated that neighborhoods chosen via design criteria mean-square
prediction error (MSPE) often yield superior LAGP predictions; see their paper for 
more information on neighborhoods chosen using MSPE. We expect their superior performance 
to extend to the OLAGP context, but investigating that represents future work.

Other potential opportunities target MJGP more holistically. The
first involves extending MJGP to noisy functions. The main concern
in this case is whether the signal-to-noise ratio is large enough for the GP
to accurately capture jumps in the data. The second involves extending 
MJGP to allow more than two jump levels. Here we assumed that the
marginal response of data from jump processes could be described as a mixture
distribution with two components. However, it is possible for the response to
have more than two levels. 
We believe MJGP could be easily extended to this case. The first two 
steps of clustering and classifying would be very similar to what we described 
in Section \ref{sec:mod}. First, cluster $Y_N$ to get $C_N = (c_1,\dots,c_N)^{\top}$ 
for $c_i \in \{1, \dots, \ell \}$, where $\ell$ is the total number of levels. Then 
use a multiclass classification method -- both the CGP from {\tt kernlab} 
and the RF from {\tt randomForest} have this capability -- to fit
 $\hat{f}(\cdot)=\mathrm{class}(X_N,C_N)$. 
 Extract $\mathbb{P}(y(x)=j)$ for $j=1,\dots,\ell-1$. Let
  $\hat{f}_j(\cdot) = \mathbb{P}(y(x)=j)$ and use each to fit 
 $\mathrm{GP}([\mathcal{X}, \hat{f}_1(\mathcal{X}), \dots,
 \hat{f}_{\ell-1}(\mathcal{X})];[X_N, \hat{f}_1(X_N),\dots,\hat{f}_{\ell-1}(X_N)], Y_N )$.

Although we limit our focus to GPs for computer experiments, 
 there is extensive literature connecting GPs and geospatial statistics
 \citep{cressie1993statistics,gelfand2016spatial}. Methods exist which fit GPs
 on errors after accounting for mean structure \citep{matheron1969krigeage} or
 which can account for shifting local mean trends \citep{haas1990lognormal}.
 Methods even exist which similarly modify the distance-based kernels
 according to the covariance between points \citep{dumelle2023spmodel}. Some
 of these approaches/problems are reminiscent of our jump processes. However,
 geostatistical methods are typically restricted to 2 or 3 inputs. Computer
 experiments often contain more input variables.

\subsection*{Acknowledgements}
ARF and RBG are grateful for support from NSF 2152679. CP is grateful for 
support from NSF 2420358. Part of this research was initiated within the frame 
of the consortium in Applied Mathematics CIROQUO, gathering partners in 
technological research and academia in the development of advanced methods 
for Computer Experiments.

\subsection*{Declaration of Interest}
The authors report no conflict of interest.

\singlespacing
\bibliographystyle{apalike}
\bibliography{references}

\begin{thebibliography}{}

\bibitem[Banerjee et~al., 2004]{banerjee2004hierarchical}
Banerjee, S., Carlin, B., and Gelfand, A. (2004).
\newblock {\em Hierarchical Modeling and Analysis for Spatial Data}.
\newblock Chapman and Hall/CRC.

\bibitem[Banerjee et~al., 2008]{banerjee2008gaussian}
Banerjee, S., Gelfand, A.~E., Finley, A.~O., and Sang, H. (2008).
\newblock Gaussian predictive process models for large spatial data sets.
\newblock {\em Journal of the Royal Statistical Society Series B: Statistical
  Methodology}, 70(4):825--848.

\bibitem[Barnett, 1979]{barnett1979matrix}
Barnett, S. (1979).
\newblock Matrix methods for engineers and scientists.
\newblock {\em McGraw-Hill}.

\bibitem[Bengtsson, 2022]{bengtsson2022rmatlab}
Bengtsson, H. (2022).
\newblock {\em R.matlab: Read and Write MAT Files and Call MATLAB from Within
  R}.
\newblock R package version 3.7.0.

\bibitem[Bornn et~al., 2012]{bornn2012modeling}
Bornn, L., Shaddick, G., and Zidek, J.~V. (2012).
\newblock Modeling nonstationary processes through dimension expansion.
\newblock {\em Journal of the American Statistical Association},
  107(497):281--289.

\bibitem[Breiman, 2001]{breiman2001random}
Breiman, L. (2001).
\newblock Random forests.
\newblock {\em Machine learning}, 45:5--32.

\bibitem[Cole et~al., 2021]{cole2021locally}
Cole, D.~A., Christianson, R.~B., and Gramacy, R.~B. (2021).
\newblock Locally induced {G}aussian processes for large-scale simulation
  experiments.
\newblock {\em Statistics and Computing}, 31(3):33.

\bibitem[Cressie, 1993]{cressie1993statistics}
Cressie, N. (1993).
\newblock {\em Statistics for spatial data (Revised edition)}.
\newblock Wiley: Hoboken, NJ.

\bibitem[Damianou and Lawrence, 2013]{damianou2013deep}
Damianou, A. and Lawrence, N. (2013).
\newblock Deep {G}aussian processes.
\newblock In {\em Artificial Intelligence and Statistics}, pages 207--215.

\bibitem[Datta et~al., 2016a]{datta2016hierarchical}
Datta, A., Banerjee, S., Finley, A.~O., and Gelfand, A.~E. (2016a).
\newblock Hierarchical nearest-neighbor {G}aussian process models for large
  geostatistical datasets.
\newblock {\em Journal of the American Statistical Association},
  111(514):800--812.

\bibitem[Datta et~al., 2016b]{datta2016nearest}
Datta, A., Banerjee, S., Finley, A.~O., and Gelfand, A.~E. (2016b).
\newblock On nearest-neighbor {G}aussian process models for massive spatial
  data.
\newblock {\em Wiley Interdisciplinary Reviews: Computational Statistics},
  8(5):162--171.

\bibitem[Dempster et~al., 1977]{dempster1977maximum}
Dempster, A.~P., Laird, N.~M., and Rubin, D.~B. (1977).
\newblock Maximum likelihood from incomplete data via the {EM} algorithm.
\newblock {\em Journal of the royal statistical society: series B
  (methodological)}, 39(1):1--22.

\bibitem[Dumelle et~al., 2023]{dumelle2023spmodel}
Dumelle, M., Higham, M., and Ver~Hoef, J.~M. (2023).
\newblock spmodel: Spatial statistical modeling and prediction in {R}.
\newblock {\em PLoS One}, 18(3):e0282524.

\bibitem[Garnett et~al., 2010]{garnett2010sequential}
Garnett, R., Osborne, M.~A., Reece, S., Rogers, A., and Roberts, S.~J. (2010).
\newblock Sequential {B}ayesian prediction in the presence of changepoints and
  faults.
\newblock {\em The Computer Journal}, 53(9):1430--1446.

\bibitem[Gelfand and Schliep, 2016]{gelfand2016spatial}
Gelfand, A.~E. and Schliep, E.~M. (2016).
\newblock Spatial statistics and {G}aussian processes: A beautiful marriage.
\newblock {\em Spatial Statistics}, 18:86--104.

\bibitem[Gramacy et~al., 2014]{gramacy2014massively}
Gramacy, R., Niemi, J., and Weiss, R. (2014).
\newblock Massively parallel approximate {G}aussian process regression.
\newblock {\em SIAM/ASA Journal on Uncertainty Quantification}, 2(1):564--584.

\bibitem[Gramacy, 2016]{gramacy2016lagp}
Gramacy, R.~B. (2016).
\newblock la{GP}: large-scale spatial modeling via local approximate {G}aussian
  processes in {R}.
\newblock {\em Journal of Statistical Software}, 72:1--46.

\bibitem[Gramacy, 2020]{gramacy2020surrogates}
Gramacy, R.~B. (2020).
\newblock {\em Surrogates: {G}aussian process modeling, design, and
  optimization for the applied sciences}.
\newblock CRC press.

\bibitem[Gramacy and Apley, 2015]{gramacy2015local}
Gramacy, R.~B. and Apley, D.~W. (2015).
\newblock Local {G}aussian process approximation for large computer
  experiments.
\newblock {\em Journal of Computational and Graphical Statistics},
  24(2):561--578.

\bibitem[Gramacy and Haaland, 2016]{gramacy2016speeding}
Gramacy, R.~B. and Haaland, B. (2016).
\newblock Speeding up neighborhood search in local {G}aussian process
  prediction.
\newblock {\em Technometrics}, 58(3):294--303.

\bibitem[Gramacy and Lee, 2008]{gramacy2008bayesian}
Gramacy, R.~B. and Lee, H. K.~H. (2008).
\newblock Bayesian treed {G}aussian process models with an application to
  computer modeling.
\newblock {\em Journal of the American Statistical Association},
  103(483):1119--1130.

\bibitem[Haas, 1990]{haas1990lognormal}
Haas, T.~C. (1990).
\newblock Lognormal and moving window methods of estimating acid deposition.
\newblock {\em Journal of the American Statistical Association},
  85(412):950--963.

\bibitem[Karatzoglou et~al., 2004]{karatzoglou2004kernlab}
Karatzoglou, A., Smola, A., Hornik, K., and Zeileis, A. (2004).
\newblock kernlab-an {S4} package for kernel methods in {R}.
\newblock {\em Journal of statistical software}, 11:1--20.

\bibitem[Kaufman et~al., 2011]{kaufman2011efficient}
Kaufman, C.~G., Bingham, D., Habib, S., Heitmann, K., and Frieman, J.~A.
  (2011).
\newblock Efficient emulators of computer experiments using compactly supported
  correlation functions, with an application to cosmology.

\bibitem[Kim et~al., 2005]{kim2005analyzing}
Kim, H.-M., Mallick, B.~K., and Holmes, C.~C. (2005).
\newblock Analyzing nonstationary spatial data using piecewise {G}aussian
  processes.
\newblock {\em Journal of the American Statistical Association},
  100(470):653--668.

\bibitem[Liaw and Wiener, 2002]{liaw2002rf}
Liaw, A. and Wiener, M. (2002).
\newblock Classification and regression by random{F}orest.
\newblock {\em R News}, 2(3):18--22.

\bibitem[Luo et~al., 2023]{luo2023nonstationary}
Luo, Z.~T., Sang, H., and Mallick, B. (2023).
\newblock A nonstationary soft partitioned {G}aussian process model via random
  spanning trees.
\newblock {\em Journal of the American Statistical Association}, pages 1--12.

\bibitem[Matheron, 1969]{matheron1969krigeage}
Matheron, G. (1969).
\newblock {\em Le krigeage universel}, volume~1.
\newblock {\'E}cole nationale sup{\'e}rieure des mines de Paris Paris.

\bibitem[McKay et~al., 1979]{mckay:1979}
McKay, M.~D., Beckman, R.~J., and Conover, W.~J. (1979).
\newblock A comparison of three methods for selecting values of input variables
  in the analysis of output from a computer code.
\newblock {\em Technometrics}, 21(2):239--245.

\bibitem[McLachlan and Basford, 1988]{mclachlan1988mixture}
McLachlan, G.~J. and Basford, K.~E. (1988).
\newblock {\em Mixture models: Inference and applications to clustering},
  volume~38.
\newblock M. Dekker New York.

\bibitem[McLachlan and Krishnan, 1997]{mclachlan1997algorithm}
McLachlan, G.~J. and Krishnan, T. (1997).
\newblock {\em The {EM} algorithm and extensions}.
\newblock John Wiley \& Sons.

\bibitem[{Microsoft} and Weston, 2022]{microsoft2022foreach}
{Microsoft} and Weston, S. (2022).
\newblock {\em foreach: Provides Foreach Looping Construct}.
\newblock R package version 1.5.2.

\bibitem[Nychka et~al., 2015]{nychka2015multiresolution}
Nychka, D., Bandyopadhyay, S., Hammerling, D., Lindgren, F., and Sain, S.
  (2015).
\newblock A multiresolution {G}aussian process model for the analysis of large
  spatial datasets.
\newblock {\em Journal of Computational and Graphical Statistics},
  24(2):579--599.

\bibitem[Park, 2022]{park2022jump}
Park, C. (2022).
\newblock Jump {G}aussian process model for estimating piecewise continuous
  regression functions.
\newblock {\em The Journal of Machine Learning Research}, 23(1):12737--12773.

\bibitem[Park et~al., 2022]{park2022sequential}
Park, C., Qiu, P., Carpena-N{\'u}{\~n}ez, J., Rao, R., Susner, M., and
  Maruyama, B. (2022).
\newblock Sequential adaptive design for jump regression estimation.
\newblock {\em IISE Transactions}, 55(2):111--128.

\bibitem[Park et~al., 2023]{park2023active}
Park, C., Waelder, R., Kang, B., Maruyama, B., Hong, S., and Gramacy, R.
  (2023).
\newblock Active learning of piecewise {G}aussian process surrogates.
\newblock {\em arXiv preprint arXiv:2301.08789}.

\bibitem[Payne et~al., 2020]{payne2020conditional}
Payne, R.~D., Guha, N., Ding, Y., and Mallick, B.~K. (2020).
\newblock A conditional density estimation partition model using logistic
  {G}aussian processes.
\newblock {\em Biometrika}, 107(1):173--190.

\bibitem[Picheny et~al., 2019]{picheny2019ordinal}
Picheny, V., Vakili, S., and Artemev, A. (2019).
\newblock Ordinal bayesian optimisation.
\newblock {\em arXiv preprint arXiv:1912.02493}.

\bibitem[Pope et~al., 2021]{pope2021gaussian}
Pope, C.~A., Gosling, J.~P., Barber, S., Johnson, J.~S., Yamaguchi, T.,
  Feingold, G., and Blackwell, P.~G. (2021).
\newblock Gaussian process modeling of heterogeneity and discontinuities using
  {V}oronoi tessellations.
\newblock {\em Technometrics}, 63(1):53--63.

\bibitem[Rasmussen and Williams, 2006]{williams2006gaussian}
Rasmussen, C.~E. and Williams, C.~K. (2006).
\newblock {\em Gaussian processes for machine learning}, volume~2.
\newblock MIT press Cambridge, MA.

\bibitem[Sacks et~al., 1989]{sacks1989design}
Sacks, J., Welch, W., Mitchell, T., and Wynn, H. (1989).
\newblock Design and analysis of computer experiments.
\newblock {\em Statistical Science}, 4(4):409--423.

\bibitem[Sampson and Guttorp, 1992]{sampson1992nonparametric}
Sampson, P. and Guttorp, P. (1992).
\newblock Nonparametric estimation of nonstationary spatial covariance
  structure.
\newblock {\em Journal of the American Statistical Association},
  87(417):108--119.

\bibitem[Santner et~al., 2003]{santner2003design}
Santner, T.~J., Williams, B.~J., Notz, W.~I., and Williams, B.~J. (2003).
\newblock {\em The design and analysis of computer experiments}, volume~1.
\newblock Springer.

\bibitem[Sauer et~al., 2023a]{sauer2023non}
Sauer, A., Cooper, A., and Gramacy, R.~B. (2023a).
\newblock Non-stationary {G}aussian process surrogates.
\newblock {\em arXiv preprint arXiv:2305.19242}.

\bibitem[Sauer et~al., 2023b]{sauer2023vecchia}
Sauer, A., Cooper, A., and Gramacy, R.~B. (2023b).
\newblock Vecchia-approximated deep {G}aussian processes for computer
  experiments.
\newblock {\em Journal of Computational and Graphical Statistics},
  32(3):824--837.

\bibitem[Schabenberger and Gotway, 2017]{schabenberger2017statistical}
Schabenberger, O. and Gotway, C.~A. (2017).
\newblock {\em Statistical methods for spatial data analysis}.
\newblock Chapman and Hall/CRC.

\bibitem[Schmidt and O'Hagan, 2003]{schmidt2003bayesian}
Schmidt, A. and O'Hagan, A. (2003).
\newblock Bayesian inference for non-stationary spatial covariance structure
  via spatial deformations.
\newblock {\em Journal of the Royal Statistical Society: Series B (Statistical
  Methodology)}, 65(3):743--758.

\bibitem[Scrucca et~al., 2023]{scrucca2023model}
Scrucca, L., Fraley, C., Murphy, T.~B., and Raftery, A.~E. (2023).
\newblock {\em Model-Based Clustering, Classification, and Density Estimation
  Using {mclust} in {R}}.
\newblock Chapman and Hall/CRC.

\bibitem[Snelson and Ghahramani, 2005]{snelson2005sparse}
Snelson, E. and Ghahramani, Z. (2005).
\newblock Sparse gaussian processes using pseudo-inputs.
\newblock {\em Advances in neural information processing systems}, 18.

\bibitem[Stein, 2012]{stein2012interpolation}
Stein, M.~L. (2012).
\newblock {\em Interpolation of spatial data: some theory for kriging}.
\newblock Springer Science \& Business Media.

\bibitem[Stein et~al., 2004]{stein2004approximating}
Stein, M.~L., Chi, Z., and Welty, L.~J. (2004).
\newblock Approximating likelihoods for large spatial data sets.
\newblock {\em Journal of the Royal Statistical Society Series B: Statistical
  Methodology}, 66(2):275--296.

\bibitem[Sung et~al., 2018]{sung2018exploiting}
Sung, C.-L., Gramacy, R.~B., and Haaland, B. (2018).
\newblock Exploiting variance reduction potential in local {G}aussian process
  search.
\newblock {\em Statistica Sinica}, pages 577--600.

\bibitem[Surjanovic and Bingham, 2013]{simulationlib}
Surjanovic, S. and Bingham, D. (2013).
\newblock Virtual library of simulation experiments: Test functions and
  datasets.
\newblock Retrieved February 1, 2024, from \url{http://www.sfu.ca/~ssurjano}.

\bibitem[Toms and Lesperance, 2003]{toms2003piecewise}
Toms, J.~D. and Lesperance, M.~L. (2003).
\newblock Piecewise regression: a tool for identifying ecological thresholds.
\newblock {\em Ecology}, 84(8):2034--2041.

\bibitem[Vecchia, 1988]{vecchia1988estimation}
Vecchia, A.~V. (1988).
\newblock Estimation and model identification for continuous spatial processes.
\newblock {\em Journal of the Royal Statistical Society Series B: Statistical
  Methodology}, 50(2):297--312.

\bibitem[Vieth, 1989]{vieth1989fitting}
Vieth, E. (1989).
\newblock Fitting piecewise linear regression functions to biological
  responses.
\newblock {\em Journal of {A}pplied {P}hysiology}, 67(1):390--396.

\bibitem[Williams and Barber, 1998]{williams1998bayesian}
Williams, C.~K. and Barber, D. (1998).
\newblock Bayesian classification with {G}aussian processes.
\newblock {\em IEEE Transactions on {P}attern {A}nalysis and {M}achine
  {I}ntelligence}, 20(12):1342--1351.

\bibitem[Zimmerman and Ver~Hoef, 2024]{zimmerman2024spatial}
Zimmerman, D.~L. and Ver~Hoef, J.~M. (2024).
\newblock {\em Spatial linear models for environmental data}.
\newblock Chapman and Hall/CRC.

\end{thebibliography}

\newpage
\begin{center}
{\large\bf SUPPLEMENTARY MATERIAL}
\end{center}
\appendix

\section{Data Generation Details}\label{sup:datagen}

Here we describe the process for generating the Phantom, Star, and Modified Michalewicz data 
sets.

\subsection{Phantom}\label{sup:phantom}

\begin{figure}[ht!]
\centering
\includegraphics[scale=0.45, trim =  50 30 50 50 clip]{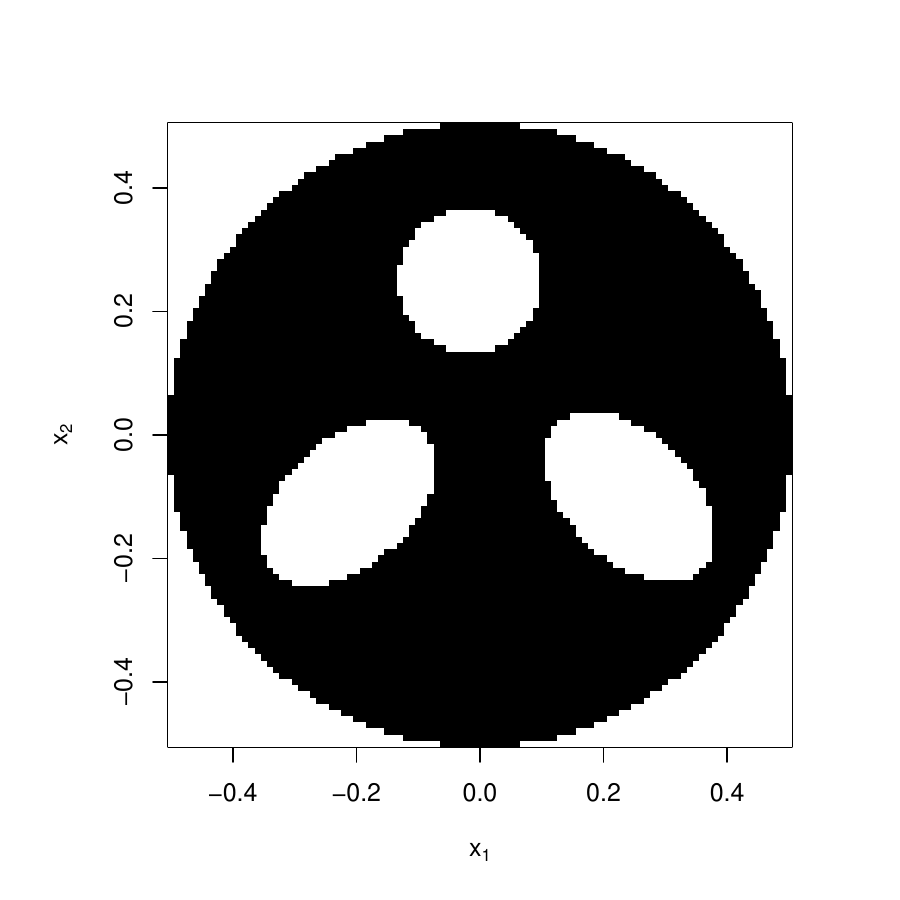}
\hspace{1.5cm}
\includegraphics[scale=0.45, trim = 50 30 50 50 clip]{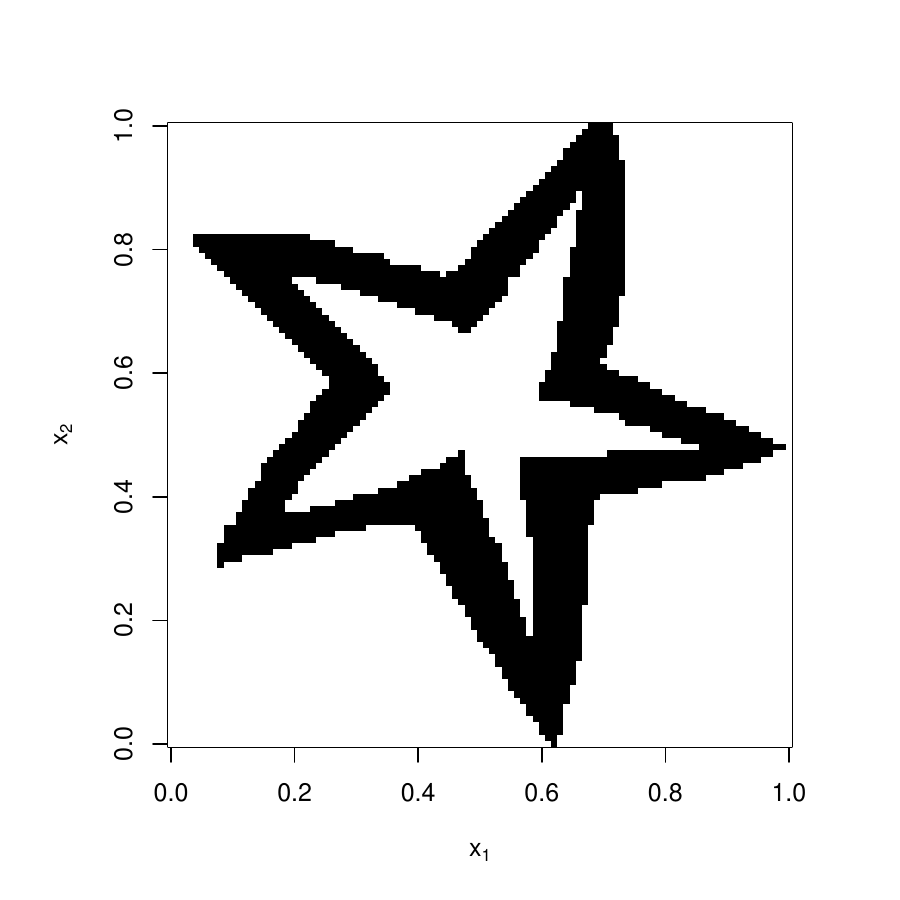}
\caption{{\em Left:} Image describing component membership for Phantom data set; 
{\em Right:} Image describing component membership for Star data set.}
\label{fig:datagen}
\end{figure}

The Phantom data set is generated in the same way as the Phantom data set in \citet{park2022jump}, but with 
different hyperparameter values. This version of the Phantom is evaluated on an evenly spaced $101 \times 101$ grid 
in $[-0.5,0.5]^2$. The response is generated using the partitioning scheme shown in the left panel of Figure \ref{fig:datagen}. 
In this plot, the color represents the true component membership of every input combination: white 
pixels belong to component 1, and black pixels belong to component 2. Data are drawn from a
 $\mathcal{N}_{N_i}(\mu_i, \Sigma(X_{N_i}))$, where $N_i$ is the number of observations in component
  $i$, $\mu_i$ is the mean of component $i$, and $X_{N_i}$ is the subset of observations which belong
   to component $i$. The mean vector is $\mu = [27,0]^{\top}$ and the covariance matrix
    $\Sigma(X_{N_i})$ is calculated using the exponential covariance function
 (\ref{eq:2gp}) with $\tau^2=9$ and $\theta=[0.1, 0.1]^{\top}$. After sampling all the data, the 
 response was standardized so that $Y_N \in [0,1]$.

\subsection{Star}\label{sup:star}
The Star data set is inspired by the Star data set in \citet{park2022jump}, but uses a different 
image for its partitioning structure. This version of the Star is evaluated on an evenly spaced
 $101 \times 101$ grid in $[0,1]^2$. The response is generated using the partitioning scheme 
 shown in the right panel of Figure \ref{fig:datagen}. Like with the Phantom, the color represents 
 the true component membership: white for component 1, and black for component 2. Data are drawn 
 from a $\mathcal{N}_{N_i}(\mu_i, \Sigma(X_{N_i}))$. The mean vector is $\mu = [-10,10]^{\top}$ 
 and the covariance matrix $\Sigma(X_{N_i})$ is calculated using the exponential covariance function
  (\ref{eq:2gp}) with $\tau^2=3$ and $\theta=[0.01, 0.01]^{\top}$. After sampling all the data, the 
  response was standardized so that $Y_N \in [0,1]$.

\subsection{Modified Michalewicz}\label{sup:michal} 
The Modified Michalewicz is a modification of the Michalewicz function from the Virtual Library 
of Simulation Experiments \citep{simulationlib}. The original Michalewicz function is defined as 
$f(x) = -\sum_{i=1}^d \sin (x_i) \sin^{2m} \bigl(\frac{ix_i^2}{\pi}\bigr)$ for $x \in [0,\pi]^4$ and
 $d$ in arbitrary dimension. We use $d=4$ and $m=10$ (the recommended default) with inputs coded into
  $[0,1]^4$. We first evaluate $f(x)$ using an LHS of size 100,000. The Modified Michalewicz 
  is then defined as:
\begin{equation}
y(x) = 
\begin{cases}
f(x)  &f(x) \leq -0.01 \\
f(x) + 0.5 & f(x)>-0.01.\\
\end{cases}
\label{eq:michal}
\end{equation}
We use this modification because the response surface of $f(x)$, without modifications, contains 
no jumps. It is comprised of many local minima that are surrounded by a nearly flat surface around 
0. Consequently, the marginal response is comprised of two ``levels'': a point mass around 0, and 
the rest of the surface. To create the jumps (i.e.~discontinuities) that we expect, we raise the flat 
area of the surface by 0.5. This creates a manifold of discontinuity where the response jumps from 
the point mass to the rest of the surface.

\section{Global and LAGP Fits}\label{sup:fits}

We provide the mean fits given by a global GP and LAGP for the Phantom and Star data sets to highlight 
the improvements made by OLAGP and MJGP. We also provide the difference between the 
predicted mean fit and the truth for each data set.

\subsection{Phantom}\label{sup:phanfits}

\begin{figure}[ht!]
\centering
\begin{subfigure}[c]{.3\textwidth}
\centering
\includegraphics[scale = 0.4, trim =  20 10 50 20, clip]{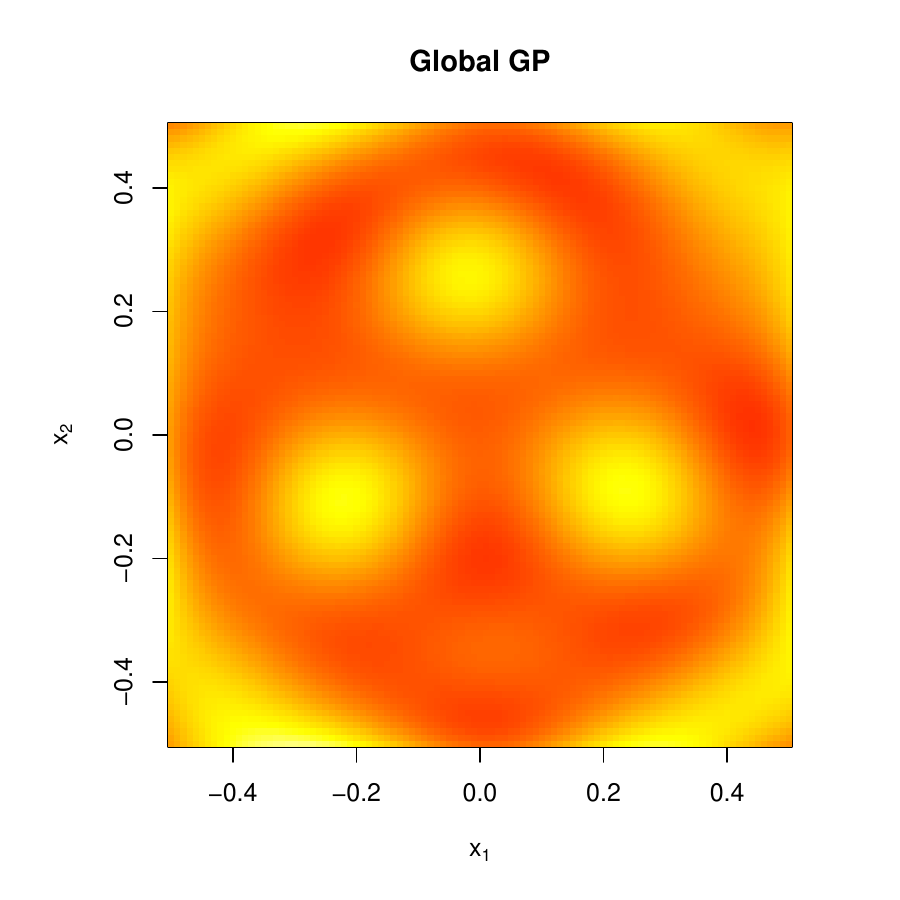}
\end{subfigure}
\hfill
\begin{subfigure}[c]{.3\textwidth}
\centering
\includegraphics[scale = 0.4, trim = 40 10 30 20, clip]{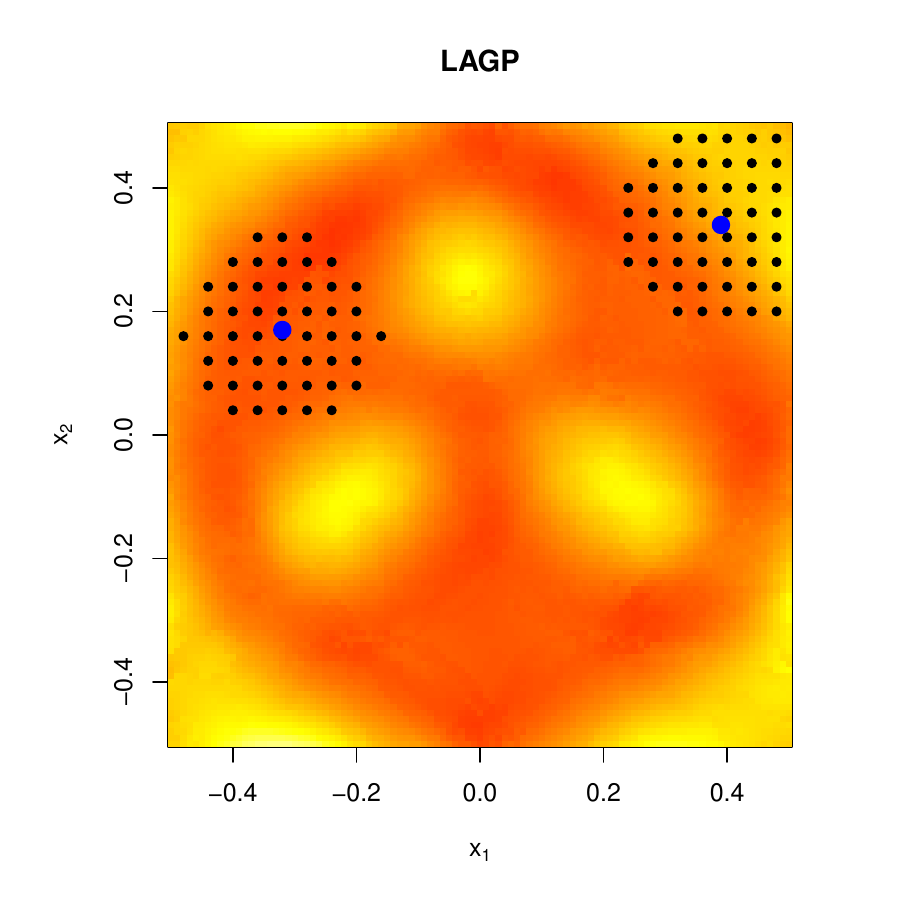}
\end{subfigure}
\hfill
\begin{subfigure}[c]{.3\textwidth}
\centering
\begin{tabular}{|c|c|}
\hline
\textbf{Method} & \textbf{RMSPE} \\
\hline
Global GP & 0.1543\\
\hline
LAGP & 0.1514 \\
\hline
OLAGP & 0.1316 \\
\hline
MJGP & 0.1207 \\
\hline
\end{tabular}
\end{subfigure}
\caption{{\em Left}: Global GP fit.
{\em Center}: LAGP fit. The blue dots
are two predictive locations and the surrounding black dots are the selected
neighborhood for that predictive location. {\em Right}: RMSPE of global GP, LAGP, 
OLAGP, and MJGP fits.}
\label{fig:phanorig}
\end{figure}

\begin{figure}[ht!]
\centering
\includegraphics[scale = 0.28, trim =  20 20 30 25, clip]{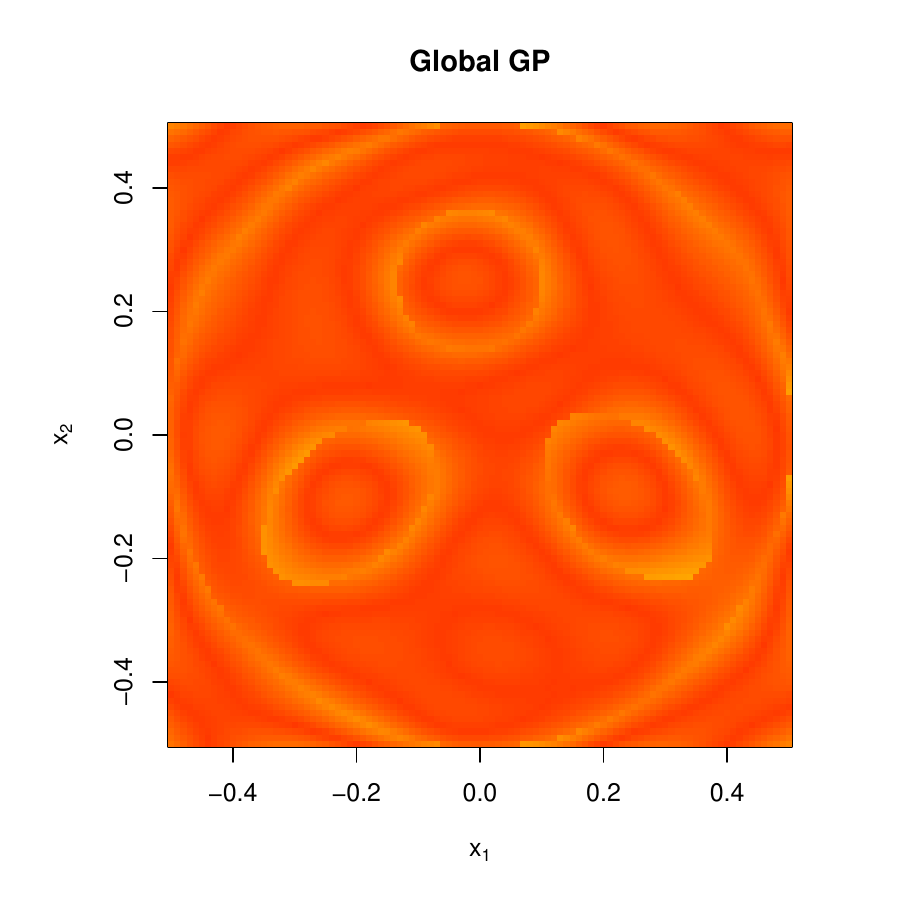}
\hspace{0.5cm}
\includegraphics[scale = 0.28, trim = 40 20 30 25, clip]{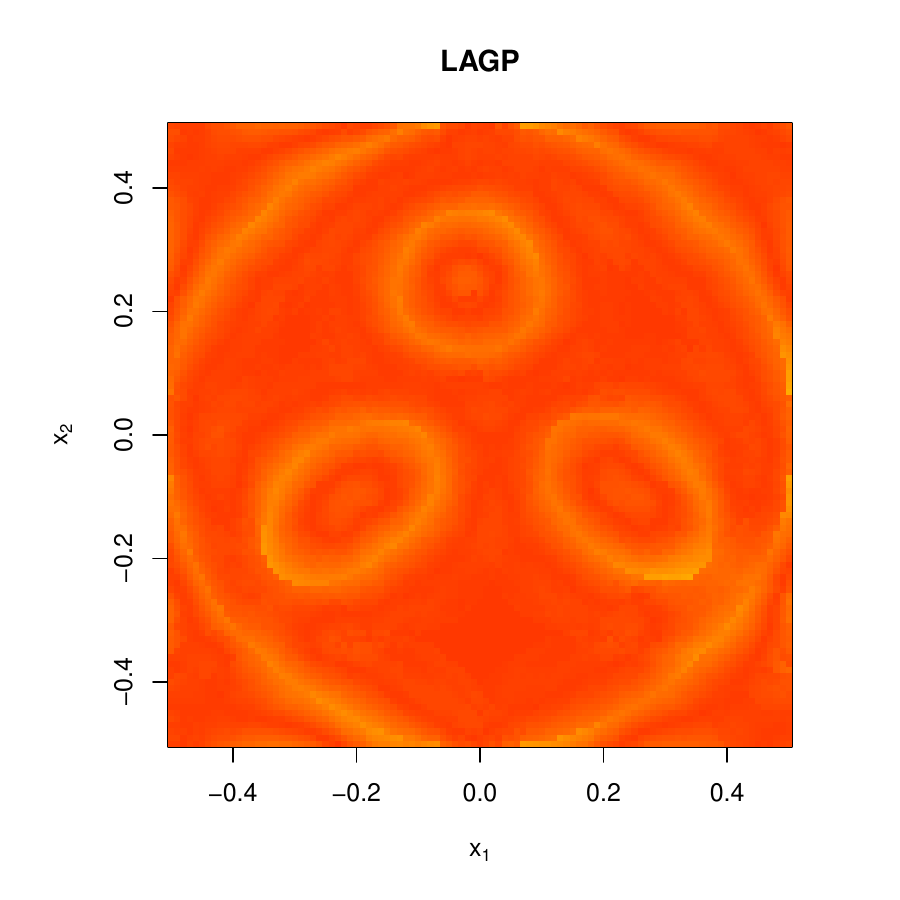}
\hspace{0.5cm}
\includegraphics[scale = 0.28, trim = 40 20 30 25, clip]{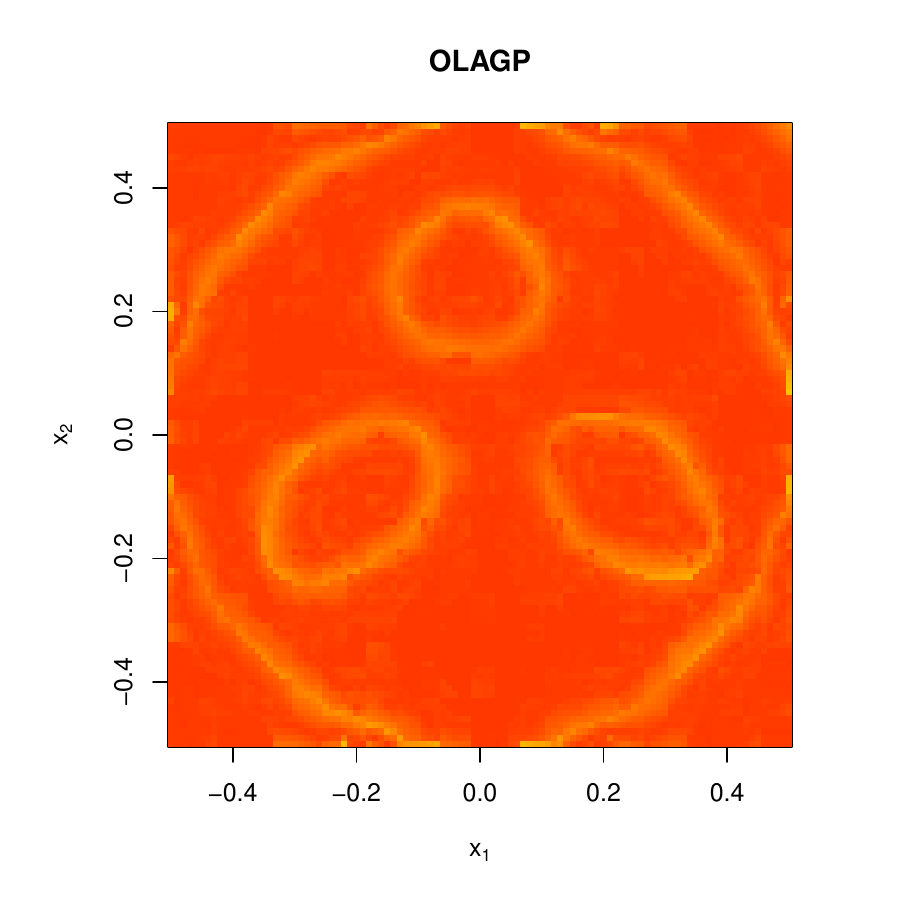}
\hspace{0.5cm}
\includegraphics[scale = 0.28, trim = 40 20 20 25, clip]{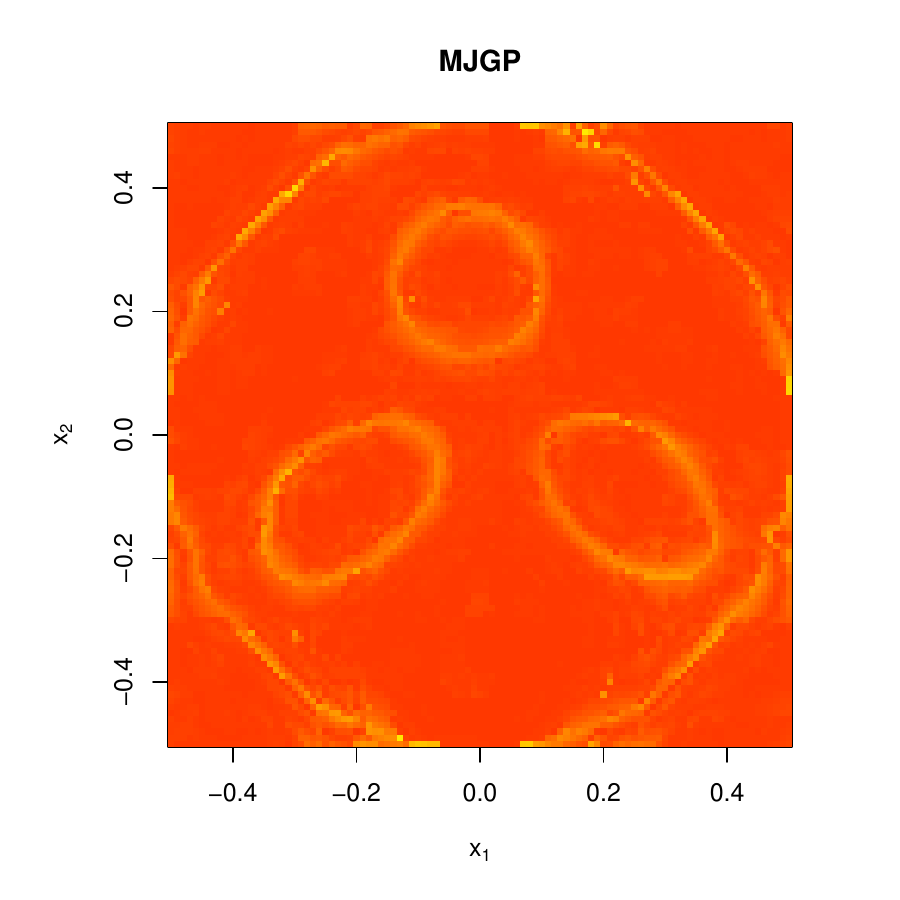}
\caption{{\em Left:} Difference between global GP fit and truth;
{\em Center Left:} Difference between LAGP fit and truth; 
{\em Center Right:} Difference between OLAGP fit and truth;
{\em Right:} Difference between MJGP fit and truth.}
\label{fig:phantom_diff}
\end{figure}

The left two panels of Figure \ref{fig:phanorig} show the predicted mean fits for the 
Phantom data set using global GP and LAGP models. 
Both fits are visually less accurate than the 
OLAGP fit in Figure \ref{fig:olagp} and MJGP fit in Figure \ref{fig:mjgp}. 
This is verified by the table of RMSPE values in the right panel, which shows that both OLAGP and  
MJGP produce more accurate predictions than either method. The LAGP fit 
in the center panel also shows the selected neighborhoods for the same two predictive locations used 
in Sections \ref{sec:opt} and \ref{sec:mod}, $(-0.32,0.17)$ and $(0.39,0.34)$. Notice in particular 
that the neighborhood for $(0.39,0.34)$ contains points in both the red and yellow regions, which 
explains why our prediction is inaccurate. We are able to select smaller and differently shaped neighborhoods 
using OLAGP and MJGP so that our predictions are more accurate. We can verify that OLAGP and  
MJGP make more accurate predictions by looking at Figure \ref{fig:phantom_diff}, which shows the absolute 
difference between the mean predictions and the truth. 
Notice that the difference is largest along the manifold of discontinuity.

\subsection{Star}\label{sup:starfits}

\begin{figure}[ht!]
\centering
\begin{subfigure}[c]{.3\textwidth}
\centering
\includegraphics[scale = 0.4, trim =  20 10 40 20, clip]{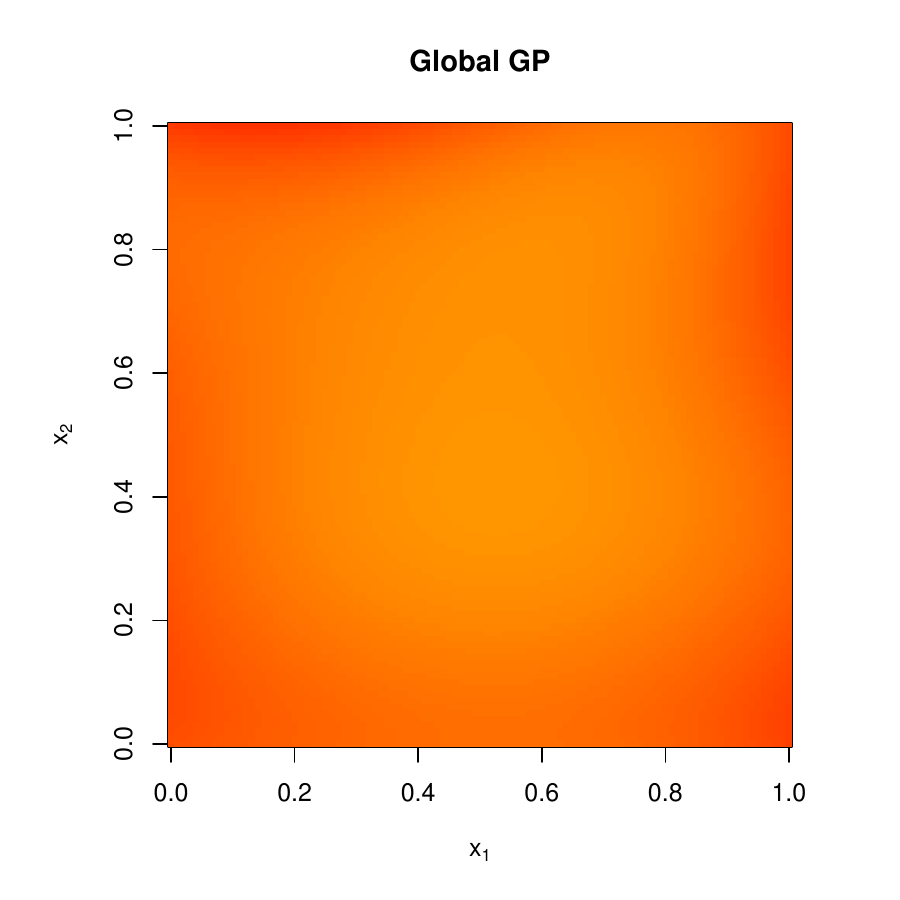}
\end{subfigure}
\hfill
\begin{subfigure}[c]{.3\textwidth}
\centering
\includegraphics[scale = 0.4, trim = 40 10 30 20, clip]{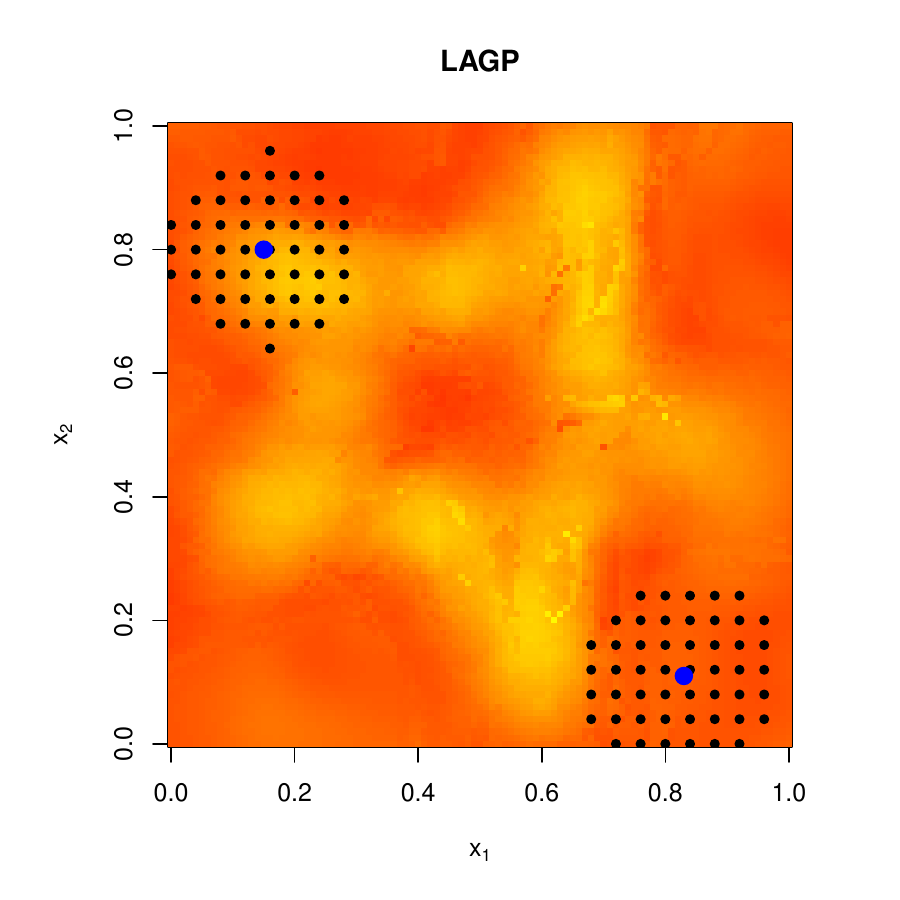}
\end{subfigure}
\hfill
\begin{subfigure}[c]{.3\textwidth}
\centering
\begin{tabular}{|c|c|}
\hline
\textbf{Method} & \textbf{RMSPE} \\
\hline
Global GP & 0.2716 \\
\hline
LAGP & 0.1967 \\
\hline
OLAGP & 0.1599 \\
\hline
MJGP & 0.1462 \\
\hline
\end{tabular}
\end{subfigure}
\caption{{\em Left}: Global GP fit.
{\em Center}: LAGP fit. The blue dots
are two predictive locations and the surrounding black dots are the selected
neighborhood for that predictive location. {\em Right}: RMSPE of global GP, LAGP, 
OLAGP, and MJGP fits.}
\label{fig:starorig}
\end{figure}

\begin{figure}[ht!]
\centering
\includegraphics[scale =  0.28, trim =  20 20 30 25, clip]{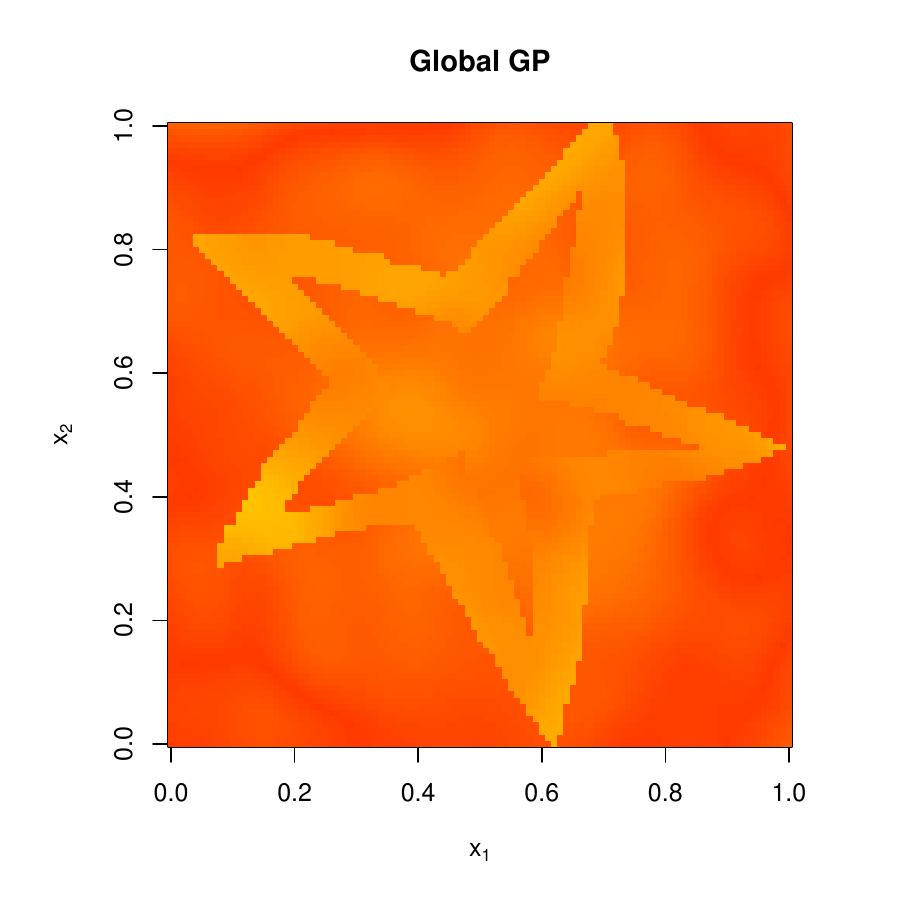}
\hspace{0.5cm}
\includegraphics[scale =  0.28, trim =  40 20 30 25, clip]{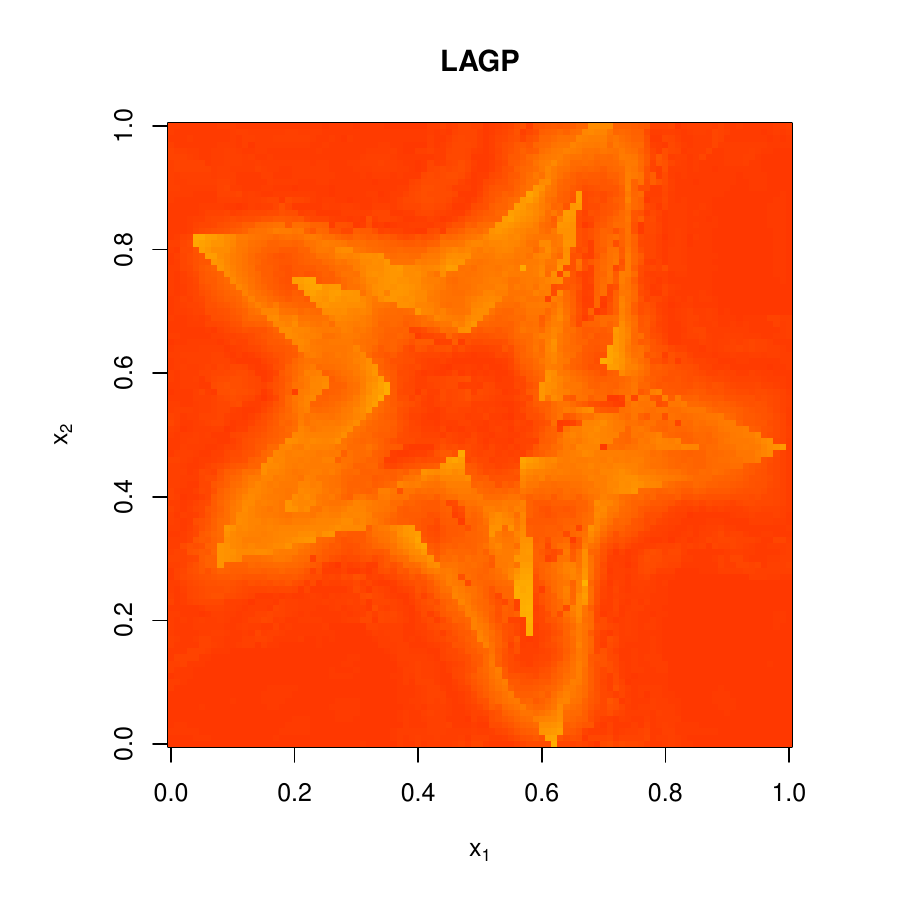}
\hspace{0.5cm}
\includegraphics[scale = 0.28, trim =  40 20 30 25, clip]{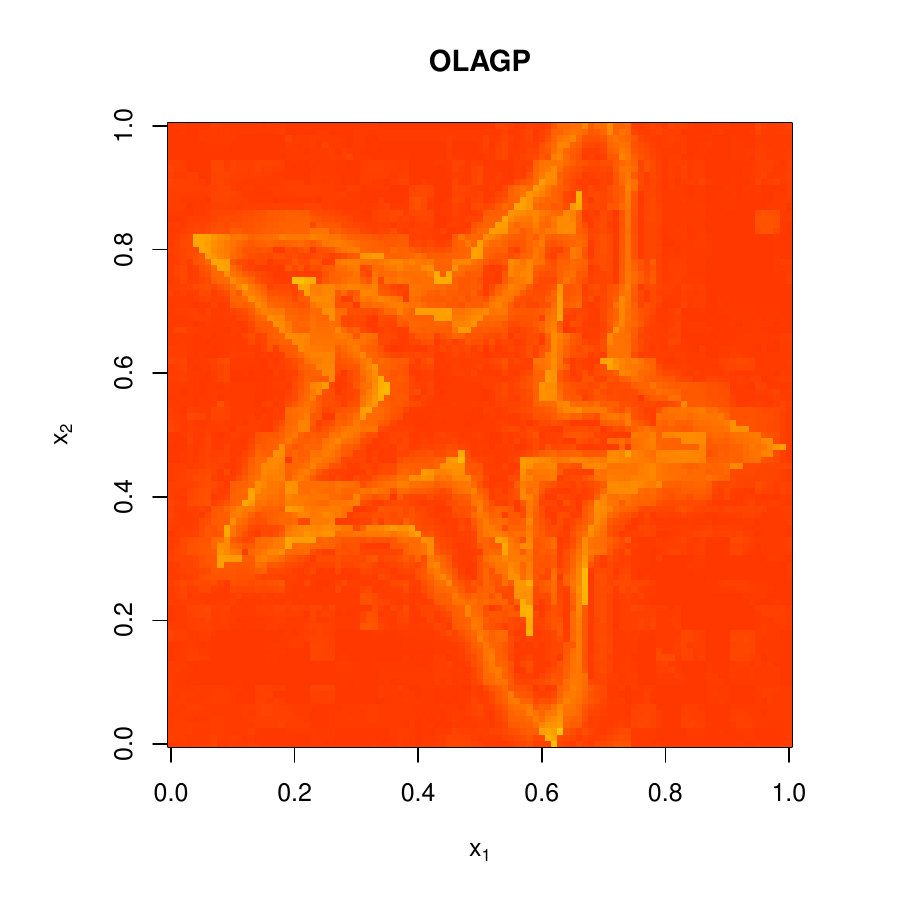}
\hspace{0.5cm}
\includegraphics[scale = 0.28, trim =  40 20 30 25, clip]{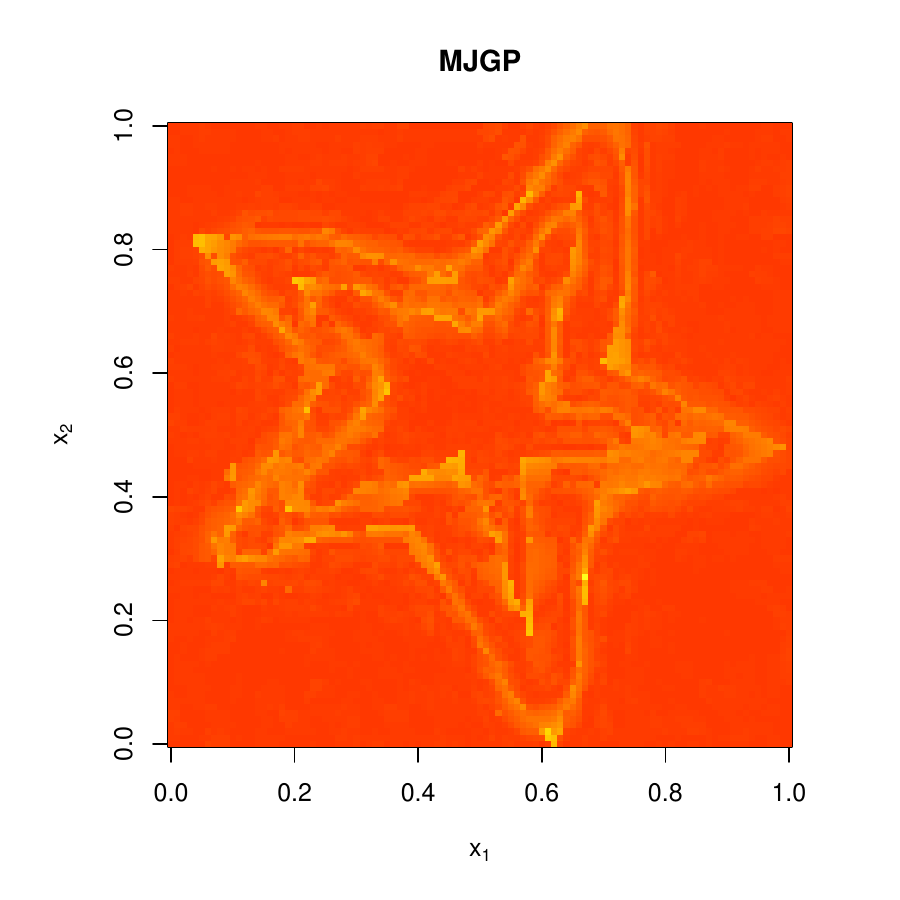}
\caption{{\em Left:} Difference between global GP fit and truth;
{\em Center Left:} Difference between LAGP fit and truth; 
{\em Center Right:} Difference between OLAGP fit and truth;
{\em Right:} Difference between MJGP fit and truth.}
\label{fig:star_diff}
\end{figure}

The left two panels of Figure \ref{fig:starorig} show the predicted mean fits for the 
Star data set using global GP and LAGP models. 
Both fits are visually less accurate than the 
OLAGP and MJGP fits in Figure \ref{fig:starfit}. The global GP performs 
particularly bad in this case, likely because the proportion of observations in component 1 is much 
larger than the proportion of values in component 2. The table of RMSPE values in the right panel 
verifies our intuition, showing that OLAGP and MJGP produce more accurate 
predictions than either method, particularly the global GP. The LAGP fit in the center panel also 
shows the selected neighborhoods for the same two predictive locations used in Section \ref{sec:starsim},
 $(0.15,0.8)$ and $(0.83, 0.11)$. Notice in particular that the neighborhood for $(0.15, 0.8)$ 
 contains points in both the red and yellow regions, which is why our prediction is inaccurate. 
 We are able to select smaller and differently shaped neighborhoods using OLAGP 
 and MJGP so that our predictions are more accurate. We can verify that OLAGP and 
MJGP make more accurate predictions by looking at Figure \ref{fig:star_diff}, which shows the absolute 
difference between the mean predictions and the truth.
Notice that the difference is largest along the manifold of discontinuity.

\section{MJGP -- Special Cases}\label{sup:special}

Here we show the performance of MJGP on two special 
cases -- data with small jumps and data with no jumps.

\subsection{Data with Small Jumps}\label{sup:smalljumps}

\begin{figure}[ht!]
\includegraphics[scale=0.4, trim = 0 15 30 20, clip]{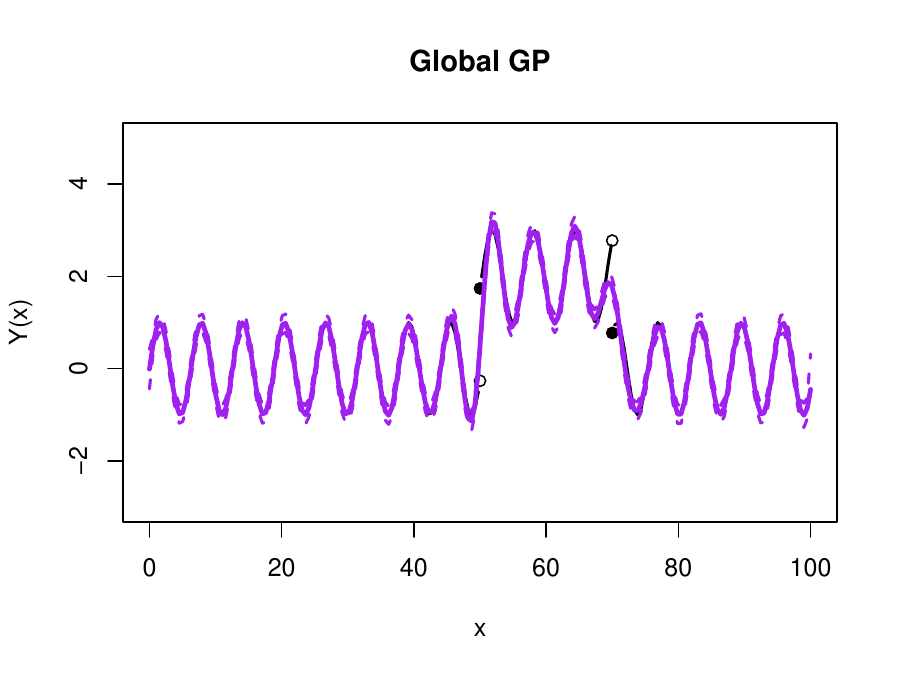}
\hspace{0.5cm}
\includegraphics[scale=0.4, trim = 30 15 30 20, clip]{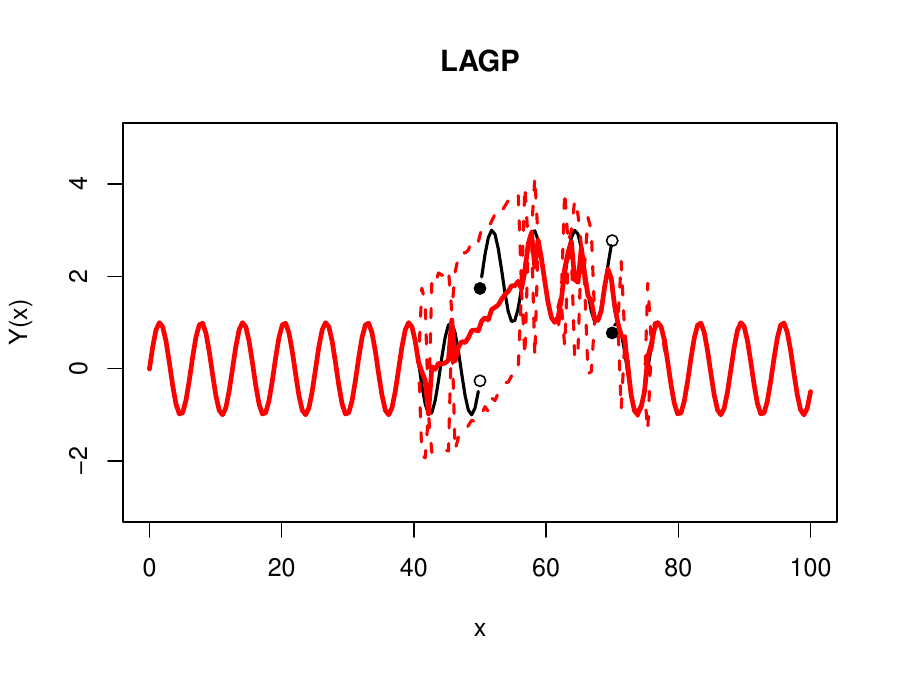}
\hspace{0.5cm}
\includegraphics[scale=0.4, trim = 30 15 30 20, clip]{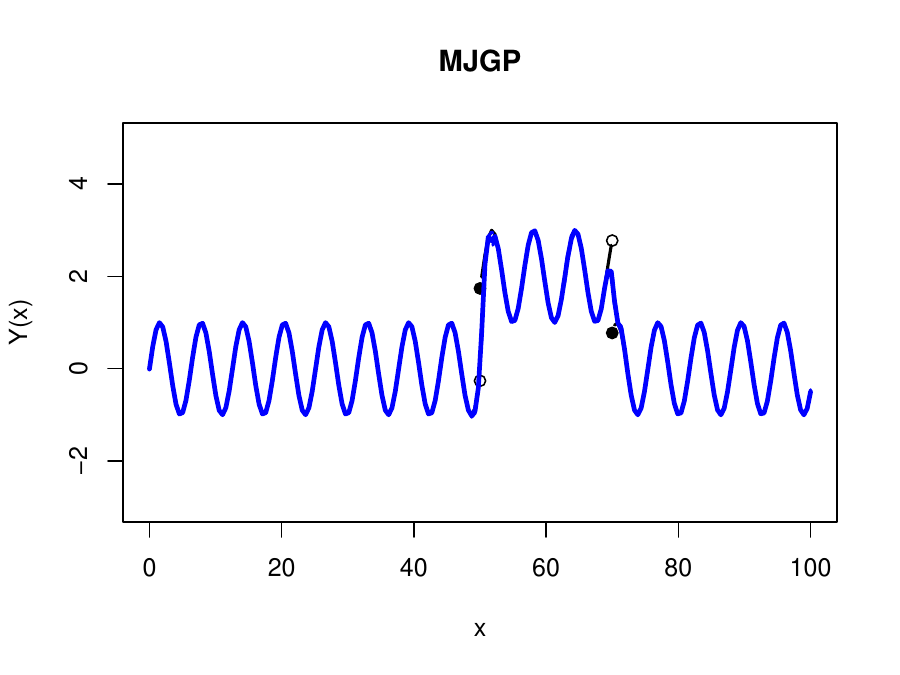}
\caption{Predictive mean and 95\% PI for the small jump 
case using a {\em (left)} global GP, {\em (center)} LAGP, and {\em (right)} 
MJGP.}
\label{fig:smalljump}
\end{figure}

For the case with small jumps, we modify the function used in Section 
\ref{sec:1}. We define the same jumping sine curve, but this time with a jump of 
2 instead of 10. In other words, we use the function in Eq.~\ref{eq:1ex}, but replace 
the 10 with a 2. We consider modeling this function using three of the four methods 
in Figure \ref{fig:plot1}; we exclude MJGP with a global GP. The results 
are shown in Figure \ref{fig:smalljump}.
The black lines represent the true function, while the colors (purple, red, and blue) 
represent the predictive mean and 95\% PI. We see that 
MJGP is still able to accurately capture the jumps, and offers improvement over the global 
GP and LAGP methods. Calculating RMSPE for each method confirms our findings; the global 
GP, LAGP, and MJGP produce RMSPEs of 0.1327, 0.4139, and 0.1038, respectively.

\subsection{Data with No Jumps}\label{sup:nojumps}
\begin{figure}[ht!]
\centering
\includegraphics[scale = 0.4, trim =  0 15 30 20, clip]{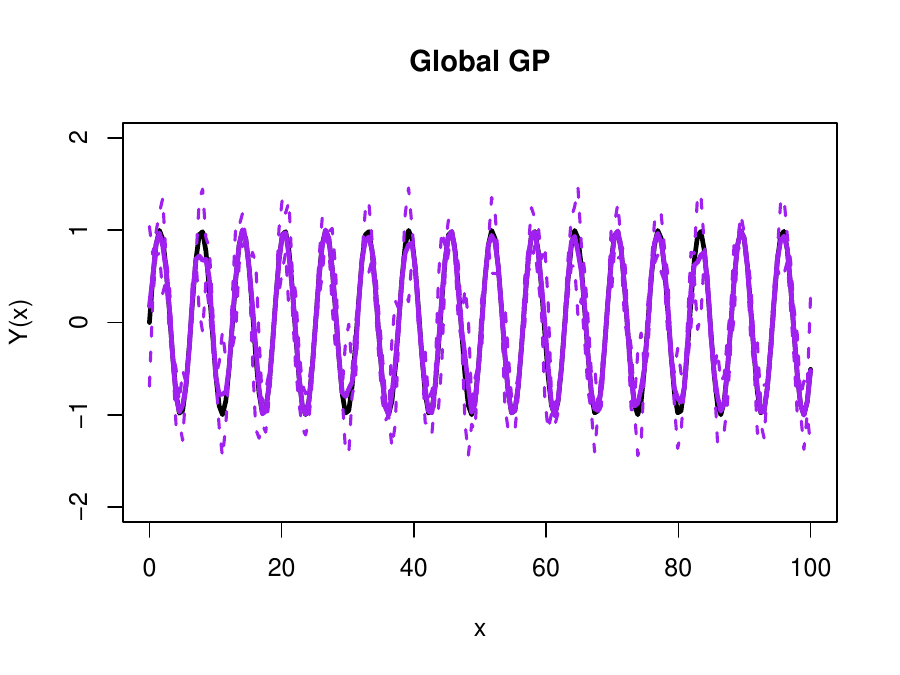}
\hspace{0.5cm}
\includegraphics[scale = 0.4, trim =  30 15 30 20, clip]{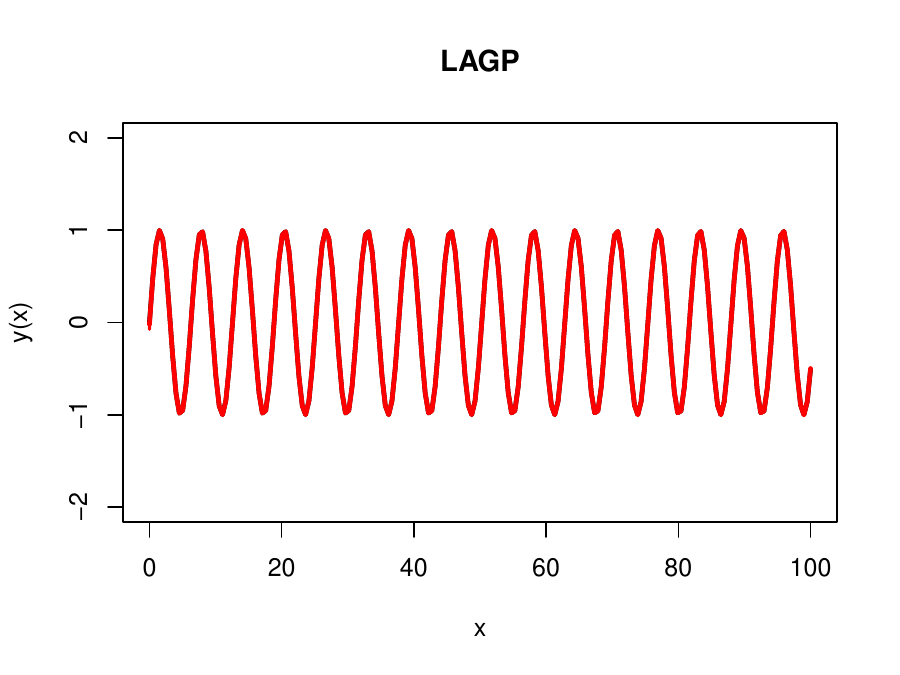}
\hspace{0.5cm}
\includegraphics[scale = 0.4, trim =  30 15 30 20, clip]{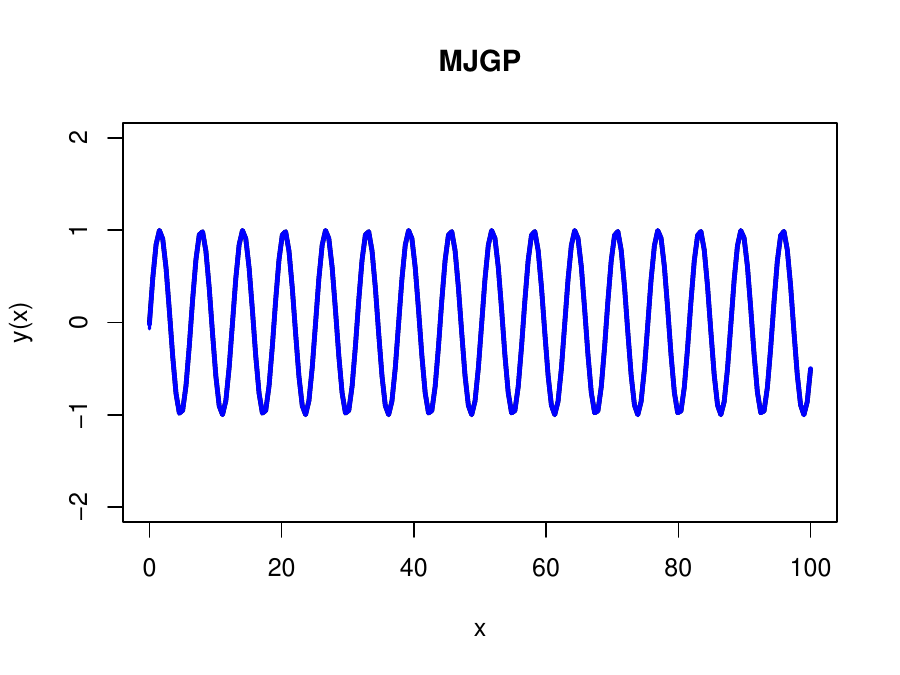}
\caption{Predictive mean and 95\% PI for the no jump 
case using a {\em (left)} global GP, {\em (center)} LAGP, and {\em (right)} 
MJGP.}
\label{fig:nojump}
\end{figure}

For the case with no jumps, we use a sine curve defined on the same 
range as our previous functions (i.e., $x \in [0,100]$) to test MJGP. 
We use the same three methods for comparison as in the previous section. 
The results are shown in Figure \ref{fig:nojump}.
Here we see that the results do not look much different for any 
of the three methods. Since there are no jumps, we do not 
expect using MJGP to offer any improvements. However, it 
is worth noting that using MJGP did not hurt us; the predictions 
made by MJGP are just as accurate as those made by LAGP. We confirm this 
statement using RMSPE; the global GP, LAGP, and MJGP produce RMSPEs of 
0.0678, 0.0017, and 0.0015, respectively. This tells us that the method 
can be used when jumps are suspected but not confirmed without risk of 
overfitting.

\section{Comparison to Partitioned GP}\label{sup:partition}
\begin{figure}[ht!]
\centering
\includegraphics[scale = 0.4, trim =  0 15 30 20, clip]{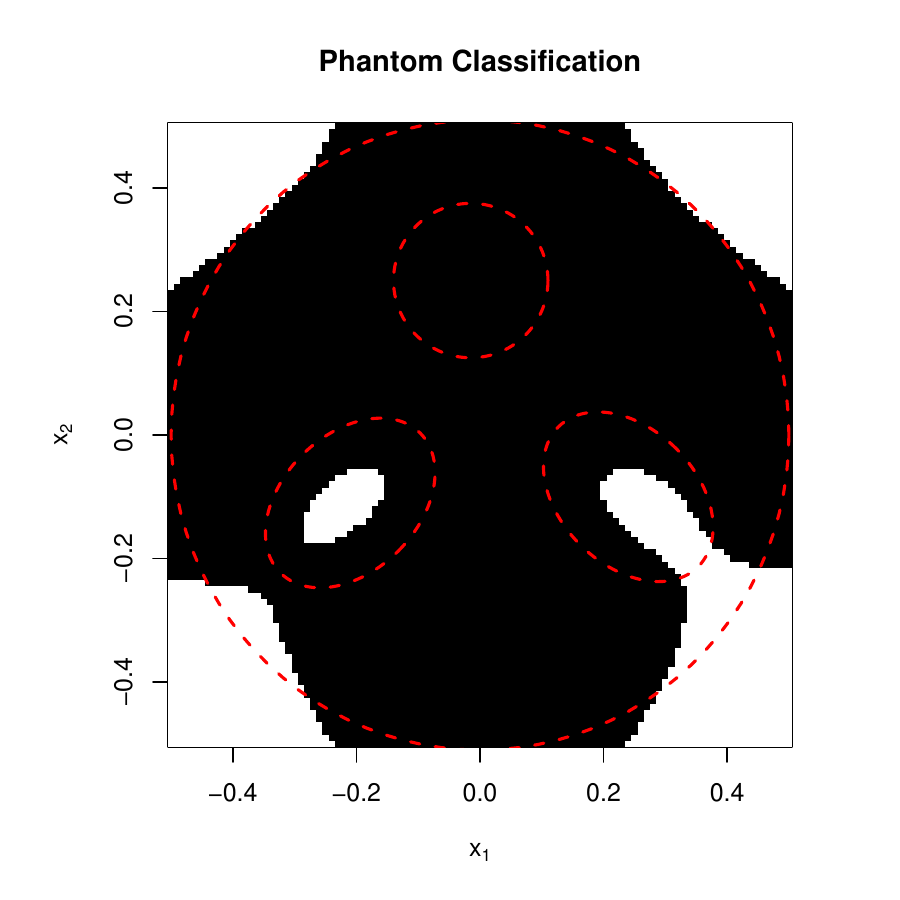}
\hspace{0,5cm}
\includegraphics[scale = 0.4, trim =  35 15 30 20, clip]{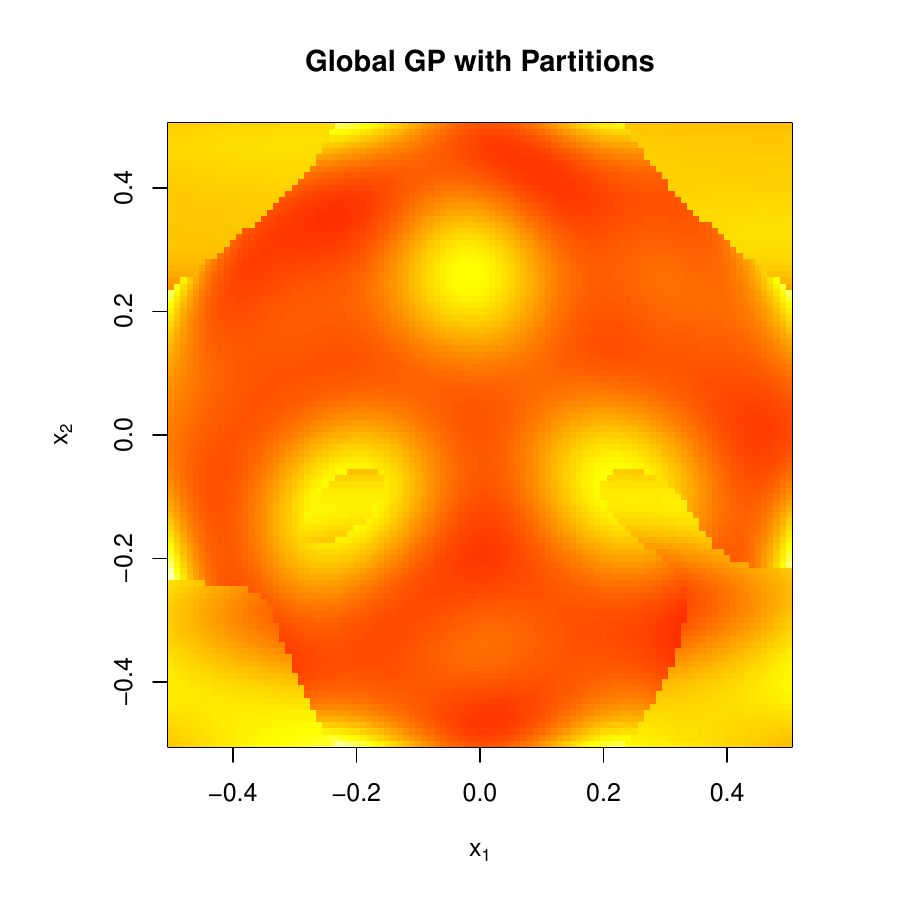}
\hspace{0.5cm}
\includegraphics[scale = 0.4, trim =  35 15 30 20, clip]{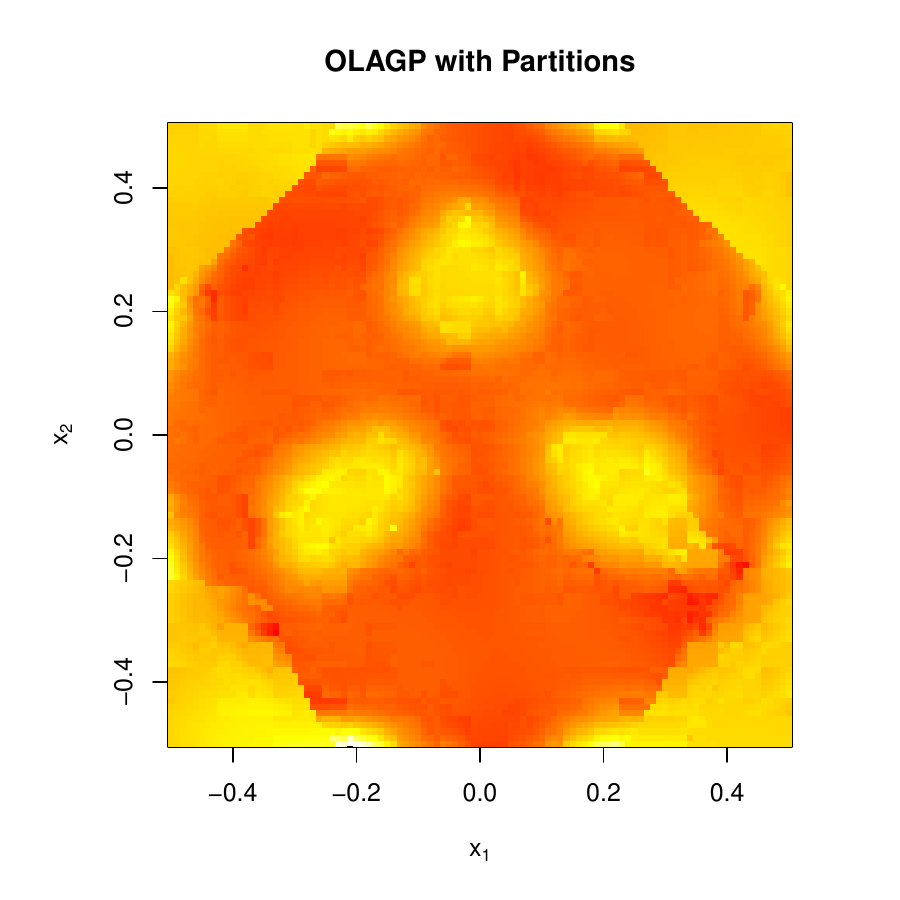}
\caption{{\em Left:} Predicted component membership by classification model; 
{\em Center:} Global GP fit on partitioned input space; {\em Right:} OLAGP fit 
on partitioned input space.}
\label{fig:partition}
\end{figure}

One alternative to MJGP is to partition the input space and then 
fit separate GPs to each partition. However, we prefer MJGP because 
these partitions are often difficult for a classifier to learn. Consider again 
the Phantom data set as an example. In Section \ref{sec:lev} we fit a classifier 
mapping $X_N$ to $C_N$, the likely component membership. Rather than using 
the predicted probability of component membership as a new feature, we could have 
partitioned the input space into two regions according to their predicted component. 
The results of such a classification are shown in the left panel of Figure \ref{fig:partition}.
The manifold of discontinuity is overlaid as a dashed red line. Here we see that 
the classification does a decent job of matching the outer circle, but struggles to classify 
the inner circle and ellipses. The top circle is completely misclassified.

Say we were to fit separate GPs based on this partitioning. The results of 
this method using a global GP and OLAGP are shown in the center and right panels of 
Figure \ref{fig:partition}, respectively.  In both fits we see the effects of fitting separate GPs; 
there are jumps along the contours of the classification, even for points on the same side 
of the manifold of discontinuity. OLAGP offers some improvements over a global GP, but 
neither performs as well as MJGP; the global GP and OLAGP fits have RMSPEs of 0.1404 
and 0.1290, respectively. This is because MJGP is able to incorporate the uncertainty 
of the classification model, while partitioned GPs ignore this uncertainty.

\section{Computational Time}\label{sup:comptime}

Table \ref{tab:time} shows the time (in minutes) of each of the 
simulations in Section \ref{sec:results}.
\begin{table}[ht!]
\begin{tabular}{c|c|c|c|c}
\textbf{Model} & \textbf{Phantom} & \textbf{Star} & \textbf{Modified Michalewicz} & \textbf{AMHV Transport}\\
\hline
Global GP & 4.92 & 3.75 & N/A & 12.87 \\
\hline
LAGP & 0.02 & 0.02 & 0.93 & 0.06 \\
\hline
OLAGP & 0.33 & 0.32 & 4.85 & 0.43 \\
\hline
OJGP & 22.32 & 22.22 & 352.95 & 30.57 \\
\hline
MJGP & 10.42 & 13.56 & 6.13 & 29.44 \\
\end{tabular}
\caption{Computation times for all data sets.}
\label{tab:time}
\end{table}
 We report the computation time 
of a single Monte Carlo iteration since the computation time is not overly 
variable between reps. Times were recorded on an 8-core, Intel i7 CPU 
at 3.5GHz. We see that in general, the two JGP methods take the longest, 
with OJGP consistently taking the longest. This is especially 
apparent for the Modified Michalewicz function, although we recognize that 
this could likely be improved by parallelizing the code. 
We also see from this table that the time required for MJGP is highly dependent 
on the classification model used. Recall that part of MJGP 
is a GP fit, either global (AMHV Transport) or OLAGP (all other data 
sets). So we might expect MJGP to have similar computational 
efficiency to its corresponding GP method. However, this is not the case 
for all data sets. MJGP takes quite a bit longer to run than 
its GP counterparts for all data sets except the Modified Michalewicz 
function. This extra time is due to the classification model used; we 
see that MJGP takes longer, in general, when a 
CGP is used than when a RF is used.

\section{Prediction Interval Coverage}\label{sup:pred}

\begin{figure}[ht!]
\centering
\includegraphics[scale = 0.28]{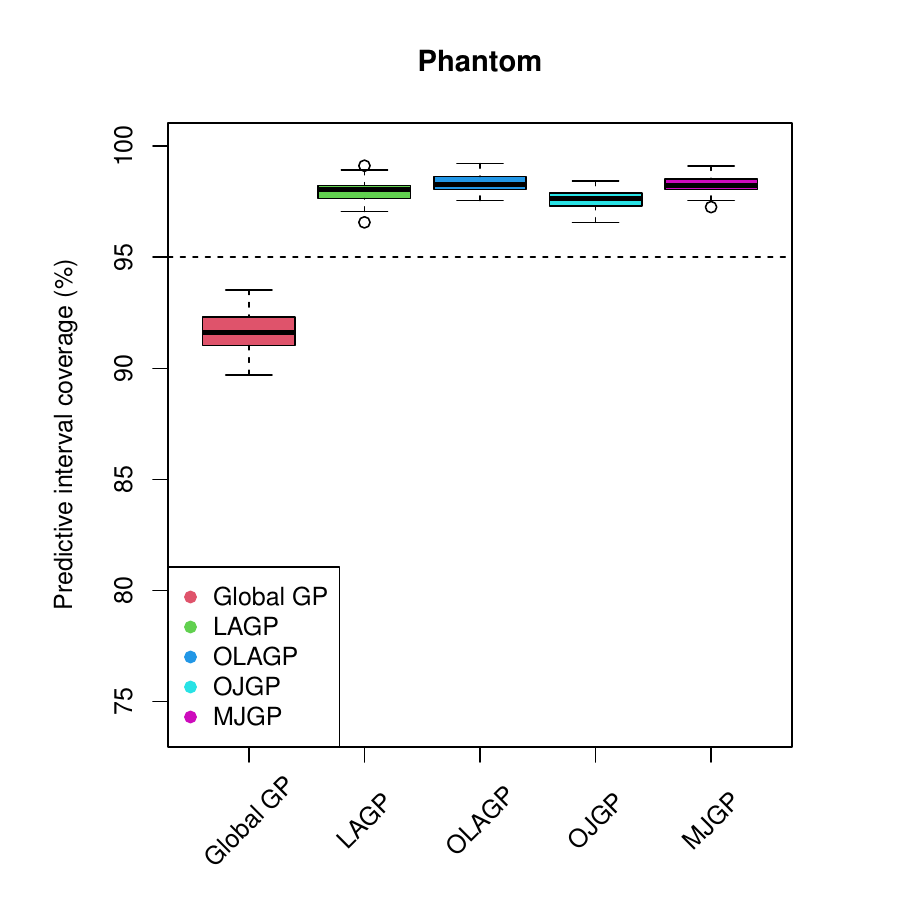}
\hfill
\includegraphics[scale = 0.28]{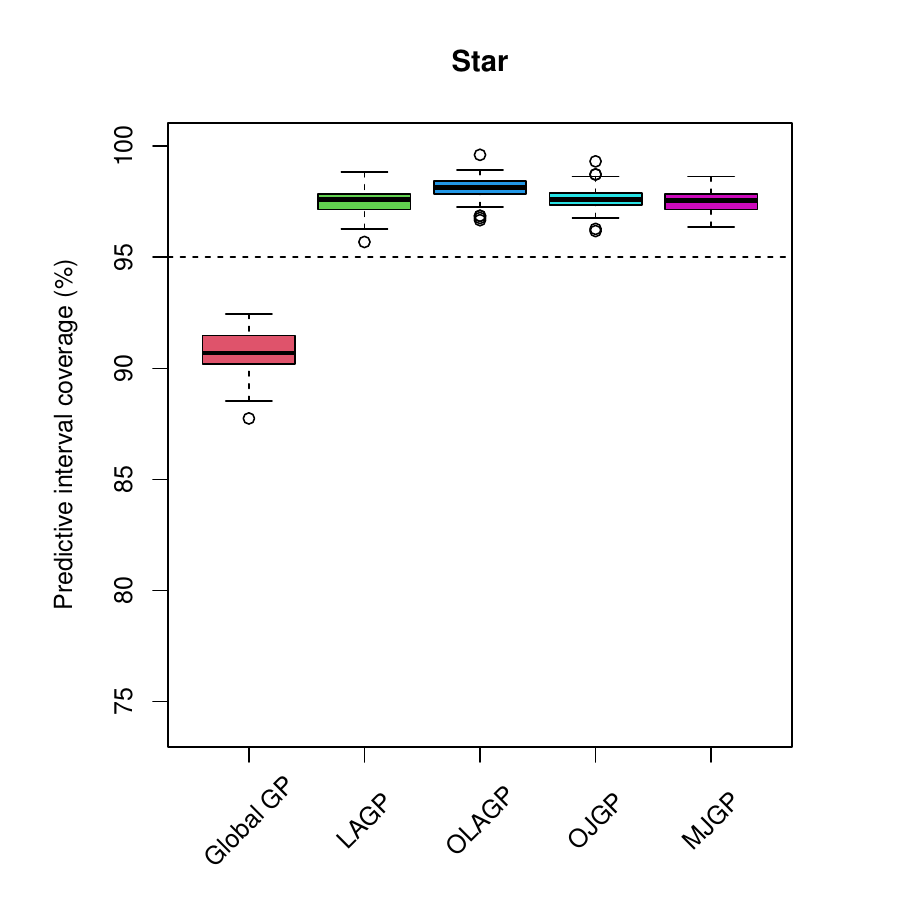}
\hfill
\includegraphics[scale = 0.28]{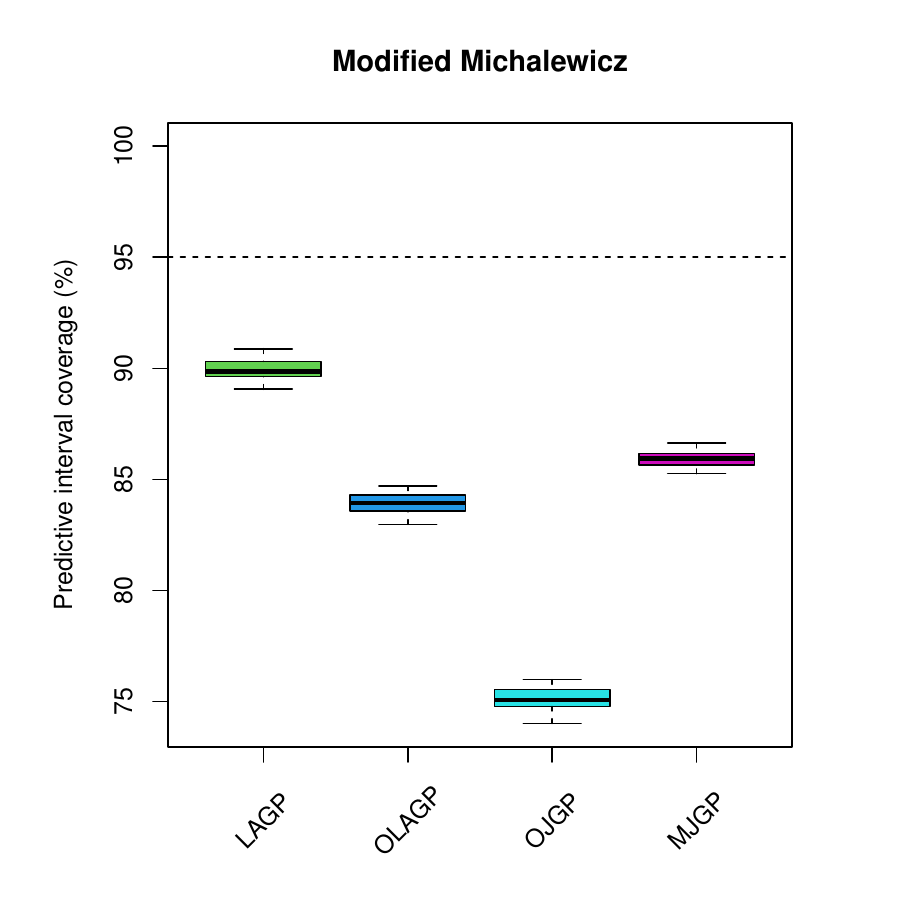}
\hfill
\includegraphics[scale = 0.28]{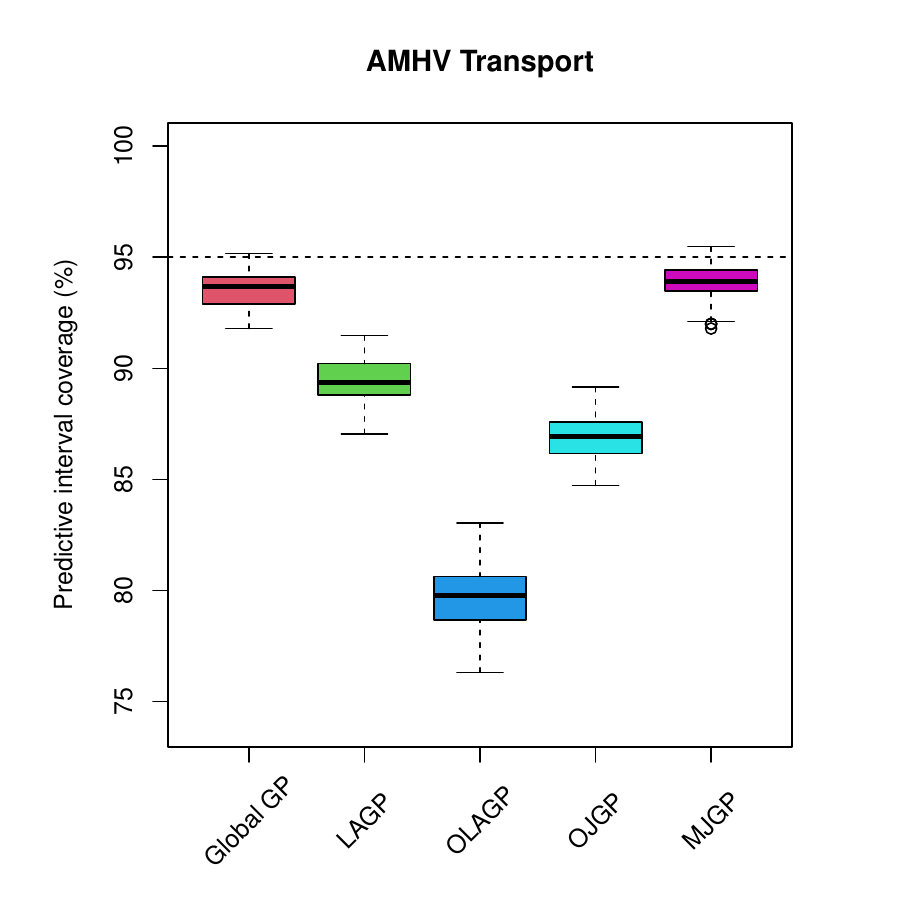}
\caption{Prediction interval coverage for the {\em (left)} Phantom, {\em (left 
center)} Star, {\em (right center)} Modified Michalewicz, and {\em (right)} 
AMHV transport data sets.}
\label{fig:pred}
\end{figure}

Figure \ref{fig:pred} shows the prediction interval coverage for 
each of the simulations in Section \ref{sec:results}.
For each rep of each simulation, we recorded the percentage of true 
responses that fell within a 95\% prediction interval of the predicted mean. 
We used those percentages to create the boxplots shown, with a dashed line 
at 95\% to represent nominal coverage. For both the Phantom and Star data sets, 
shown in the left two panels, 
all methods except the global GP show coverage that is similar to each other but slightly higher 
than the nominal rate. Higher than nominal coverage indicates that prediction intervals 
are wider than necessary. This is a potential cause for concern, but since coverage 
is similar for all methods we prefer the method that produces the lowest RMSPEs,
 MJGP.

The results for the Modified Michalewicz and AMHV Transport data sets, shown 
in the right two panels, show a different trend. Here we see that all methods have lower 
than nominal coverage. Starting with the Modified Michalewicz data set, we see that OJGP, 
in addition to having the highest RMSPEs, also has the worst interval coverage of all methods. 
Similarly, OLAGP, which had the second highest RMSPEs, also has the second lowest interval 
coverage. LAGP, although having higher RMSPEs than MJGP, has better interval coverage. So 
LAGP may be preferable to MJGP for the Modified Michalewicz data set depending on the task 
at hand.

For the AMHV Transport data set, we again see that all methods undercover, 
but to a lesser extent than for the Modified Michalewicz data set. Here we see OLAGP 
produces the lowest interval coverage, which again matches the RMSPE trends we saw. 
We suspect this is caused by the aforementioned issues of OLAGP with sparse data. 
LAGP is less affected by these issues, showing better coverage than OJGP, which had 
the second lowest RMSPEs but second worst interval coverage. Both the global GP and 
MJGP have the best coverage, producing (on average) slightly lower the nominal coverage. 
Since MJGP also produced the lowest RMSPEs, we find that it was the best performing model 
overall.

\end{document}